\documentclass[iop]{emulateapj}
\slugcomment{{\sc Accepted to ApJ:} Jan 2, 2013}
\usepackage{epsf}
\usepackage{graphicx}

\newcommand{\etal}{{et al.~}}
\newcommand{\masyr}{ \ {\rm{mas \ yr^{-1}}}\>}
\newcommand{\kms}{ \ {\rm{km \ s^{-1}}}\>}
\newcommand{\PM}{{\rm PM}}

\begin{document}
\bibliographystyle{apj}
\pagenumbering{arabic}
\title{Third-Epoch Magellanic Cloud Proper Motions I:\\
HST/WFC3 data and Orbit Implications}   
\shorttitle{Third-Epoch Magellanic Cloud Proper Motions I}
\author{Nitya Kallivayalil\altaffilmark{1,2}, Roeland P. van der Marel\altaffilmark{3}, Gurtina Besla\altaffilmark{4}, Jay Anderson\altaffilmark{3},  Charles Alcock\altaffilmark{5}}
\altaffiltext{1}{YCAA Prize Fellow, Yale Center for Astronomy \& Astrophysics,
260 Whitney Ave, New Haven, CT 06511}
\altaffiltext{2}{Dept. of Astronomy, University of
  Virginia, 530 McCormick Road, Charlottesville, VA 22904}
\altaffiltext{3}{Space Telescope Science Institute, 3700 San Martin Drive, Baltimore, MD 21218}
\altaffiltext{4}{Hubble Fellow, Columbia Astrophysics Laboratory, 1027 Pupin Hall, MC 5247, New York, NY 10027}
\altaffiltext{5}{Harvard-Smithsonian Center for Astrophysics, 60 Garden Street,
Cambridge, MA 02138}
\email{nitya.kallivayalil@yale.edu}

\begin{abstract}
We present proper motions for the Large \& Small Magellanic Clouds
(LMC \& SMC) based on three epochs of \textit{Hubble Space Telescope}
data, spanning a $\sim 7$ yr baseline, and centered on fields with
background QSOs. The first two epochs, the subject of past analyses,
were obtained with ACS/HRC, and have been reanalyzed here. The new
third epoch with WFC3/UVIS increases the time baseline and provides
better control of systematics. The three-epoch data yield proper
motion random errors of only 1--2\% per field. For the LMC this is
sufficient to constrain the internal proper motion dynamics, as will
be discussed in a separate paper. Here we focus on the implied
center-of-mass proper motions: $\mu_{W, {\rm LMC}} = -1.910 \pm 0.020
\masyr$, $\mu_{N, {\rm LMC}} = 0.229 \pm 0.047 \masyr$, and $\mu_{W,
  {\rm SMC}} = -0.772 \pm 0.063 \masyr$, $\mu_{N, {\rm SMC}} = -1.117
\pm 0.061 \masyr$. We combine the results with a revised understanding
of the solar motion in the Milky Way to derive Galactocentric
velocities: $v_{\rm tot,LMC} = 321 \pm 24 \kms$ and $v_{\rm tot,SMC} =
217 \pm 26 \kms$. Our proper motion uncertainties are now dominated by
limitations in our understanding of the internal kinematics and
geometry of the Clouds, and our velocity uncertainties are dominated
by distance errors. Orbit calculations for the Clouds around the Milky
Way allow a range of orbital periods, depending on the uncertain
masses of the Milky Way and LMC. Periods $\lesssim 4$ Gyr are ruled
out, which poses a challenge for traditional Magellanic Stream
models. First-infall orbits are preferred (as supported by other
arguments as well) if one imposes the requirement that the LMC and SMC
must have been a bound pair for at least several Gyr.
\end{abstract}


\section{Introduction}
\label{sec:intro}

High-precision proper motion (PM) measurements of the Large and Small
Magellanic Clouds made by our group with two epochs of Hubble Space
Telescope's (\textit{HST}) ACS High Resolution Camera (HRC) data
\citep[hereafter {\bf K2} \& {\bf K1}]{NK2, NK1}, and confirmed by
\citet[hereafter {\bf P08}]{P08}, have revolutionized the field of
Magellanic Clouds research. The implied tangential velocities were
high enough, approximately $100$ $\kms$ higher than in previous
theoretical models, and the observational errors were small enough
($\sim 0.07 \masyr$ for LMC), that an orbital solution in which the
Clouds are either only now on their first infall about the Milky Way
(MW) \citep[hereafter {\bf B07}]{Besla07}, or are on an eccentric,
long period ($> 6$ Gyr) orbit \citep[B07]{Shattow09} are the favored
solutions. In addition, the observed relative velocity between the
Clouds was of order the escape speed of the SMC from the LMC ($105 \pm
42$ ${\rm km \ s^{-1}}$; see K2) leaving open the possibility that the
Clouds may not be bound to each other, although bound orbits were
still allowed within the relatively large error bars which came from
the less-precise SMC PM determination ($\sim 0.18 \masyr$). These
results received much attention from the community in part because
they require a new formation mechanism for the Magellanic Stream, a
young coherent stream of H {\small I} gas that trails the Clouds $\sim
150^\circ$ across the sky \citep{Wannier72, Mathewson74, Putman03,
  Bruns05, Nidever10}. Most models, be they tidal or ram-pressure in
nature, require multiple pericentric passages in order to be viable
stripping mechanisms \citep{Ruzicka09, Bekki08, Connors06, Mastro05,
  GN96, Moore94, Murai80}.

We have recently put forth an alternative formation mechanism for the
Stream in which the material is removed by LMC tides acting on the SMC
before the system falls into the MW for the first time
\citep{Besla10}. A firm prediction of this model is that the Clouds
have been bound to each other and further that the SMC is on a highly
eccentric, prograde orbit about the LMC ($e=0.7$; apocenter $\sim 100$
kpc). This orbital configuration prevents the Clouds from merging and
also leaves the dispersion-supported material within the SMC's disk
radius relatively unaffected, while resonances aid in removing
rotationally supported material from the outskirts of the SMC's
disk. Thus, despite the large parameter space involved in modeling a
feature such as the Stream, there are two pieces of information that
can dramatically reduce the uncertainty in the models: 1) knowing
whether the Clouds are on a short or long period orbit about the MW,
and 2) the orbital eccentricity of the SMC's orbit about the LMC.  We
recently expanded on this work, building on earlier ideas of
\cite{Yoshizawa03} and \cite{Bekki05}, arguing that the internal
kinematics and structure of the LMC strongly favor a scenario in which
the MCs have recently experienced a direct collision
\citep{Besla12}. We also looked more generally at the implications of
such dwarf-dwarf interactions for Magellanic Irregulars, a class of
dwarf galaxies for which the LMC is a prototype. We found that
prograde dwarf-dwarf tidal interactions can efficiently remove baryons
from the lower mass companion and that structures such as off-centered
bars and one-armed spirals may be hallmarks of ongoing or recent
interactions with a low mass companion (mass ratio 1:10).

Apart from our own work on the formation of the Stream, these PM
measurements have spurred many other new theoretical investigations
into the Clouds' origin, whether from M31 or the far reaches of the
Local Group \citep{Shattow09, NK4, Yang10}. Attempts have been made to
measure the Milky Way mass from the kinematic properties of the LMC
\citep{NK4, Busha11a, Boylan-Kolchin11}. Recent studies have attempted
to place the orbits of the Magellanic Clouds in a cosmological
context, looking at the expected infall times in a statistical sense,
and favoring a scenario in which the Clouds are likely on their first
infall into the MW \citep[][see also \cite{Bekki11}]{Boylan-Kolchin11,
  Busha11b}. \cite{Rocha11} analyzed subhalos in the Via Lactea II
cosmological simulation and find that present day orbital energies are
tightly correlated with the time at which subhalos crossed into the
host's virial radius. Their analysis indicates that the LMC entered
the MW virial radius $\sim 4$ Gyr ago, in agreement with the
\cite{Busha11a} and \cite{Boylan-Kolchin11} results.

Overall, the \cite{Busha11a} and \cite{Boylan-Kolchin11} works find
different conclusions for the mass of the MW.  This is because Busha
\etal also folded in constraints on the separation between the Clouds
and the MW host which yields posterior distributions that favor lower
mass MW halos.  The Boylan-Kochin \etal study favors MW models with a
high mass because the K1 velocities imply the LMC is moving close to
the escape speed of a $10^{12} M_{\odot}$ halo. Since subhalos are
rarely ever accreted on unbound orbits, this means that the Bayesian
analysis will be weighted towards a higher mass. Consequently, they
favor a MW mass in excess of $2\times10^{12} M_{\odot}$. Both studies
represent a novel approach and new avenue for near-field cosmology,
i.e., using accurate orbital kinematics/histories of MW satellites to
constrain the properties of the MW.  Because the speed is so close to the
escape speed it is crucial that the velocity measurement be accurate for the LMC.
 A lower speed for the LMC would favor MW mass estimates
between 1--2 $\times 10^{12} M_{\odot}$.

In order to better address these questions about the Clouds'
velocities and likely orbits, we obtained an additional epoch of {\it
  HST} data with the Wide Field Camera 3 (WFC3).  With measurement
errors similar to those in epochs 1 \& 2, and an increased time
baseline from 2 to 7 yrs, this allows for decreased PM errors by a
factor of a few. In turn, this yields improved knowledge of the past
orbit. Moreover, a third epoch provides a valuable check on systematic
errors: there is always a straight line between two points, but if a
third point, particularly obtained with a completely different
detector, fails to line up then that is a clear sign of errors in the
analysis.

In this paper we present the results from the analysis of the third
epoch of {\it HST} data of the Magellanic Clouds, and the implications
for their space motion and past orbit. In $\S$\ref{sec:data} we
discuss the details of the observations, the analysis of the WFC3
data, an improved reanalysis of the ACS data, and the resulting
two-epoch and three-epoch PMs. In $\S$\ref{sec:PMresults} we derive
the PMs of the LMC and SMC Center of Mass (COM) from the measurements
of the individual fields. In $\S$\ref{sec:compPM} we compare the new
results to previously reported measurements and estimates. In
$\S$\ref{sec:spacemotions} we discuss the corresponding space motions
of the Clouds. In $\S$\ref{sec:orbit} we discuss the implications for
the orbits of the Clouds, and the new insights and improvements thus
obtained.  The main conclusions are summarized and discussed in
$\S$\ref{sec:conc}. Appendices discuss some additional \textit{HST}
data that we obtained, but which are not included in the present
study.

In a companion Paper~II we use the new LMC data to study its PM
rotation field. This provides new insights into the distance, center,
orientation, and rotation of the LMC disk. The COM motions derived in
the present paper use the field dependent corrections derived in
Paper~II.

\section{Data \& Analysis}
\label{sec:data}

\subsection{Description of Observations}
\label{sec:obs}

\cite{Geha03} identified a total of 54 QSOs behind the Clouds from
their optical variability in the MACHO database. These QSOs provided
the inertial reference frame against which the PM measurements were
made over a 2 yr baseline in K1 \& K2, based on
observations obtained between August 2002 and June 2005.  For most
efficient use of {\it HST} resources we decided to observe in SNAPSHOT
mode for both epochs. Our two SNAP programs of imaging with the HRC
achieved an overall completion rate of 48\% yielding a final sample of
21 QSOs behind the LMC and 5 behind the SMC.

We obtained an additional epoch of data during the period July 2007 to
November 2008. These observations executed with the HST WFPC2 camera,
due to the failure of ACS at that time. We briefly describe these data in
Appendix A, but due to their limited quality, we don't use them in the
analysis presented here.

We obtained another epoch of data with the HST WFC3 camera during the
period October 2009 to July 2010. These observations also executed in
SNAPSHOT mode, and we refer to them as the ``third epoch''. The final
yield from this program was 11 observed QSO fields for the LMC and 4
observed QSO fields for the SMC. However, two QSO fields, one in each
galaxy, proved unsuitable for PM determinations. As discussed in
Appendix B, in one field the QSO showed a bright extended host galaxy,
which complicates astrometry, and in the other field the nature of the
QSO was not confirmed. These two fields are omitted from the
subsequent discussion and analysis. Therefore, the final number of
QSOs used for the three-epoch analysis is 10 for the LMC and 3 for the
SMC.

The HRC data, which makes up the first two epochs of this study, is
discussed in detail in K1. Briefly, for our main astrometry
goals we chose the F606W filter which is a broad $V$ filter with high
throughput. Eight dithered exposures were taken in each epoch to
minimize pixelization-related systematic errors, and the dither
pattern was kept the same over the two epochs. Exposure times were
chosen so as to achieve a S/N of at least 100 for the QSOs based on
the known MACHO magnitudes at the time.

For the third epoch with WFC3/UVIS (PID 11730), we again used the
F606W filter and used the 4-point DITHER-BOX pattern which is made for
optimal sampling of the PSF. We aimed at $S/N\sim200$ for the QSOs in
order to match the astrometric quality of the ACS data.  Even though
the WFC3/UVIS pixels are bigger than ACS/HRC pixels, with adequate
dithering this should not degrade the astrometry because the
astrometric error is proportional to the $(FWHM/SNR)$ of the
target. With QSO brightnesses in the epoch 1 sample ranging from $16.5
\le V \le 22.0$, and an average brightness of $V=19.5$, this yielded
total science exposure times for epoch 3 ranging from 2.6 to 17.7 
minutes, with an average of 12 minutes. Tables~\ref{tab:LMCobs} \&
\ref{tab:SMCobs} describe the combined three-epoch dataset, including
RA/DECs for the QSOs, whether they were observed with WFC3 or not, the
total WFC3 exposure times, and the corresponding time baseline for the PM
measurement.

\begin{figure*}
\centerline{
\epsfxsize=0.54\hsize
\epsfbox{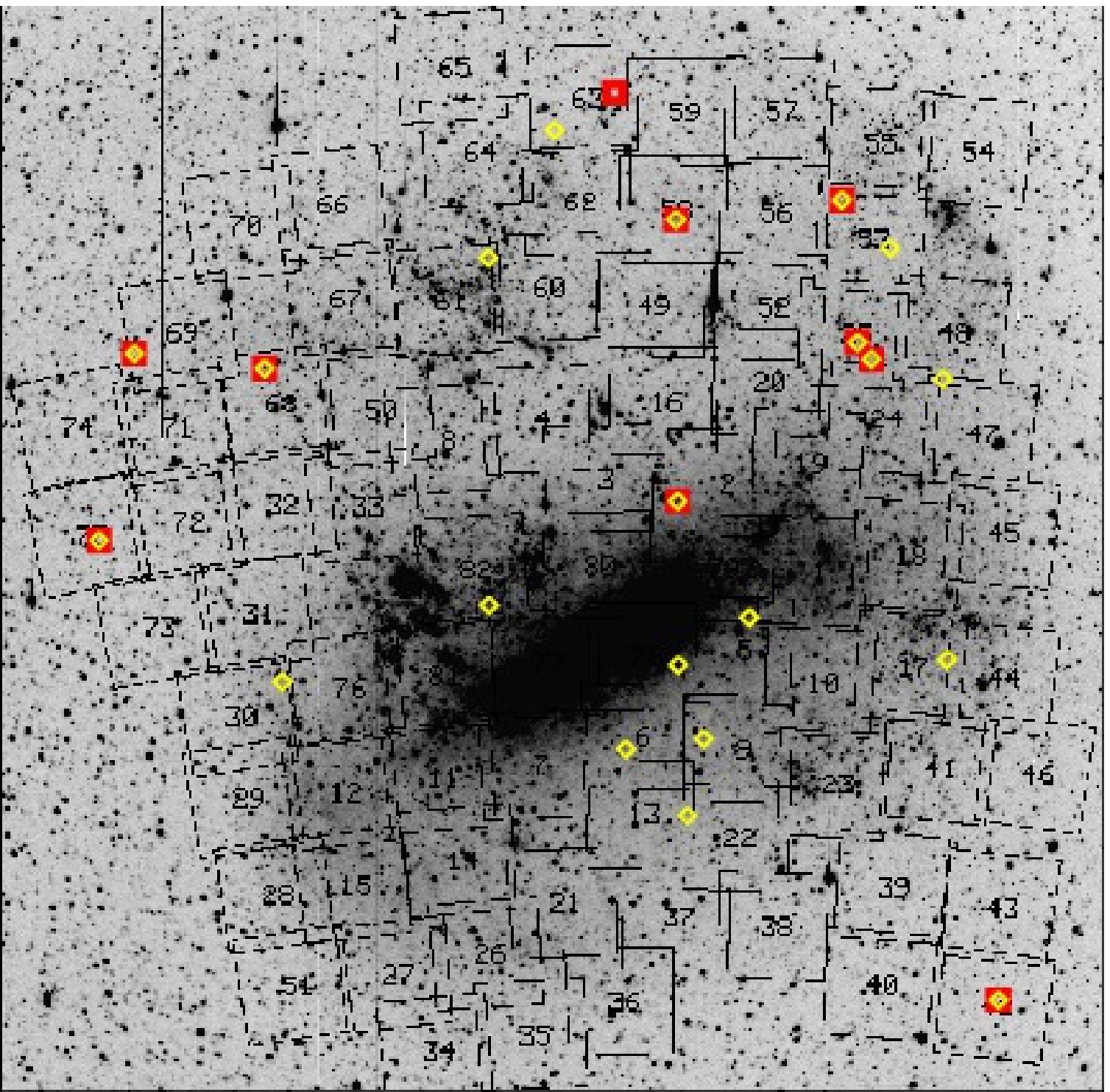}
\epsfxsize=0.4\hsize
\epsfbox{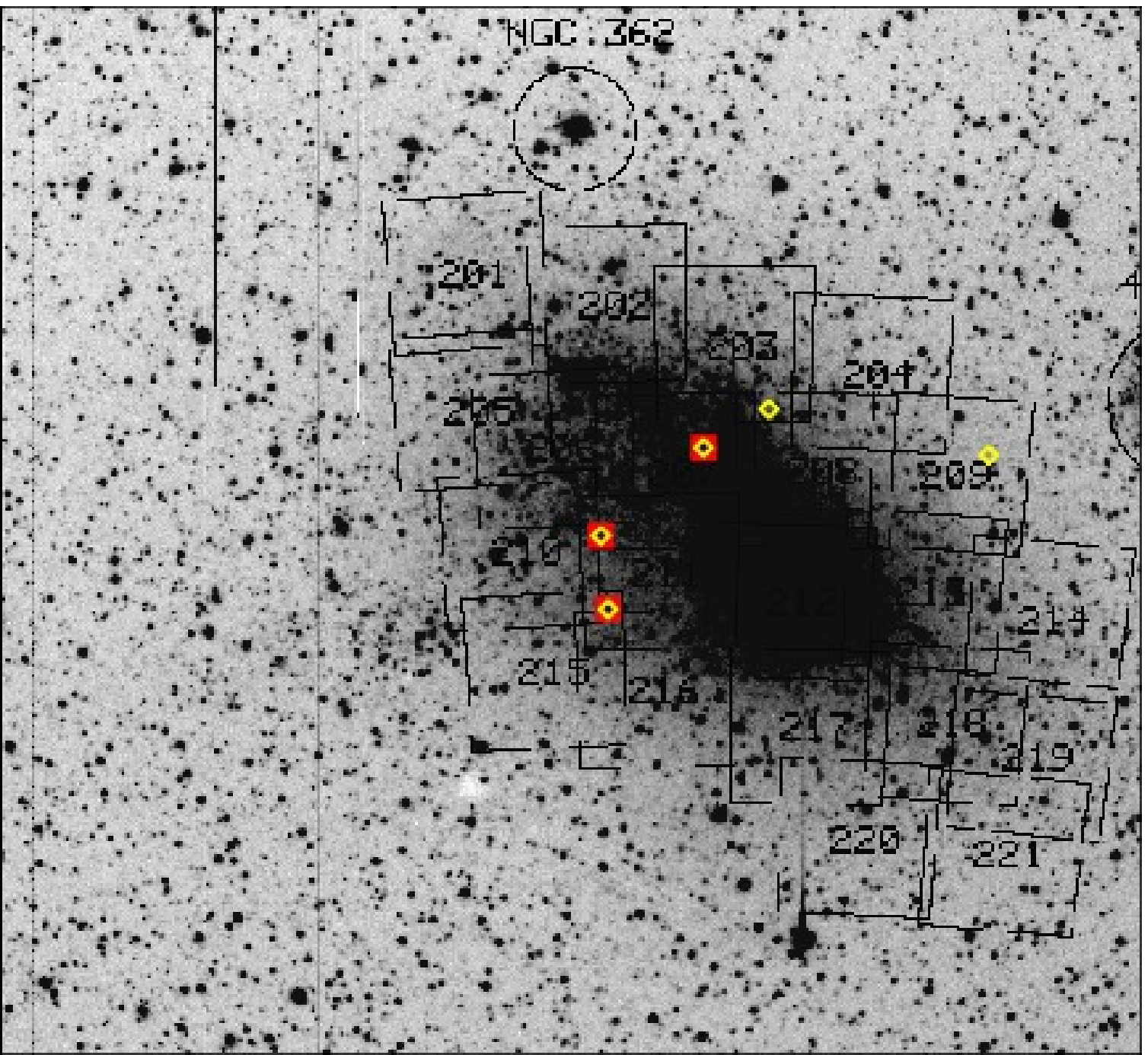}}
\caption{$R$-band image of the LMC ($8^{\circ} \times 8^{\circ}$), and
  the SMC ($3^{\circ} \times 5^{\circ}$). The MACHO photometric
  coverage is indicated. Red squares indicate reference QSOs with WFC3
  data and yellow diamonds indicate those with two epochs of ACS
  data. }
\label{fig:qsos}
\end{figure*}

Figure~\ref{fig:qsos} shows the locations of the QSOs behind the LMC and
SMC. QSOs for which we obtained two epochs of data are marked with
yellow diamonds. Their distribution behind the LMC is sparse but
reasonably uniform. QSOs observed with the WFC3 are marked with red
squares. Note that there is 1 ``new'' LMC field for which there is only
first epoch ACS and third epoch WFC3 data.

\subsection{Analysis of WFC3/UVIS Data}
\label{sec:WFC3}

As in K1 \& K2, we used the bias-subtracted,
dark-subtracted, flat-fielded images (\_flt.fits) provided by the
STScI data reduction pipeline for our basic reduction purposes.
\cite{AK04} found that these images are well-behaved astrometrically.
They are not, however, corrected for geometric distortion.  We did not
use the geometrically corrected products created by the MultiDrizzle
software (\_drz.fits) since they involve a re-sampling that might
degrade the astrometry.  Instead of geometrically correcting the
images, we geometrically corrected the positions that we measured on
the images. The positions were measured by means of empirical PSFs
constructed from other data sets (similar to the approach discussed by 
Anderson \& King 2006 in ACS/ISR 06-01). The geometric corrections
were performed according to \cite{Bellini11}. In their study, the
geometric distortion is modeled with two components: a third-order
polynomial and an empirically-derived look-up table for the residuals,
one per chip and filter combination. The method is akin to that used
in \cite{AK04} and is shown in \cite{Bellini11} to be stable over time
to better than 0.008 pixels (RMS) in each coordinate.

Since we are performing relative astrometry we have freedom in
constructing the masterframe (per QSO field) into which all relevant
star positions are to be transformed. While PSF-fitting was performed
on the flt images, we did use stars in the drz image as a starting
point to defining our masterframe, because their WCS header
information is the most accurate. Simple aperture photometry was
performed on the drz image to obtain initial positions and magnitudes
for real sources in the region of overlap of all 4 individual
exposures with orientation North up and East to the left. We then did
a simple transformation of these initial positions to an $(x,y)$-frame
of identical orientation, but with a chosen center and $25$ mas pixels
(for simplicity). A six-parameter linear transformation for the 4
individual flt WFC3 exposures into this frame was then
calculated. Based on previous experience, we only use stars brighter
than an instrumental magnitude, $M_{\rm INSTR} = -8$ ($S/N \sim
40$). The instrumental magnitude is defined as $-2.5 \log(counts)$ in
the F606W flt image (without application of a zero-point). We then
calculated a new masterframe position for each source by averaging the
linearly transformed WFC3 positions. At this stage we iterated on the
linear transformations, using only stars with measurement errors
smaller than $\sim 0.1$ pixels, that are common to all 4 WFC3
exposures. The positional errors are calculated as the RMS scatter
between multiple measurements of the same source.

In general, we can routinely center WFC3 stars to 0.02 pixels. Because
our exposures are deeper and the FOV is much larger than for ACS/HRC,
we identify on average $\sim200 - 800$ sources in the WFC3 images,
depending on the field. By contrast there are far fewer sources in
the HRC fields.

\subsection{Re-analysis of ACS/HRC Data}
\label{sec:ACS}

In our original K1 analysis we argued, based on detailed
calculations and consistency checks, that the degrading charge
transfer efficiency (CTE) of the HRC was not expected to
systematically affect our analysis. The main reason was that the SNAP
nature of the program yielded random roll angles of the telescope for
each QSO field.  Given that we had $N=21$ QSO fields with more or less
random detector orientations on the sky, any CTE induced astrometric
shift along columns would average down to zero $\propto
N^{-1/2}$. There is no explicit model for the underlying HRC CCD
detector physics and charge trap properties, as there now is for other
cameras \citep{Anderson10}. The residual CTE effects, given the random
roll angles of our exposures, would be expected to puff up the final
RMS of the measurements. We included this effect by using the RMS
scatter between fields as our final error estimates, instead of simply
propagating the per-field weighted errors (which would have suggested
a factor of $2$ higher accuracy). 

In K1, our final PM value was based on 13 fields that we deemed as
high quality for various reasons, and because of this reduced number
of fields, and the small number of total SMC fields, CTE issues
remained a legitimate concern. P08 subsequently performed an
independent re-analysis of our HRC data and used a simple correction
for CTE that was a function of $Y$ coordinate of the object, the time
since installation, and $\propto S/N^{-0.42}$. This latter exponent
came from the ACS Handbook. They found, as expected, that correcting
for CTE did not have a large systematic effect on the analysis (they
obtained a COM PM for the LMC within $1\sigma$ from that of
K1). However, they did obtain better field-to-field agreement. This
yielded smaller error bars on the COM motion, and the possibility to
derive a rudimentary PM rotation curve.

For the present study we reanalyzed the HRC data using our own new
prescription for CTE degradation as well as other improvements. The
main differences in this reanalysis versus the analysis in K1 \&
K2 are: (1) we use new codes for the implementation of
6-parameter linear transformations between the two epochs that employ
error-weighting for the stars. In the previous analysis we only used
rejection. (2) We employ a prescription to correct for CTE, the form
for which is as follows:

\begin{eqnarray}
y_{\rm NEW} = y_{\rm RAW} - (\alpha \times \frac{y}{1024.0} \times
\frac{t}{365.25}
   \times \nonumber \\\frac{{\rm MAX}[(M_{\rm INSTR}-M_{\rm INSTR,lim}),0]}{{\rm norm}}),
\label{eq:CTE}
\end{eqnarray}

where $y$ is the measured position of a star on the detector, $M_{\rm
  INSTR}$ is the instrumental magnitude, typically ranging from
$M_{\rm INSTR}=-6$ (faintest) to $-12$ (brightest) in our data, and
$t$ is the time since installation of ACS (Modified Julian Date =
52333).  CTE affects faint stars more than bright stars, and stars
brighter than $M_{\rm INSTR, lim}$ are assumed to be unaffected. Norm
is simply a normalizing factor discussed below. The slope, $\alpha$,
was fit in the following fashion: we sequentially corrected averaged
detector $y$ positions by a range of slopes.  After application of the
correction of a given slope, error-weighted transformations were
calculated for stars from epoch 2 into the epoch 1 frame. The input
error per epoch is the RMS/$\sqrt{N}$ obtained from centering the star
over $N$ exposures in that epoch. We then calculated what slope gave
the minimum total $\chi^2/DOF$ for these transformations summed over
all 21 QSO fields, where DOF stands for degrees of freedom, and is
defined as follows: $N_{\rm DOF} = N_{\rm data} - N_{\rm param}$,
where $N_{\rm data}$ is twice the number of fields and $N_{\rm param}$
is the number of free parameters optimized in the fit. The QSO is not
included in these transformations.

\begin{figure}
\begin{center}
\plotone{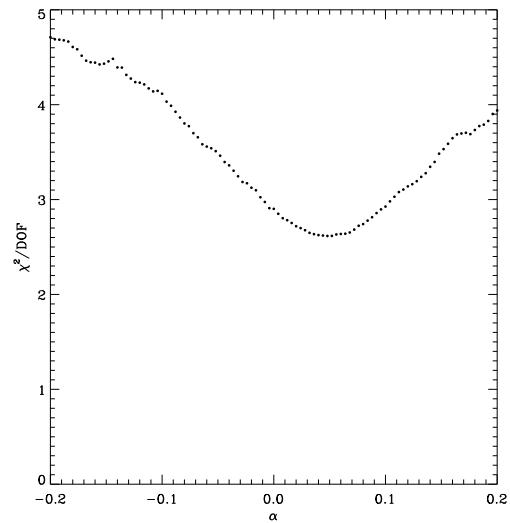}
\caption{The total $\chi^2/DOF$ for all the QSO fields as a function
  of CTE slope $\alpha$ (see Equation~\ref{eq:CTE}), for  
  $M_{\rm INSTR, lim} = -10$ and $norm = 4$. The minimum indicates 
  the value of $\alpha$ that provides the best correction.}
\label{fig:chisq}
\end{center}
\end{figure}

Figure~\ref{fig:chisq} shows the total $\chi^2/DOF$ for all the QSO
fields as a function of $\alpha$. We tried several $M_{\rm INSTR, lim}$
limits, $M_{\rm INSTR, lim}=(-10, -11, -12)$, where $-12$ corresponds roughly
to the brightest stars in the field. We normalized the magnitude
correction such that it is the greater of zero, when a star is brighter
than or equal to $-12$, or a fractional correction up to a size of 
approximately $1$, for $M_{\rm INSTR}=-6$ to $-12$. Therefore, for $M_{\rm
  INSTR,lim}=(-10, -11, -12)$, the corresponding values for norm are $(4,
5,6)$. This range of $M_{\rm INSTR,lim}$ did not affect the final PM values
by more than $1\sigma$.  In addition to fitting for CTE-slope, we
tried a number of different faint end magnitude cut-offs for the stars
used in the transformations, settling on only using stars brighter
than $M_{\rm INSTR}$ = -8. We also employed rejection in the form of
($\mid$residual/error$\mid< 3$), i.e. a cut based on the size of the residual
of the transformation of a star divided by its measurement error.

\begin{figure}
\begin{center}
\plotone{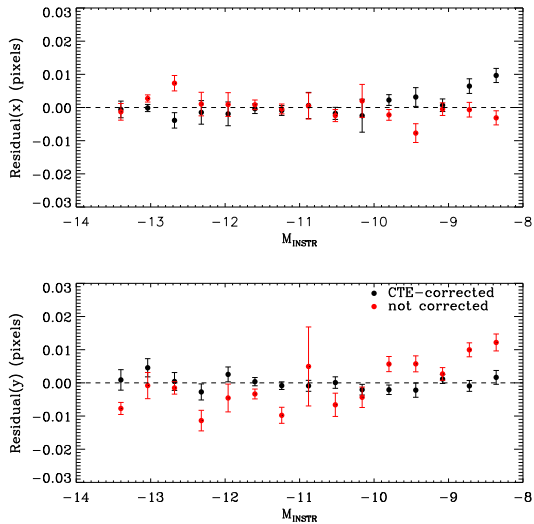}
\caption{Residual star PMs (pix) after CTE correction via
  Equation~(\ref{eq:CTE}) as a function of $M_{\rm INSTR}$ (black
  symbols) and plotted in the ACS e1 frame, i.e. $x$ and $y$ are in
  the native detector directions. The pixel scale is $28$
  mas/pix. Stars from all fields were binned together. The average PM
  should be zero by construction. Remaining trends are indicative of
  systematic errors. Points in red show the results without
  application of CTE corrections.}
\label{fig:acsresiduals}
\end{center}
\end{figure}

As a consistency check, we calculated the residual in pixels for each
Magellanic Cloud star between epoch 2 and epoch 1. This PM should be
zero on average, since differential streaming motions {\it within}
each field are too small to be resolved by our measurements. We
averaged the $y$-residuals over all stars in all QSO fields, and
binned the results by $M_{\rm INSTR}$. The QSO was again
excluded. Figure~\ref{fig:acsresiduals} shows the results. Points in
red correspond to $\alpha=0$ (i.e., no CTE-correction applied). A
clear trend with magnitude in the $y$-residuals reveals CTE-induced
astrometric shifts. By contrast the black points are for our best-fit
CTE-correction parameters in eq.~(\ref{eq:CTE}), which are $\alpha =
0.048$ for $M_{\rm INSTR, lim} = -10$ and $norm = 4$. There is no
evidence for any remaining $y$-CTE trend larger than $0.003$ pixels
(i.e., smaller than our random errors). There is no indication that
serial-transfer CTE in the $x$-direction might have significantly
affected the analysis, so we do not apply any corrections for it.

\subsection{Two-epoch Proper Motions}
\label{sec:2ePMs}
From hereon we define the proper motion in the west ($\mu_W$) and
north ($\mu_N$) directions in terms of the change in right ascension
($\alpha$) and declination ($\delta$) on the sky: $ \mu_W \equiv
-(d\alpha/dt)\,{\rm cos}(\delta), \mu_N \equiv d\delta/dt$. With the
final ACS to ACS transformation terms in hand, we determined the PM
for each source, including the QSO. A linear function with time was
fit to the available positions. The best-fit slope gives the PM. The
average PM of the stars in each field is zero by
construction. However, the QSO appears to move due to its reflex
motion with respect to the LMC/SMC stars in the foreground. Hence, the
average absolute PM of the stars in the field is simply minus the PM
of the QSO. The final error in this PM has two components. One
component is the error in the average PM of all the stars:
$\sigma_{\rm{\langle PM \rangle}}$\footnote{Although the average is
  formally zero by construction, the precision of this zeropoint is
  limited by the measurement errors of the stars at the two epochs.}.
This error quantifies how accurately we were able to align the
star-fields between the epochs. This was added in quadrature to the
formal error of the fit to the QSO's motion, $\delta \rm
{PM_{QSO}}$. So the error of a given field in either direction is
$\delta \mu_W$ or $\delta \mu_N = \sqrt{\delta \rm{PM_{QSO}}^{2}+
  \sigma_{\rm{\langle PM \rangle}}^{2}}$.

Columns 7--10 of Tables~\ref{tab:LMCobs} \& \ref{tab:SMCobs} list the
PMs thus inferred for all fields with two epochs of ACS data, and the
associated errors, $(\delta \mu_W, \delta \mu_N)$. These results are
very consistent with those of P08 even though we have used different
methodologies for PSF-fitting, linear transformations and
CTE-correction. The average residual between our reanalysis and that
of P08 is $0.08 \masyr$ lower in the West direction and $0.06 \masyr$
lower in the North direction, while the $\chi^2$ summed over all 21
fields is 12 in the West direction and 8 in the North direction. This
is acceptable compared to $N_{\rm DOF}$. The $1\sigma$ difference
between our final values and those of P08 arises from the inclusion of
error-weighting in the transformations, and not the specifics of the
CTE correction. If we exclude error-weighting, we recover values
closer to P08. Since there are some sparse fields in which only a
small number of stars are used in the transformations, it is
especially important to include error-weighting in the analysis.

\subsection{Three-epoch Proper Motions}
\label{sec:WFC3-ACS}

The inclusion of the WFC3 data gives us 20 images per QSO field for 15
fields: 8 ACS first epoch (e1), 8 ACS second epoch (e2), and 4 WFC3
third epoch (e3). One LMC field has only e1 and e3 data, and hence 12
images. To calculate PMs we first correct the ACS positions for CTE as
in $\S$\ref{sec:ACS}. We then perform preliminary transformations for
stars (brighter than $M_{\rm INSTR}=-8$) in each of the 20 images into
the masterframe defined in $\S$\ref{sec:WFC3}. All 20 resulting
$(x,y)$ centers of each star are averaged and the RMS is calculated to
make up the final masterframe and error (the QSO positions are not
averaged). We then treat this as the initial target frame for
transformations. As the input frame, we use the individual positions
of the stars in each of the 20 images, with input error the RMS per
epoch. Again, as in $\S$\ref{sec:ACS}, error-weighting, rejection, and
iteration are used in the determination of the final transformation
terms into the masterframe.

At this stage we lose the possible advantages of having such a large
FOV and high number of sources in e3 because we are limited by the
relatively sparse (and small) HRC fields. When we perform final
transformations to put e1, e2 and e3 into the masterframe, we only use
the sources that are common to the HRC. The number of sources used in
the transformations, after all cuts, is listed in column~6 of
Tables~\ref{tab:LMCobs} \& \ref{tab:SMCobs}. The variation in this
number reflects the real variation in stellar density at different
locations in the Clouds and is not an artifact of the analysis.

\begin{figure*}
\begin{center}
\epsfxsize=0.39\hsize
\epsfbox{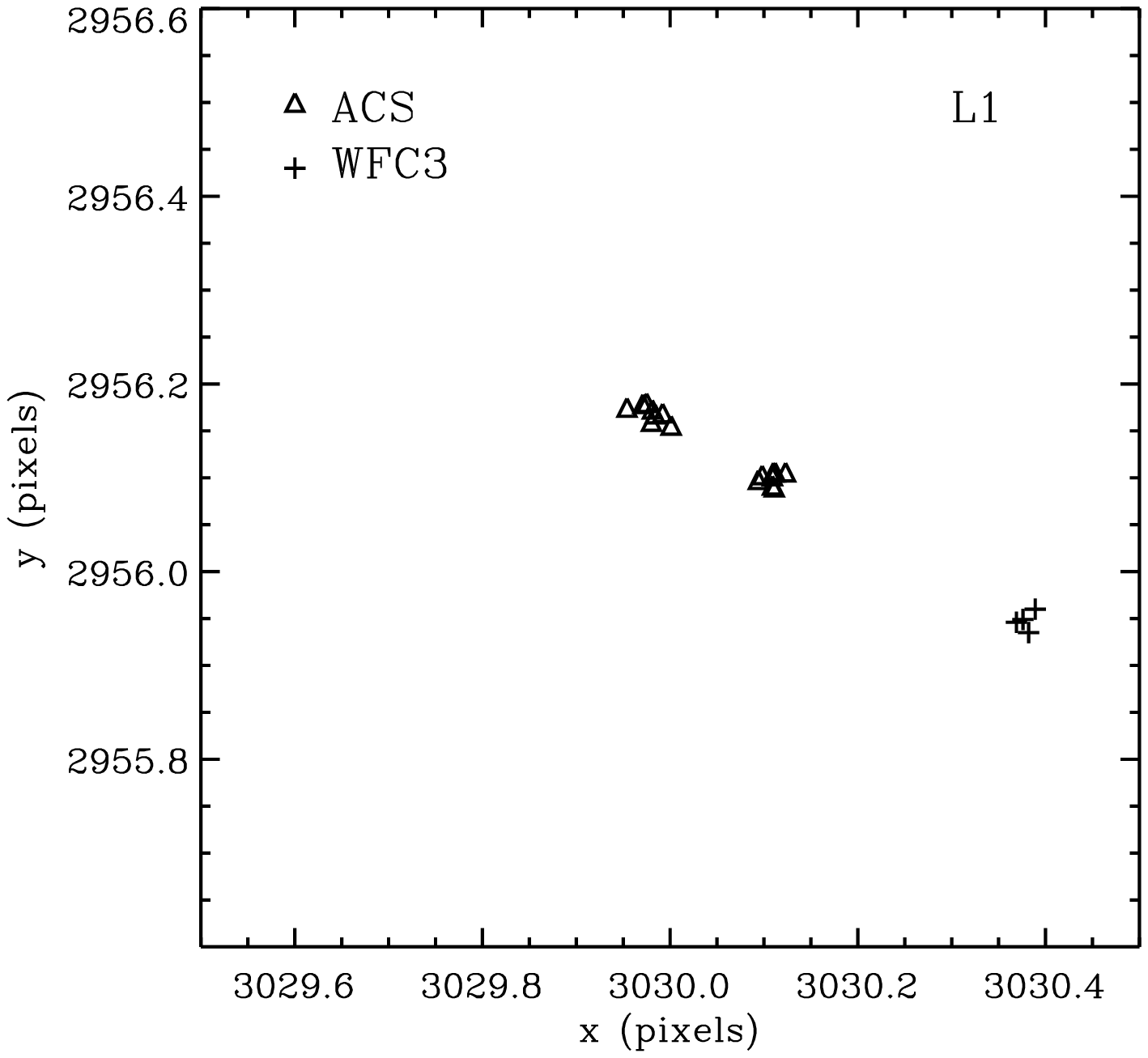}
\epsfxsize=0.39\hsize
\epsfbox{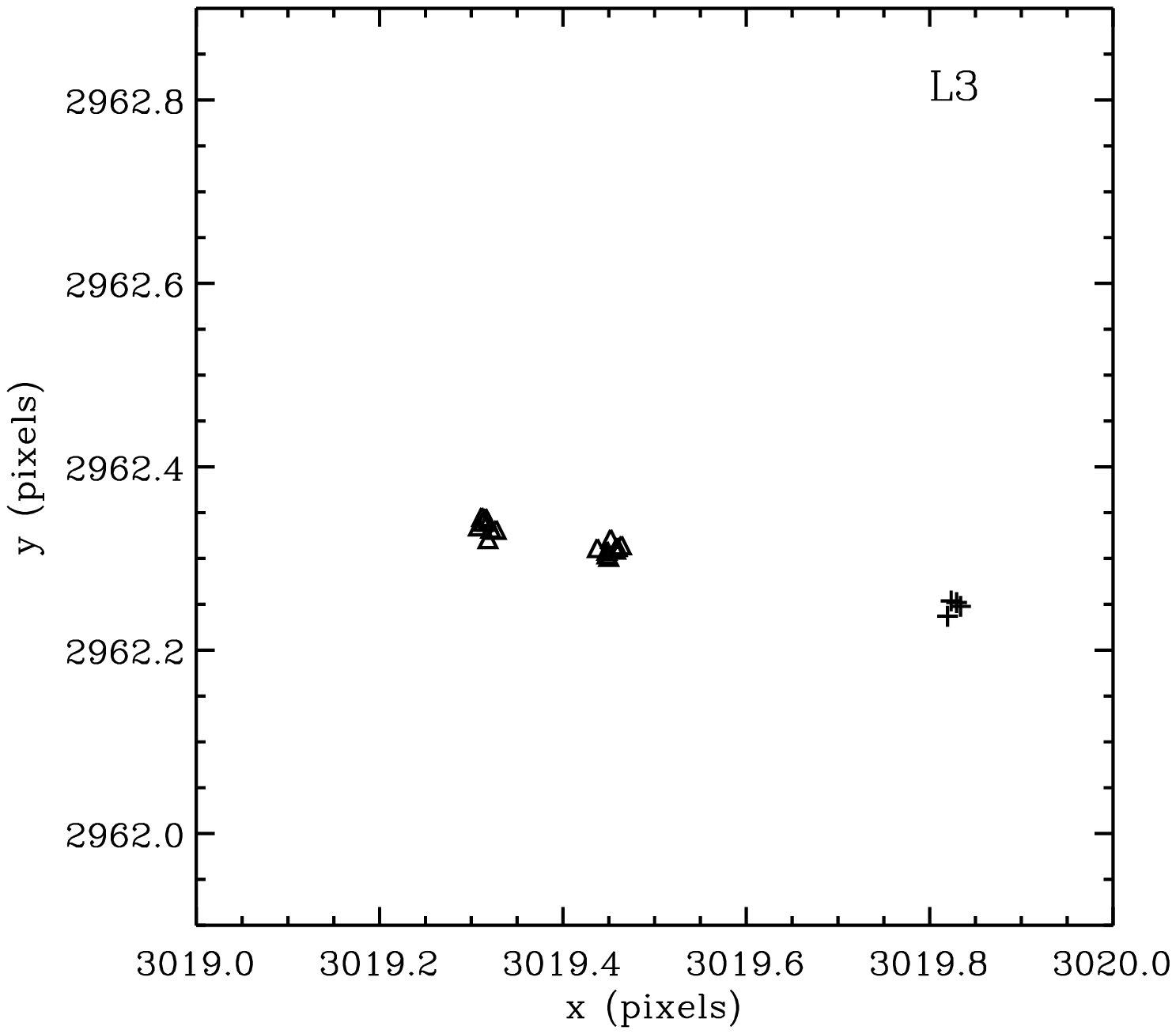}
\end{center}
\begin{center}
\epsfxsize=0.39\hsize
\epsfbox{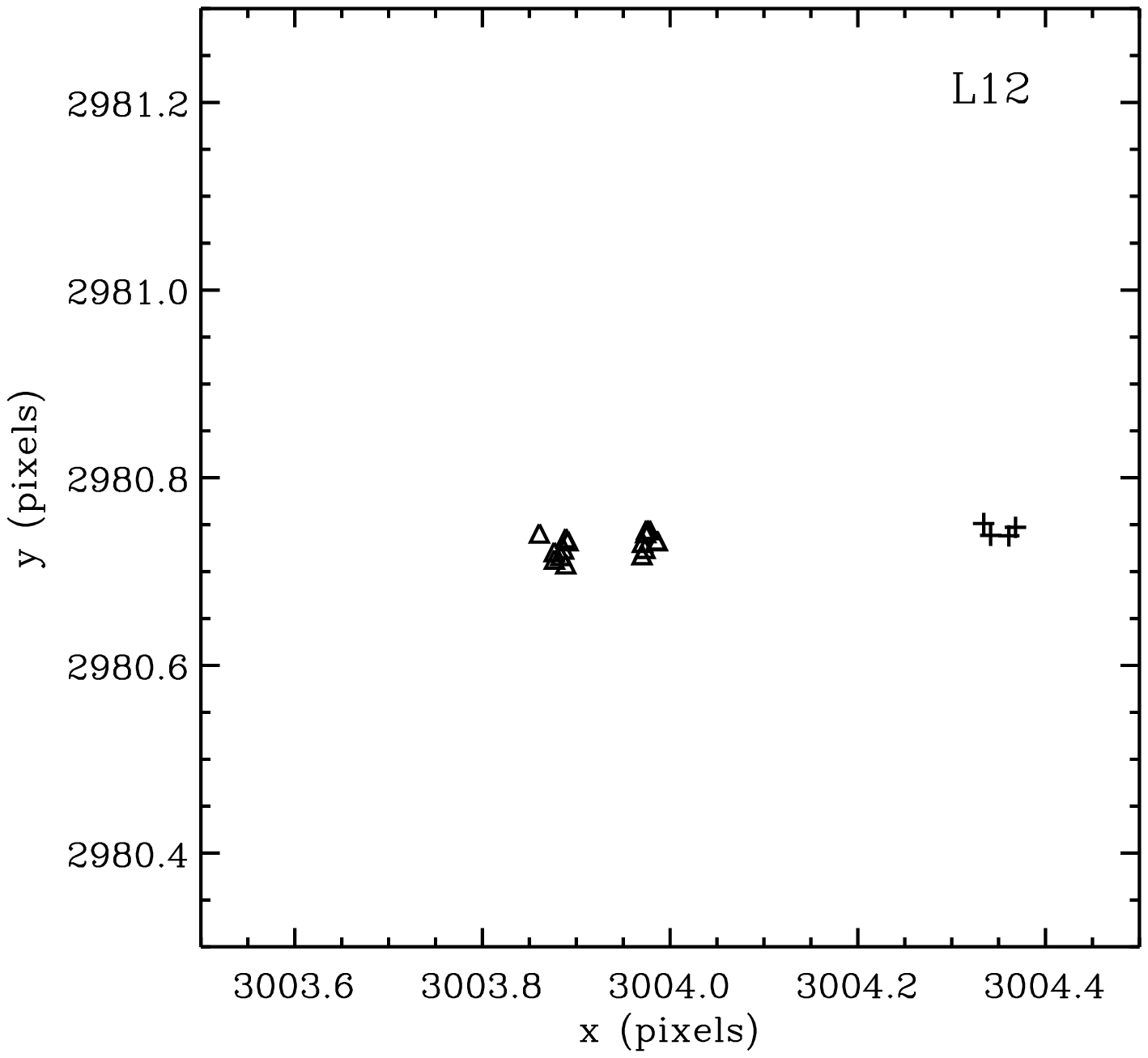}
\epsfxsize=0.39\hsize
\epsfbox{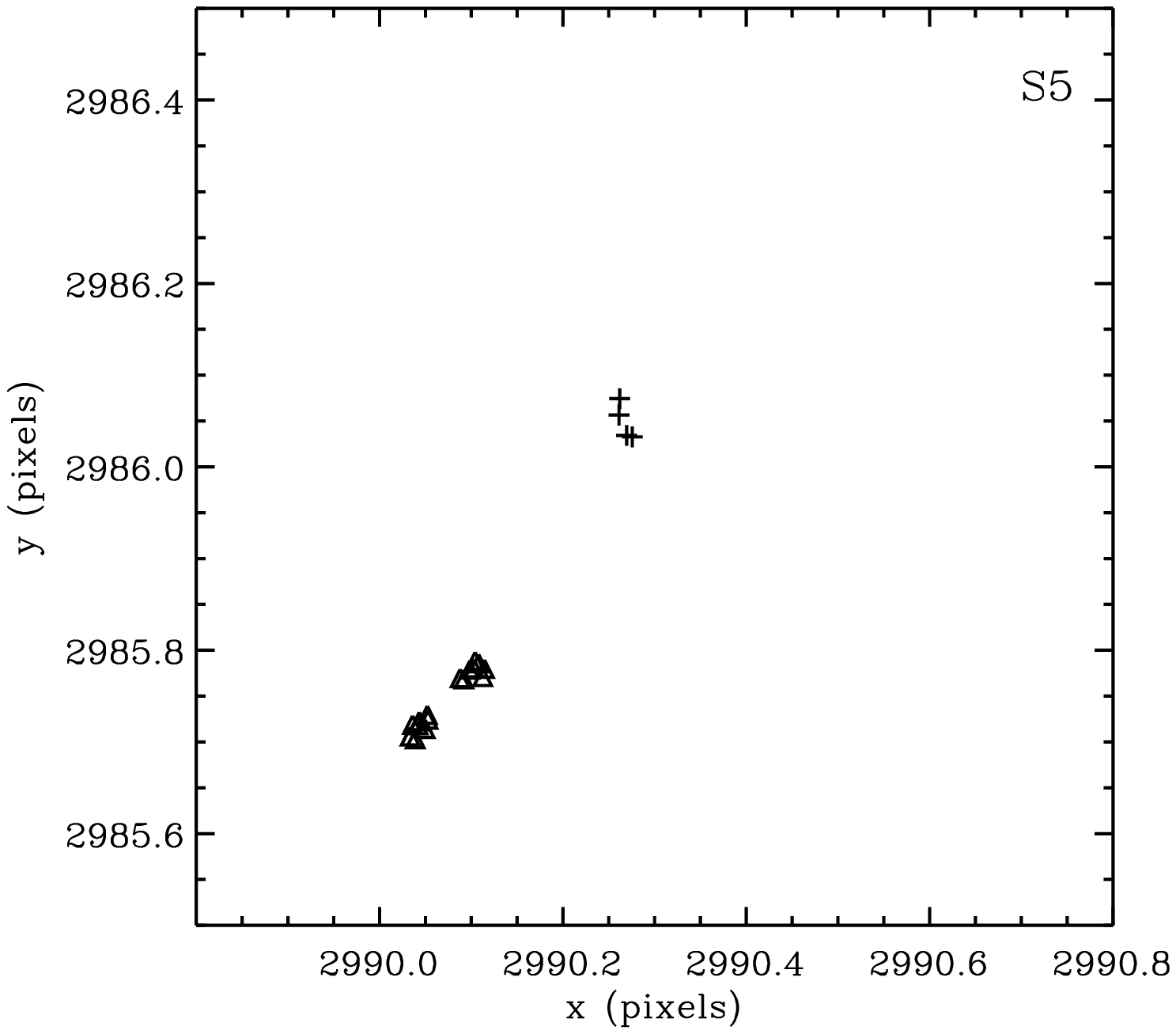}
\end{center}
\caption{The $x$ vs. $y$ (equivalent to $W,N$) positions in pixels of
  the QSO, after transformation into the masterframe, for 4 randomly
  chosen QSO fields for all 3 epochs of data. The fields are labeled
  according to Table~\ref{tab:LMCobs}. Linear motion (reflex motion of
  the L/SMC) is clearly visible, as is the high $S/N$ consistency
  across the three epochs. Each symbol represents 1 of the 20
  positions of the QSO relative to the starfield, and are
  distinguished here using triangles for ACS data and plus signs for
  WFC3 data. The uncertainty in centering the QSO within a given epoch
  is visible as scatter between the points. The first two ACS epochs
  are the ones that are most closely spaced in time (and QSO
  position). The QSO's reflex motion is predominantly towards the
  West, because the Magellanic Clouds move predominantly towards the
  East. Each panel is $1\times1$ pixel. 1 pixel = 25 mas.}
\label{fig:pmpix}
\end{figure*}

\begin{figure*}
\begin{center}
\plottwo{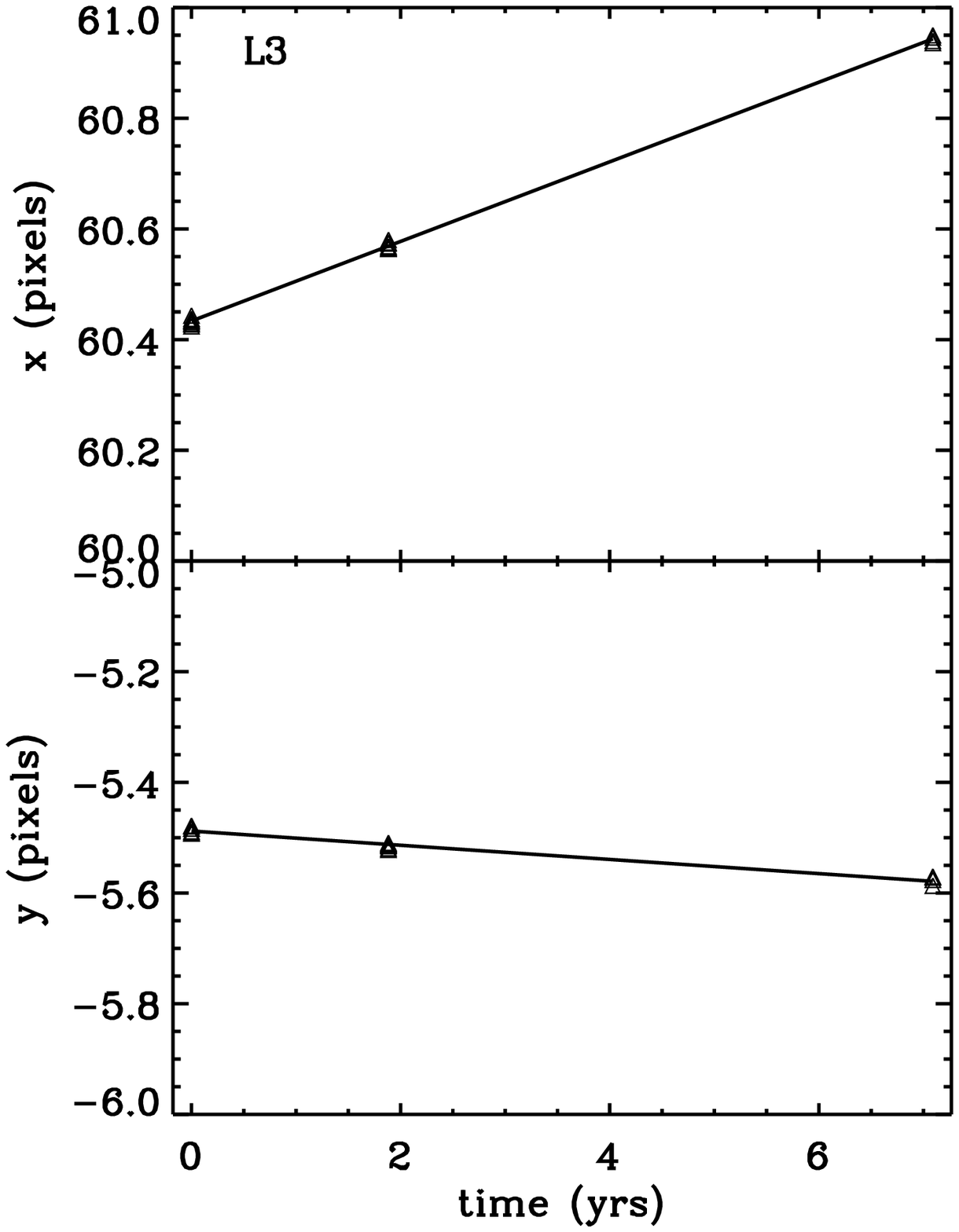}{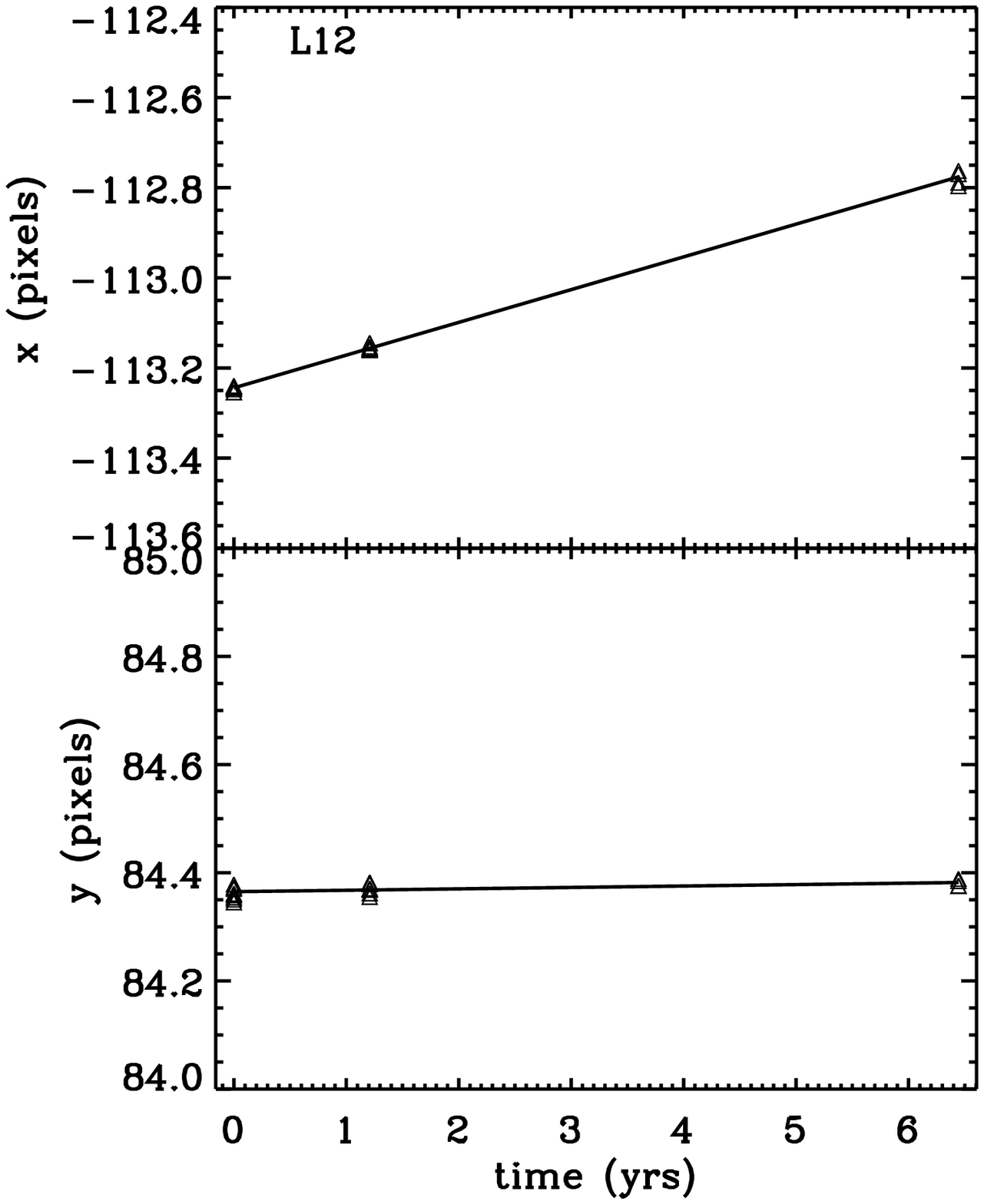}
\caption{Analogous to Figure~\ref{fig:pmpix} but now showing $x$ and
  $y$ positions of the QSO in pixels as a function of time for two of
  the LMC fields also in Figure~\ref{fig:pmpix}. The fitted linear
  motion is depicted by the solid lines. 1 pix = 25 mas.}
\label{fig:pmtime}
\end{center}
\end{figure*}

Figure~\ref{fig:pmpix} shows the $x$ vs. $y$ (equivalent to $W,N$)
positions in pixels of the QSO, after transformation into the
masterframe, for 4 randomly chosen QSO fields. The fields are labeled
according to Tables~\ref{tab:LMCobs} \& \ref{tab:SMCobs}. Linear
motion is clearly visible, as is the consistency across the three
epochs. Each triangle represents 1 of the 20 positions of the QSO
relative to the starfield. The scatter reflects the real error in
centering the QSO within a given epoch. In most cases the quality
(RMS) of the WFC3 data is comparable to that of HRC as we expected
based on the fact that we aimed for higher $S/N$ in e3 to account for
the larger pixels of the WFC3. This plot represents the reflex motion
of the L/SMC. The actual PM and its random error were determined as
described in $\S$\ref{sec:2ePMs} by fitting linear motion with time to
all sources. The high quality of the results is visible by eye in
Figure~\ref{fig:pmpix}. For two of the LMC fields in
Figure~\ref{fig:pmpix} we also show the $x$ and $y$ position in pixels
of the QSO as a function of time in Figure~\ref{fig:pmtime}. The
corresponding linear fits are also shown.

Columns 11--14 of Tables~\ref{tab:LMCobs} \& \ref{tab:SMCobs} list the
PMs thus inferred for all fields with both ACS and WFC3 data, and the
associated errors, $(\delta \mu_W, \delta \mu_N)$.

\section{Proper Motion Results}
\label{sec:PMresults}

\subsection{Comparison of Two-Epoch and Three-epoch PMs}
\label{subsec:comparison}

\begin{figure}
\begin{center}
\epsfxsize=0.9\hsize
\epsfbox{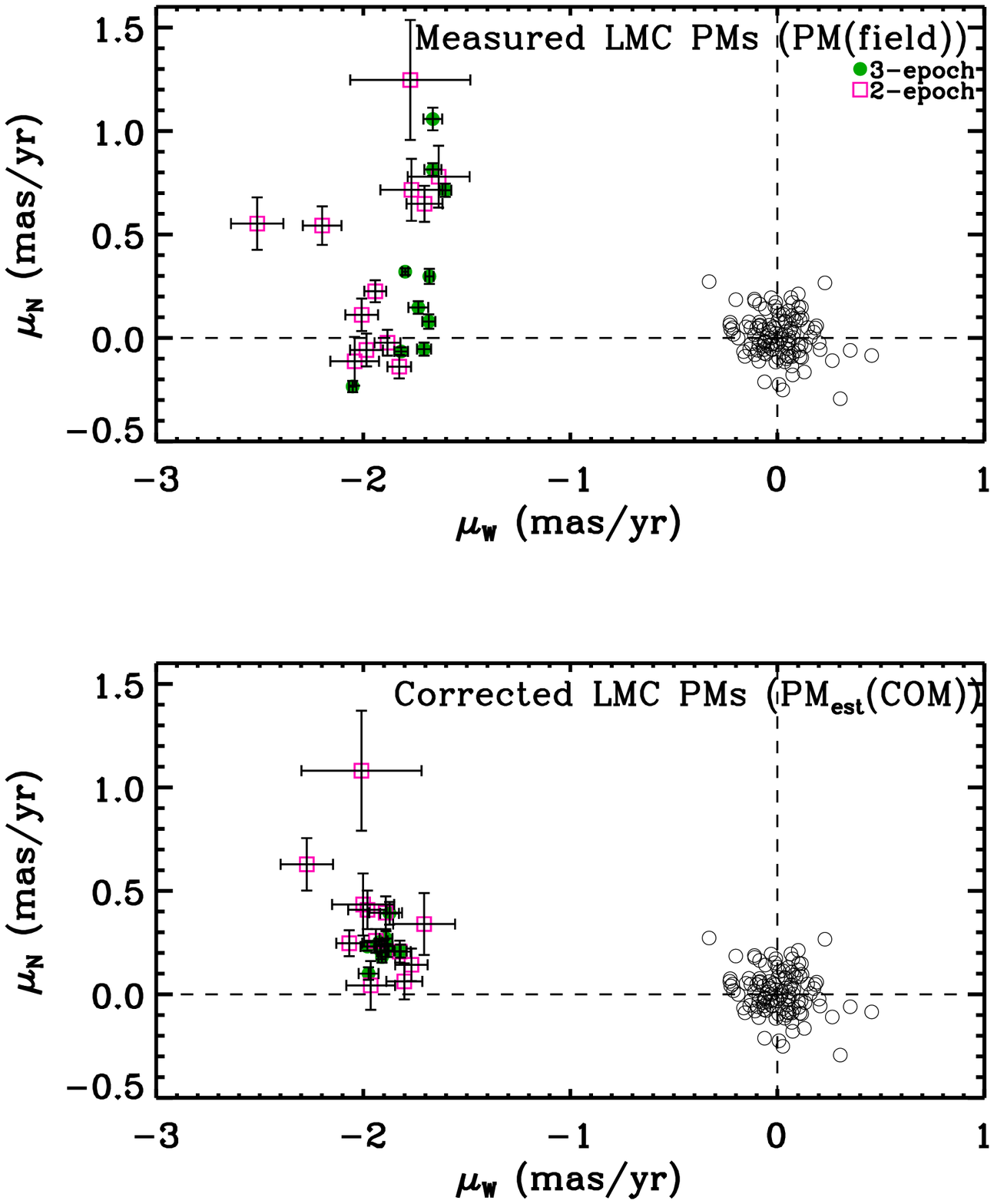}
\caption{({\it Top}) PMs for the QSO fields in the LMC (colored symbols) 
  in the ($\mu_W,\mu_N$)-plane, compared to the relative star PMs
  within the fields (open
  circles; centered around zero by construction).  The
  green filled circles show the results from linear fits to all three
  epochs of data and the pink squares show results from \S~\ref{sec:ACS} for
  e1 and e2 only.  ({\it Bottom}) The estimates $\PM_{\rm est}({\rm
    COM})$ after subtraction of the field-dependent contributions from
    viewing perspective and internal motions in the LMC determined in
    Paper~II.}
\label{fig:LMCPMs}
\end{center}
\end{figure}

\begin{figure}
\begin{center}
\epsfxsize=0.9\hsize
\epsfbox{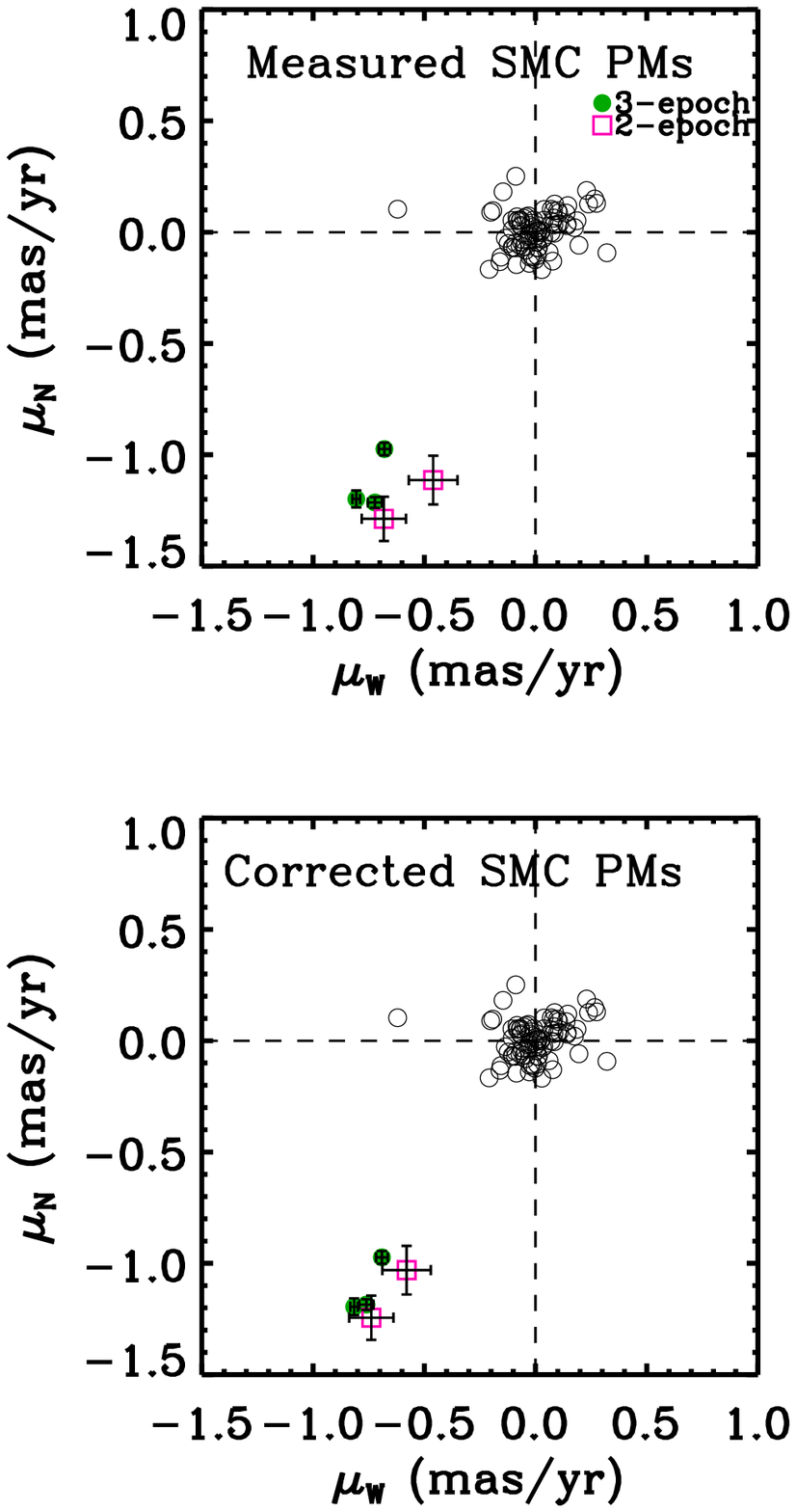}
\caption{({\it Top}) Measured QSO motions (colored symbols) with
  respect to stellar motions (open circles), similar to
  Figure~\ref{fig:LMCPMs}, but now for the SMC. Green filled circles
  show results from all three epochs of data, and the pink squares for
  e1 and e2 only. The bottom panel assumes $V_{\rm rot} = 0 \kms$.}
\label{fig:SMCPMs}
\end{center}
\end{figure}

In Figures~\ref{fig:LMCPMs} and \ref{fig:SMCPMs}({\it top}) we show
the measured QSO motions (colored symbols) in the
($\mu_W,\mu_N$)-plane compared to the star motions (open circles;
centered around zero by construction) for the LMC \& SMC
respectively. The stellar motions clearly separate from the QSO
motions. The green filled circles show the results from linear fits to
all three epochs of data, and the pink squares show results from
$\S$\ref{sec:ACS} for e1 and e2 only.  The fact that the averages are
consistent with each other is visible by eye. The addition of a third
epoch with a different {\it HST} instrument not only improves the accuracy
of the measured PMs, but also demonstrates that
there were no fundamental systematic problems with the earlier
two-epoch analysis of the observations in K1, K2, and P08.

\subsection{LMC COM Motion}
\label{subsec:LMCPM}

As discussed in K1, given the large size the LMC subtends on the
sky and the fact that we know it possesses internal rotation, the
motion observed for an individual QSO field must be written as: $
\PM({\rm field}) = \PM({\rm COM}) + \PM_{\rm res}({\rm field})$.  Here
$\PM({\rm COM})$ is the proper motion of the LMC COM and $\PM_{\rm
  res}({\rm field})$ is a field-dependent residual. The latter
contains contributions from different effects: (a) variations as a
function of position in the components of the three-dimensional
velocity vector of the COM that are seen in the plane of the sky
(``viewing perspective''), and (b) the internal rotation of the LMC
\citep[see formulae in][hereafter {\bf vdM02}]{vdM02}. Both
corrections make a similar contribution to the observed PMs. Given the
location of our fields, the viewing perspective contributes $\lesssim
0.4 \masyr$ and the internal LMC rotation contributes $\lesssim 0.3
\masyr$. This is less than $\sim 20$\% of the overall PM, but that is
significant given the accuracy that we are trying to achieve.

The field-dependent residuals depend on the COM PM and vice-versa and
hence we used an iterative procedure to evaluate them in K1.
For the internal rotation model of the LMC we used the model of 
  vdM02, based on the line-of-sight (LOS) velocity data for 1041
carbon stars. This was because our PMs themselves were not accurate
enough to significantly constrain the LMC rotation model. P08 took a
similar approach by leaving the LMC center and disk plane orientation
fixed at the values found by vdM02. However, they did vary and fit the
amplitude of the rotation curve.

With the addition of the WFC3 epoch, the per-field, per coordinate PM
errors are now typically $0.03 \masyr$ (see Table~\ref{tab:LMCobs},
columns 13 \& 14), which is only $\sim 7 \kms$. We show in Paper II
that this provides enough information to independently constrain all
parameters of the PM rotation field, without reference to prior
knowledge from LOS velocity studies. Table~\ref{tab:fit} lists the
results, and compares them to the values from vdM02 which were used by
K1 and P08. We also compare them to the values from a more recent
study by \cite{Olsen11}, who analyzed the LOS velocities of $\sim
6000$ massive red supergiants, oxygen-rich and carbon-rich AGB stars.
A detailed discussion of the commonalities and differences between the
results is presented in Paper II. We summarize here only the salient
features relevant to the determination of the LMC COM motion.

The inferred orientation of the LMC disk, as measured by the
inclination $i$ and the position angle $\theta$ of the line of nodes,
is consistent with the range of values previously reported using other
methods (see reviews of \cite{vdM06} and \cite{vdM09}). The rotation
curve amplitude inferred from the PM data falls between those
determined from the LOS velocities of (old) carbon stars
\citep[vdM02;][]{Olsen07} and (young) red supergiants \citep[][see
    Table~3]{Olsen11}. This is as expected, since the (bright) stars
  in our HST fields contain a mix of older and younger stars (see the
  color-magnitude diagram in Figure~6 of K1). This confirms the
  accuracy of the new PM data, both in a random and a systematic
  sense. The rotation curve amplitude inferred from our PM data is
  not, however, consistent with the result of P08. They were able to
  infer a rotation curve from the two-epoch data, but their rotation
  velocity was surprisingly high: $120 \pm 15 \kms$. This is
  approximately 30--$40 \kms$ higher than the value derived from the LOS
  velocities of H {\small I} and red supergiants \citep{Kim98,
    Olsen11}. It would be particularly hard to understand how the
  stars in the LMC could be rotating significantly faster than the H {\small I}
  gas.

The most important difference between our new results from Paper II
and the LOS study of vdM02 (used by K1 and P08) is
in the position of the LMC center. In Paper~II we find strong evidence
that the center of the stellar PM rotation field is consistent with
the position of the H {\small I} dynamical center. This makes
theoretical sense, since the stars and gas are orbiting in the same
gravitational potential. However, this result differs by
$1.12^{\circ}$ from the position advocated by vdM02, which
agrees with the brightest part of the LMC bar. The use of a new
rotation center affects the inferred PM of the LMC COM,
$(\mu_{W,LMC},\mu_{N,LMC})$, as discussed in
Section~\ref{subsec:compareHST}.

In Table~\ref{tab:LMCobs} we list for each field the estimate for the
LMC COM that is obtained after subtraction of the contributions from
viewing perspective and LMC rotation indicated by the best fit
model. In Figure~\ref{fig:LMCPMs}({\it bottom}) we show the estimates
$\PM_{\rm est}({\rm COM})$ resulting from this procedure. The scatter
is considerably reduced with respect to the top panel. The weighted
average of the estimates is $\mu_{W,LMC} = -1.910 \pm 0.008 \masyr$
and $\mu_{N,LMC} = 0.229 \pm 0.008 \masyr$. The error bars on these
values reflect only the propagation of the random errors in the
data. We show in Paper~II that there are also contributions from two
other sources. First, there is excess scatter between measurements
from different fields that is not accounted for by random errors, disk
rotation, and viewing perspective. The best-fitting model has a
reduced $(\chi^2_{\rm min}/N_{\rm DOF})^{1/2} = 1.8$; here
$\chi^2_{\rm min}$ is the $\chi^2$ value of the best-fit
model. Second, there are uncertainties in the geometry and rotation of
the best-fitting LMC disk model that need to be propagated. The latter
provide the dominant uncertainty in the final result. When these
sources of uncertainty are taken into account as described in
Paper~II, the final estimate of the LMC COM PM becomes $\mu_{W,LMC} =
-1.910 \pm 0.020 \masyr$ and $\mu_{N,LMC} = 0.229 \pm 0.047 \masyr$.

\subsection{SMC COM Motion}
\label{sec:SMCfield}

For the SMC we have PM measurements for only 5 QSO fields, of which
only 3 fields have three-epoch data. This sparse coverage cannot
provide much insight into the SMC geometry and rotation field. We
therefore fit a relatively simple model to the PM data in order to
determine the SMC COM PM $(\mu_{W,SMC},\mu_{N,SMC})$.

To calculate a PM model prediction for each observed field, we keep
the SMC center fixed at the H {\small I} kinematical center $(\alpha,
\delta) = (16.25^{\circ}, -72.42^{\circ})$ \citep{Stanimirovic04}, the
distance modulus fixed at $m-M = 18.99$ \citep{Cioni00}, and the
radial velocity fixed at $v_{\rm sys} = 145.6 \kms$
\citep{Harris06}. We account for the influence of viewing perspective
as in vdM02. We also allow for the possibility of a single
overall rotation of velocity $V_{\rm rot}$ in the plane of the sky
(i.e., as though we were viewing a face-on disk). We treat the SMC COM
PM as a free parameter that is optimized by minimizing the $\chi^2$ of
the model fit to the data.

To determine the uncertainties on the best-fit model parameters, we
create pseudo-data in a Monte-Carlo sense from the best-fitting model,
with properties similar to the real data. For this we use the
observational error bars, but multiplied by a factor $(\chi^2_{\rm
  min}/N_{DOF})^{1/2}$ to account for the observed scatter between the
measurements. Many different pseudo-data sets are created that are
analyzed similarly to the real data set. The dispersions in the
inferred model parameters are a measure of the $1\sigma$ random errors
on the model parameters. In the Monte-Carlo simulations we also
propagate the uncertainties on the model parameters that are kept
fixed. We use $\Delta (m-M) = 0.10$, $\Delta v_{\rm sys} = 0.6 \kms$,
and an uncertainty of $0.2^{\circ}$ per coordinate in the plane of the
sky on the SMC center position.

LOS velocity studies provide some insight into the importance of
rotation in the SMC. The old stellar population of the SMC, traced by
red giants, is consistent with a pressure-supported ($V < \sigma$)
spheroidal system with little evidence for rotation \citep[$V_{\rm
    rot} < 17 \kms$;][]{Harris06}. This is consistent with the fact
that many studies have found a large LOS depth in the SMC (e.g.,
\cite{Crowl01}). By contrast the H {\small I} gas in the SMC shows
rotation with an amplitude of $\sim 40 \kms$
\citep{Stanimirovic04}. This presumably indicates that the H {\small
  I} gas resides in a more disk-like distribution than the old
stars. Young stars (spectral types O, B, A) show evidence for a
rotation velocity gradient of similar magnitude as the H {\small I}
gas, but surprisingly, with a different major axis position angle
\citep{Evans08}.

The results of \cite{Harris06} suggest that it is reasonable to assume
that the SMC has limited rotation in the plane of the sky. If we
analyze the PM data with fixed $V_{\rm rot} = 0$, the inferred SMC COM
PM is $\mu_{W,LMC} = -0.772 \pm 0.033 \masyr$ and $\mu_{N,LMC} =
-1.117 \pm 0.043 \masyr$. This can be compared to the result of just
taking the weighted average of the 5 SMC PM data-points, which yields
$\mu_{W,LMC} = -0.754 \pm 0.013 \masyr$ and $\mu_{N,LMC} = -1.133 \pm
0.016 \masyr$. These results are statistically consistent, despite the
fact that the weighted average doesn't account for the influence of
viewing perspective. In Table~\ref{tab:SMCobs} we list the estimate
for the SMC COM that is obtained after subtraction of the contribution
from viewing perspective for each field. Taking the weighted average
of results for individual fields yields artificially low error bars,
because it doesn't account for the actual scatter between measurements
for different fields, or for uncertainties in the geometry and
rotation of the SMC.

To assess the importance of rotation, we have also performed model
fits in which the rotation is treated as a free parameter to be
determined from the PM measurements. This yields $V_{\rm rot} = 29.5
\pm 24.3 \kms$. The PM data are therefore reasonably consistent with
the absence of rotation, but they don't provide a very useful
constraint. The corresponding best-fit SMC COM PM estimate is
$\mu_{W,LMC} = -0.694 \pm 0.074 \masyr$ and $\mu_{N,LMC} = -1.055 \pm
0.078 \masyr$. This is consistent with the estimate obtained assuming
$V_{\rm rot} = 0 \kms$. However, the uncertainties are now significantly
increased, due to the allowed variations in the rotation model
corresponding to an amplitude variation $\Delta V_{\rm rot} = 24.3 \kms$.

As our final SMC COM PM estimate we adopt the result for $V_{\rm rot}
= 0 \kms$, but allowing for a $1\sigma$ Gaussian uncertainty  $\Delta
V_{\rm rot}$. Based on the preceeding discussion, the average star
used in our PM analysis probably rotates significantly less than the H
{\small I}. We therefore adopt somewhat arbitrarily $\Delta V_{\rm
  rot} = 15 \kms$. This yields for the SMC COM PM that $\mu_{W,LMC} =
-0.772 \pm 0.063 \masyr$ and $\mu_{N,LMC} = -1.117 \pm 0.061
\masyr$. The uncertainties are almost twice those quoted above,
obtained with $\Delta V_{\rm rot} = 0 \kms$. So as for the LMC, the quality
of the data is now such that the uncertainties in the structure of the
Cloud play an important role in how well we can establish the COM PM.

The value of $(\chi^2_{\rm min}/N_{DOF})^{1/2}$ for the best-fit SMC
model is $\sim 2.8$, independent of whether $V_{\rm rot}$ is treated
as a free parameter or not. By contrast, for the best-fit LMC model
derived in Paper~II, $(\chi^2_{\rm min}/N_{DOF})^{1/2} =
1.8$. Therefore, for both galaxies the measurements show somewhat more
scatter than what can be accounted for by the models. This is not
surprising, given the complexity of the data analysis, and the
relatively idealized nature of the models. Additional scatter might be
due, e.g., to subtle detector geometric distortion variations over
time, residual astrometric effects due to imperfect CTE, or more
complicated galaxy structures than are accounted for in the models. It
is intriguing that $(\chi^2_{\rm min}/N_{DOF})^{1/2}$ is larger for the
SMC than for the LMC. This might indicate that the structure and
kinematics of the SMC are more complex (or less in dynamical
equilibrium) than that of the LMC.

\section{Comparison to Previous PM Results}
\label{sec:compPM}

\subsection{Previous HST Results}
\label{subsec:compareHST}

PM measurements from various sources, including the present paper, are
summarized in Table~\ref{tab:PMs}. For the LMC, K1 obtained from
two-epochs of data that $\mu_{W,LMC} = -2.03 \pm 0.08 \masyr$ and
$\mu_{N,LMC} = 0.44 \pm 0.05 \masyr$. P08 obtained from the same
data that $\mu_{W,LMC} = -1.956 \pm 0.036 \masyr$ and $\mu_{N,LMC} =
0.435 \pm 0.036 \masyr$. If we analyze the new PM data with the same
{\it fixed} LMC center, inclination angle, and line-of-nodes position
angle from vdM02 used by K1 and P08, we obtain
$\mu_{W,LMC} = -1.899 \pm 0.017 \masyr$ and $\mu_{N,LMC} = 0.416 \pm
0.017 \masyr$. This shows that the addition of our new third epoch of
data, while keeping the LMC model the same, has the following effects:
(a) the random errors in the final LMC COM PM components decrease, as
expected based on the increased time baseline for about half the
fields; (b) the value of $\mu_{W,LMC}$ decreases by about $1.6\sigma$
compared to the earlier estimates; and (c) the value of $\mu_{N,LMC}$
remains unchanged compared to the earlier estimates, within the
uncertainties.

Despite this good consistency, our final LMC result of $\mu_{W,LMC} =
-1.910 \pm 0.020 \masyr$ and $\mu_{N,LMC} = 0.229 \pm 0.047 \masyr$
{\it deviates significantly} from the final K1 and P08
estimates. This is because of the different LMC model used here, and
in particular, the different position for the center. The difference
is primarily evident as a decrease by $\sim 0.20 \masyr$ in
$\mu_{N,LMC}$. By contrast, our new value for $\mu_{W,LMC}$ agrees
with that of P08 at the $1.1\sigma$ level. It is also worth
noting that our final best-fit estimate does not have uncertainties
that are much lower than those of P08, in particular in
$\mu_{N,LMC}$. However, the reason for this is that P08 and 
  K1 {\it underestimated} the uncertainties, by not propagating all
relevant uncertainties in the LMC disk model. Therefore, our new
values are in fact significantly more accurate (both in a random and
systematic sense) than the older ones.

For the SMC, our results are more similar to those of P08 than
K2. This is due primarily to the way in which we combined the results
for different fields in K2. On a field-by-field basis, K2 and P08
actually obtained rather similar results. In K2 we didn't explicitly
correct for the astrometric effects of CTE. So we reported a value
that was essentially an average of only 2 fields, because 3 of the 4
used QSO fields had been taken with the same roll angle at both
epochs. We also increased our errors to reflect a `true' systematic
plus random error, estimated from our analysis of the more numerous
LMC fields. The K2 result therefore had large errorbars. While our new
three-epoch result agrees with K2 for $\mu_{N,SMC}$, for $\mu_{W,SMC}$
it differs by $2.4\sigma$. Again, this is mainly because of the way in
which fields were combined in K2. The new three-epoch result, however,
includes CTE corrections and a longer time baseline, and is therefore
the more accurate one.

Our three-epoch result for the SMC agrees reasonably well with that of
P08. The value for $\mu_{W,SMC}$ agree within the uncertainties,
while for $\mu_{N,SMC}$ the results differ by $1.7\sigma$. This
agreement is similar as for the LMC, if one compares analyses with the
same assumed LMC center. So overall, there is acceptable internal
consistency between studies with different analysis methods,
scientific instruments, and time baselines.

\subsection{Ground-based Results}
\label{subsec:compareground}

A number of older ground-based (and Hipparcos) analyses of the LMC and
SMC PMs were discussed in K1 and K2. These studies had
large uncertainties, but within these uncertainties the results were
generally consistent with what we know now from \textit{HST} data.

There are now more recent results from two groups that are worth
discussing (see Table~\ref{tab:PMs}). \cite{Vieira10} measured the PMs
of the Clouds based on CCD and plate material from the Yale/San Juan
Southern Proper Motion program, spanning a baseline of 40 years, and
covering a large area in the inter-Cloud region as well. They
ultimately tie their reference frame to the ICRS. Within the
uncertainties of $\sim 0.3 \masyr$ per coordinate, their LMC and SMC
PM determinations agree with those presented here. They obtain a
stronger constraint on the relative motion between the Clouds, than on the
motion of either cloud individually: $(\mu_W, \mu_N)_{\rm SMC - LMC} =
(0.91, -1.49) \pm (0.16, 0.15)\masyr$. The corresponding difference between the
Clouds' motions from our new HST results is: $(\mu_W, \mu_N)_{\rm SMC -
  LMC} = (1.188, -1.383) \pm (0.039, 0.064)\masyr$. These results
agree for $\mu_N$, but differ by $1.7\sigma$ for $\mu_W$. Given the
totally different methodologies, this is a satisfactory
result. However, the uncertainties on the ground-based data are large
enough that they don't really help to validate the \textit{HST} data
at the level of its own uncertainties.

Costa \etal performed two recent studies using the 2.5m telescope at
Las Campanas and centered on background QSOs. In \cite{Costa09} they
measure 1 fairly outlying QSO in the LMC over a 5 year baseline. The
inferred LMC COM PM depends significantly on the assumed LMC rotation
velocity. In Table~\ref{tab:PMs} we quote the average of their results
for $V_{\rm rot}=50 \kms$ and $120 \kms$ (which corresponds to a
rotation velocity consistent with that derived in \cite{Olsen11} and
Paper~II). The uncertainties are 0.13--$0.15 \masyr$ per
coordinate. Their PM value differs from our new \textit{HST} result by
$1.5\sigma$ in $\mu_{W,LMC}$ and $1.8\sigma$ in $\mu_{N,LMC}$.  This
is more than expected based on random errors alone. However, we are
unable to evaluate our \textit{HST} results on the basis of this study, given that
we have many more quasars and typically higher accuracy.

\cite{Costa11} present an SMC COM PM estimate from 5 QSO fields over a
7 year baseline. This is similar to our study, but \textit{HST} of
course has better spatial resolution. Their PM value differs from our
new \textit{HST} result by $1.5\sigma$ in $\mu_{W,LMC}$, but agrees in
$\mu_{N,LMC}$ to within the uncertainties. Given the totally different
methodologies compared to our work, this too is a satisfactory
result. However, again the uncertainties on the ground-based data are
large enough that they don't really help to validate the \textit{HST}
data at the level of its own uncertainties.

\section{Galactocentric Velocities}
\label{sec:spacemotions}

The methodology for transforming an observed PM to a space motion in
the Galactocentric rest frame is described in vdM02. The correction
for the solar reflex motion requires knowledge of the Solar motion in
the Milky Way. In K1 and K2 we adopted the standard IAU value
\citep{Kerr86} for the circular velocity of the Local Standard of Rest
(LSR), $V_0 = 220 \kms$. However, models based on the PM of Sgr A*
\citep{Reid04} and masers in high-mass star-formation regions
\citep{Reid09} have suggested that the circular velocity may be
higher. \cite{Shattow09} argued that this may significantly affect the
orbit of the Magellanic Clouds. \cite{McMillan11} has presented a
Milky Way analysis that includes all relevant observational
constraints, from which he derived $V_0 = 239 \pm 5 \kms$. This is the
value that we will adopt for the present study. We also use the
\cite{McMillan11} value for the distance of the Sun from the Milky
Way, $R_0 = 8.29 \pm 0.16$ kpc. For the peculiar velocity of the Sun
with respect to the LSR we adopt the recent estimate from
\cite{Schonrich10}: $(U_{\rm pec}, V_{\rm pec}, W_{\rm pec}) = (11.1,
12.24, 7.25)$, with uncertainties of $(1.23,2.05,0.62) \kms$ (being
the quadrature sum of the random and systematic errors). Previous work
adopted the solar peculiar velocity from \cite{Dehnen98}. However,
there is now increasing evidence that $V_{\rm pec}$ from that study is
too small by $\sim 7 \kms$.  As a result, the solar velocity in the
Galactocentric $Y$-direction, $v_Y = V_0 + V_{\rm pec}$, is $251.2
\kms$ in our calculations here (consistent with the recent
\cite{Bovy12} study who find $v_Y = 242^{+10}_{-3} \kms$). By
contrast, it was $26 \kms$ lower in the calculations of K1 and K2.
This directly impacts the LMC and SMC space velocity.

In Table~\ref{tab:vels} we list the space velocity in the
Galactocentric rest frame of both the LMC (lines 1--6) and SMC (lines
7--10), implied by the various PM measurements available from
\textit{HST}.  We also list the corresponding relative velocity of the
SMC with respect to the LMC (lines 11--14). The relative velocity does
not depend on the assumed solar motion. However, the individual LMC
and SMC velocities do depend on the solar motion. For all PM
measurements we list (labeled ``IAU'') the results obtained under the
assumption of the IAU value of $V_0 = 220 \kms$ and the
\cite{Dehnen98} solar peculiar velocity. For the new three-epoch
measurements from the present paper we list also the results (labeled
``new'') obtained under the assumption of the \cite{McMillan11} value
of $V_0 = 239 \pm 5 \kms$ and the \cite{Schonrich10} solar peculiar
velocity. For all results we assume a distance uncertainty $\Delta m-M
= 0.1$ mag. For the IAU value of $V_0$ we also (arbitrarily) set the
uncertainty to be $\Delta V_0 = 5 \kms$; the actual uncertainty must
be significantly larger, if the McMillan (2011) results are indeed
correct. The space motions that we derive based on the K1, 
  K2 and P08 PMs, with the ``IAU'' solar motion, are consistent
with the values quoted in those papers. However, our errorbars are
generally larger and more accurate. This is because these previous
studies did not propagate the error in distance, which is in fact a
dominant error term (since $v \propto D \mu$).

The final results from the present paper are presented in lines 1, 7,
and 11 of Table~\ref{tab:vels}. This uses the three-epoch PM dataset.
For the LMC, it uses the LMC geometry and rotation determined from the
PMs in Paper~II. 
The results are characterized by the following properties. The
LMC has a total velocity (i.e., length of the Galactocentric velocity
vector) of $v_{\rm tot, LMC} = 321 \pm 24 \kms$. The SMC has a total
velocity $v_{\rm tot, SMC} = 217 \pm 26 \kms$. For both galaxies, the
radial velocity $v_{\rm rad}$ is much less than the tangential
velocity $v_{\rm tan}$. The value of $v_{\rm rad, LMC} > 0$, so that
the Clouds are past pericenter and moving towards apocenter. The
relative velocity of the SMC with respect to the LMC is $v_{\rm tot,
  rel} = 128 \pm 32 \kms$. The radial component of the relative
velocity is almost twice the tangential component of the relative
velocity. This implies a rather elliptical orbit, but a purely radial
orbit ($v_{\rm tan,rel} = 0$) is ruled out. The value of $v_{\rm rad,
  rel} > 0$, so that the SMC is past its pericenter with respect to
the LMC, and moving towards apocenter.

The motions derived by K1, K2, and P08 have been
used for various studies in recent years. It is therefore of interest
to examine how the new results differ from the old results, and why.

\smallskip 

\noindent {\bf Influence of solar velocity:} Comparison of lines 2 and
1 in Table~\ref{tab:vels} for the LMC, and lines 8 and 7 for the SMC,
shows that going from the old to the new values for the solar velocity
decreases all of $v_{\rm tot}$, $v_{\rm rad}$ and $v_{\rm tan}$, by
15--$22 \kms$. This is consistent with arguments made by
\cite{Shattow09} about the influence of the solar velocity on the
computation of the Clouds' orbits.

\smallskip

\noindent {\bf Influence of LMC center position:} Comparison of lines
3 and 1 shows that going from the vdM02 center (consistent with the
center of the LMC bar) to the new LMC center from Paper~II (consistent
with the H {\small I} dynamical center), reduces $v_{\rm tot}$,
$v_{\rm rad}$ and $v_{\rm tan}$ only minimally ($\sim 7
\kms$). However, $\mu_{N,LMC}$ decreases by $\sim 0.2 \masyr$ (see
Table~\ref{tab:PMs}) which equals $\sim 47 \kms$. This therefore
changes primarily the angle of the tangential velocity as seen in
projection from the Galactic Center, as previously pointed out by
B07. The change is such so as to better align the past orbit with the
location of the Magellanic Stream, but not to fully align it as we
will discuss in $\S$\ref{sec:orbit}.
 
\smallskip

\noindent {\bf Influence of third-epoch data for the LMC:} Comparison
of lines 6, 5 and 4 shows that (when using the same solar velocity and
the same parameters for the LMC geometry from vdM02) the
addition of the third epoch WFC3 data and the reanalysis of the
two-epoch ACS data decreases $v_{\rm tot}$ and $v_{\rm tan}$ for the
LMC by $\sim 30 \kms$, when going chronologically from K1, to
P08, to the present paper. These changes are due to a
progressive decrease in $|\mu_{W,LMC}|$.  The value of $v_{\rm rad}$
stays more-or-less the same.

\smallskip

\noindent {\bf Influence of third-epoch data for the SMC:} Comparison
of lines 10, 9 and 8 shows that (when using the same solar velocity)
the addition of the third epoch WFC3 data and the reanalysis of the
two-epoch ACS data decreases $v_{\rm tot}$ and $v_{\rm tan}$ for the
SMC by $\sim 65 \kms$ (a $\sim 1\sigma$ change from the K2 velocities
and errors), when going chronologically from K2, to P08, to the
present paper. These changes are due to how the K2 data from different
fields were combined, which leads to a large difference in the
inferred $\mu_W$ component compared to what is obtained from a simple
average of all the fields. The K2 combination strategy also increased
the error bars, however, which is why there is consistency at the
$1\sigma$ level. The new data also indicate a small decrease in
$|\mu_{N,SMC}|$ compared to P08. The value of $v_{\rm rad}$ changes by
less than $\sim 15 \kms$.

\smallskip

\noindent {\bf Influence of third-epoch data on velocity
  uncertainties:} Comparison of lines 6, 5, and 4, or 10, 9, and 8,
respectively, does not show much of a decrease in random errors on the
velocities, when going chronologically from K1/K2, to P08, to the
present paper. This is because the PMs are now accurate enough that
much of the velocity uncertainty is driven by distance
uncertainties. These distance uncertainties were ignored in these
previous papers, producing overly optimistic error bars.

\smallskip

Many of these individual influences change the Galactocentric
velocities in a similar direction. As a result, our new velocities are
quite different from those found by K1 \& K2.

\smallskip

\noindent {\bf New LMC velocity compared to K1:} Our new velocity has
lower $v_{\rm tot}$ and $v_{\rm tan}$ by $\sim 57 \kms$. This is due
to a combination of two effects: a $\sim 30 \kms$ decrease due to a
lower value of $|\mu_{W,LMC}|$ and a $\sim 27 \kms$ decrease to a
revised understanding of the solar velocity. The latter also decreases
$v_{\rm rad}$ by $\sim 25 \kms$. Moreover, we also derive a lower
value of $|\mu_{N,LMC}|$ due to a change in the derived LMC center.
The new $v_{\rm tot}$ is approximately halfway between the traditional
orbit required by tidal Magellanic Stream models in a logarithmic halo
\citep[][hereafter GN96]{GN96}, and the K1 result that led B07 to
suggest using $\Lambda$CDM-motivated halo models that the Clouds may
be on their first passage about the MW. We therefore perform new orbit
calculations in $\S$\ref{sec:orbit}.

\smallskip

\noindent {\bf New SMC velocity compared to K2:} Our new velocity has
lower $v_{\rm tot}$ and $v_{\rm tan}$ by $\sim 85 \kms$. This is due
to a combination of two effects: a $\sim 65 \kms$ decrease due to a
lower value of $|\mu_{W,SMC}|$ (mainly due to how the fields were
combined in K2) and a $\sim 20 \kms$ decrease to a revised
understanding of the solar velocity. The latter is also primarily
responsible for decreasing $v_{\rm rad}$ by $\sim 34 \kms$. There is
also a smaller contribution from the different SMC center used here
compared to K2.

\smallskip

\noindent {\bf New relative SMC-LMC velocity compared to K2:} The
total relative velocity $v_{\rm tot}$ has changed very
little. However, $v_{\rm rad}$ is now larger than $v_{\rm tan}$, so
the orbit is more elliptical than before. The value of $v_{\rm tot}$
is not quite as high as found by P08, but consistent within the
errors.

\section{Orbit Implications}
\label{sec:orbit}

The PMs and Galactocentric velocities from K1 \& K2 have been used in
many studies to look at the past orbit of the Magellanic Clouds with
respect to the MW, and the formation of the Magellanic Stream. As
discussed in the preceeding sections, the latest \textit{HST}
observations and data analysis, combined with the latest understanding
of the LMC geometry and solar velocity, imply decreased Galactocentric
velocities, and improved understanding of the observational errors.
We therefore rederive here the past orbit of the Magellanic Clouds
based on the new results.

\subsection{Methodology and Parameter Space}
\label{subsec:methodology}

We explore a total of 5 models for the MW: 3 static models with total
virial masses of $10^{12}$, $1.5 \times 10^{12}$, and $2\times10^{12}
M_{\odot}$; and two cases in which we allow for the mass evolution of
the MW as is expected from the hierarchical build up of mass in
$\Lambda$CDM. While it is beyond the scope of this work to provide an
exhaustive compilation of the observational determinations of MW mass,
we do want to motivate our adopted range by pointing to the work of
\cite{Gnedin10}, who obtain a MW virial mass of $1.6 \times 10^{12}
M_{\odot}$ if the middle value for their data-based circular velocity
determination is used. The uncertainty around this value is 20\%.

\begin{figure*}
\centerline{
\epsfxsize=0.35\hsize
\epsfbox{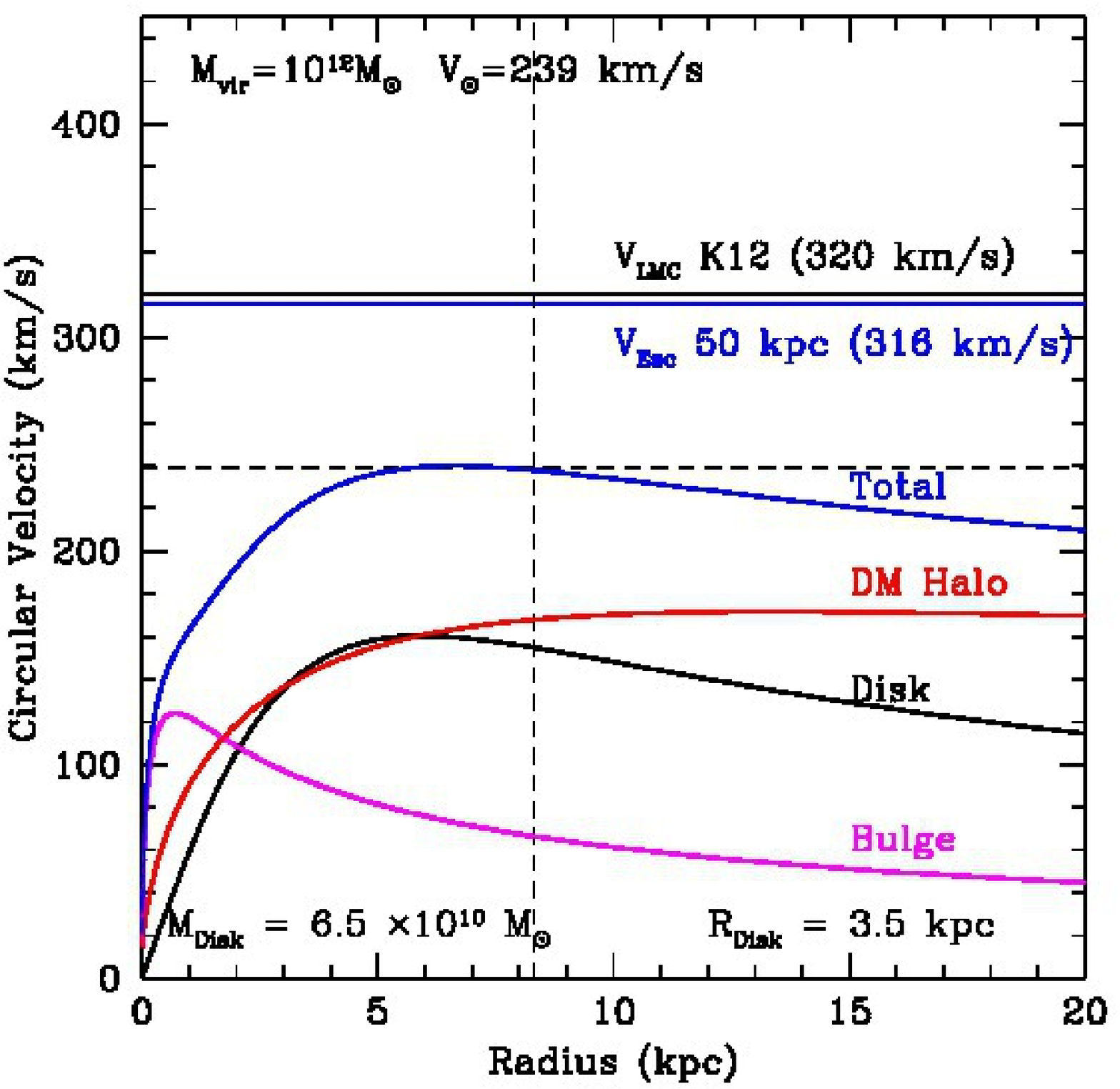}
\epsfxsize=0.35\hsize
\epsfbox{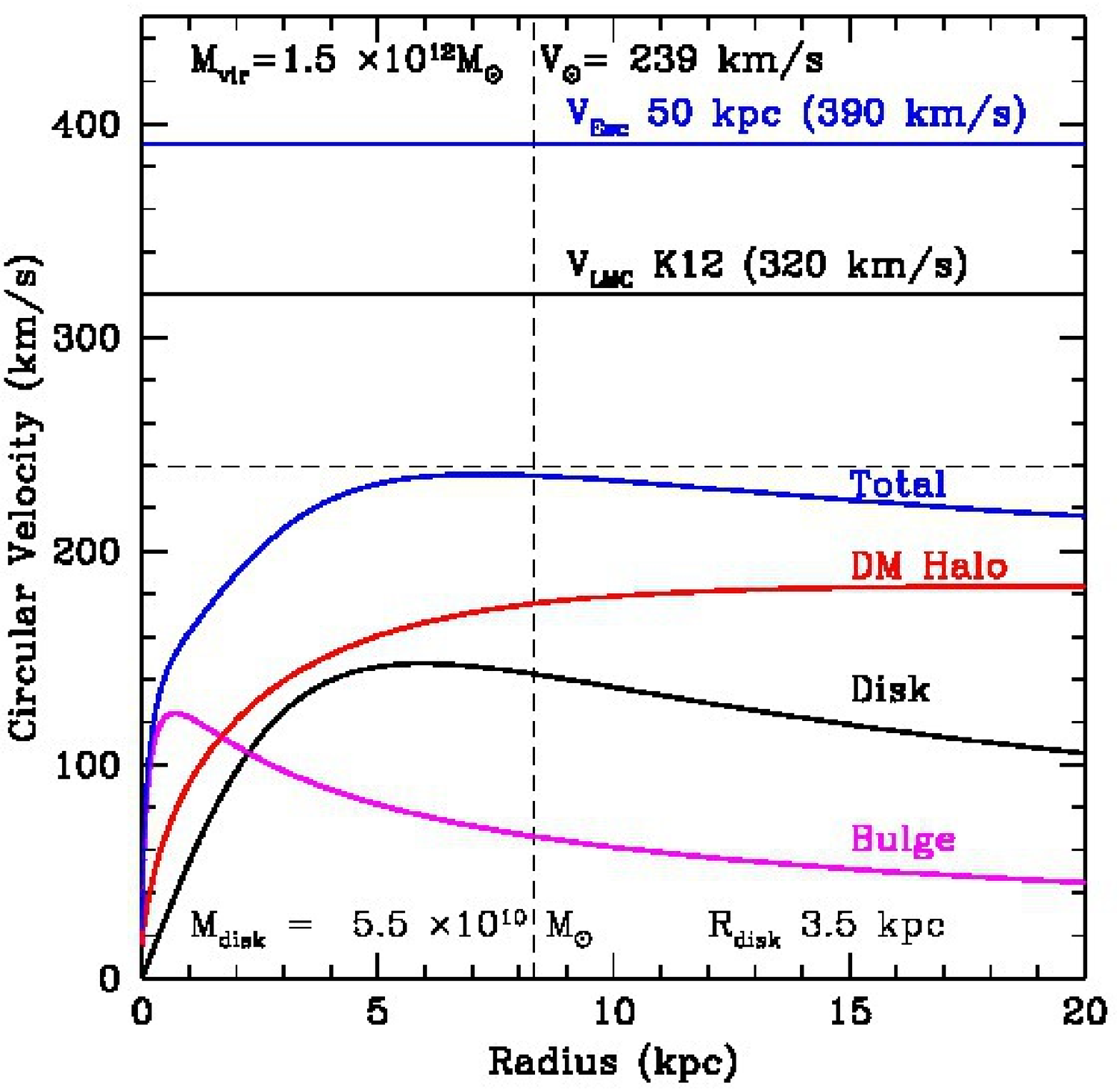}
\epsfxsize=0.35\hsize
\epsfbox{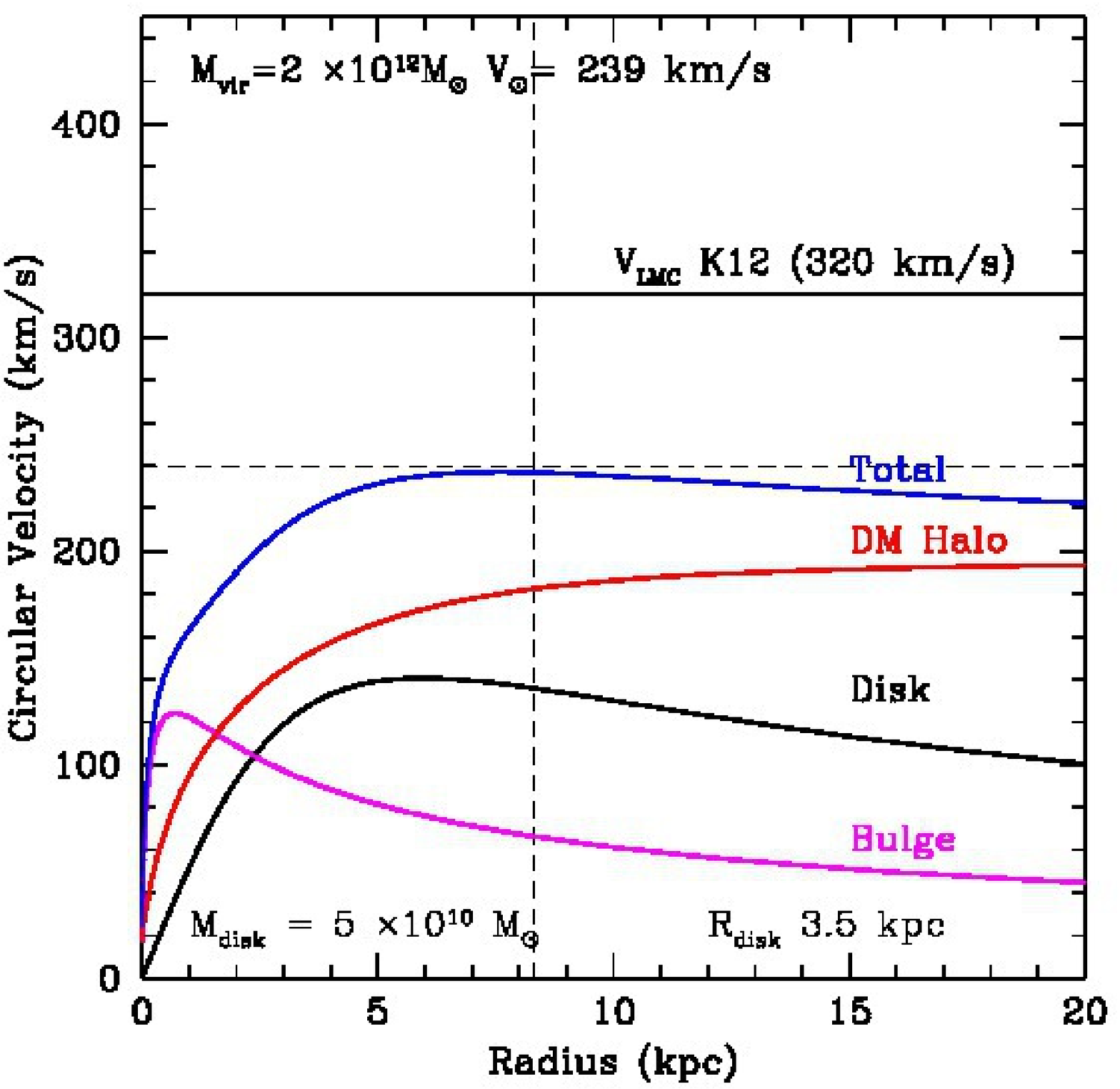}}
\caption{Rotation curves for the three static MW potentials that we
  explore, with MW virial mass = $1 \times 10^{12} M_{\odot}$
  (\textit{left}), $1.5 \times 10^{12} M_{\odot}$ (\textit{middle}),
  and $2 \times 10^{12} M_{\odot}$ (\textit{right}).  $V_{\odot}$, the
  Sun's velocity at the solar circle, is $239\kms$ in all three cases,
  as indicated indicated by the horizontal dashed line in each
  panel. $R_{\odot}$, the solar radius, is kept fixed at 8.29 kpc, as
  indicated by the vertical dashed lines. The panels also indicate the
  amount of mass in the MW disk, as well as its disk scale-length. The
  horizontal black line shows the mean third-epoch LMC velocity
  derived here. The horizontal blue line shows the value for the MW
  escape velocity at 50 kpc. In the $2\times 10^{12}M_{\odot}$ MW
  case, the escape velocity = $435 \kms$ (just outside plotting range).}
\label{fig:rotationcurves}
\end{figure*}

For the static MW cases we follow B07 and model the MW as an
axisymmetric three-component potential with a NFW halo, Miyamoto-Nagai
disk \citep{Miyamoto75} and a Hernquist bulge \citep{Hernquist90}. In
all cases the bulge is modeled with a scale radius of 0.7 kpc and
total mass of $10^{10} M_{\odot}$.  The NFW halo is adiabatically
contracted to account for the presence of the disk using the CONTRA
code \citep{Gnedin04}.  The NFW density profile is also truncated at
the virial radius (unlike in B07). While it is widely agreed that
halos should be truncated, exactly how and where the truncation is
done is arbitrary. The exact nature of this truncation does not affect
our conclusions, because we consider a one-parameter family of models
in which MW halo mass is considered a free parameter.

Halo and disk parameters are listed in Table~\ref{tab:MWLMCparams}. In
all cases the disk scale radius is kept fixed at 3.5 kpc while the
total disk mass is allowed to vary in order to reproduce the correct
circular velocity at the solar circle.  The MW circular velocity is
taken as the updated value of $239 \kms$ at the solar radius of 8.29
kpc \citep{McMillan11}, rather than the standard IAU value, in all
models, unlike B07. Rotation curves for each MW model are presented in
Figure~\ref{fig:rotationcurves}.  We follow the methodology outlined
in B07 to capture the effects of dynamical friction acting on the
Clouds owing to their passage through the MW's dark matter halo.

In the cases where the LMC is not on a first passage but has had a
past pericentric passage, the typical periods are quite large
($\gtrsim 5$ Gyr).  As such, we also want to investigate a simple
model that accounts for the MW's mass evolution over the past 10
Gyr. In this case we model the MW only as a NFW halo (neglecting the
disk, etc.). We model the mass evolution following the mean mass
growth rate with redshift from \cite{Fakhouri10}. The expected
$1\sigma$ scatter in this evolution at each redshift is roughly 20\%
\citep[from Figure~1 of][]{Boylan-Kolchin10}. We model the
concentration evolution following \cite{Klypin11}, and the scaling of
the virial radius with redshift and mass following relations in
\cite{Maller04}\footnote{The \cite{Fakhouri10} work is based on the
  Millenium II simulations and therefore uses an old value for
  $\sigma_8$, while the \cite{Klypin11} work uses the Bolshoi
  simulations which have an updated value for $\sigma_8$. However, we
  do not think that this inconsistency should affect MW mass objects
  very much, since the old value of $\sigma_8$ primarily suppresses
  power on low-mass scales.}. We consider current-day MW virial masses
of $1.5\times10^{12} M_{\odot}$ and $2\times10^{12} M_\odot$ in this
exercise.  We only explore the higher MW mass range because we are
interested in constraining whether the LMC is on a first infall (and
as we will show below, it is always on a first infall in the low MW
mass case).

Our expectation is that the LMC mass is the dominant uncertainty in
its orbital history (over MW mass evolution), since dynamical friction
will change the LMC's orbit on timescales shorter than the MW's mass
evolution\footnote{However, the MW's mass evolution can substantially
  modify solutions with orbital periods of order $\gtrsim 6$ Gyr, as
  the \cite{Fakhouri10} relations imply that the MW's halo should be
  $\sim$65\% as massive 6 Gyr ago \citep[see also][]{Moster12}.}. Therefore, for each static MW
mass model, a variety of LMC masses are explored, ranging from
$3\times 10^{10}$ to $2.5 \times 10^{11} M_{\odot}$ (the motivation
for which is discussed directly below).  This approach is different
from B07, where we only considered one LMC mass of $3 \times
10^{10}M_{\odot}$.

Our knowledge of the mass of the LMC is limited by the fact that
kinematic data are only available in the inner $\sim 9$ kpc. The
observational estimate of LMC total mass within this distance is $1.3
\times 10^{10} M_{\odot}$ \citep{vdM09}. For each mass case considered
here, the LMC is modeled as a Plummer sphere, where the softening
radius is chosen such that the total mass contained within 9 kpc is
$\sim 1.3 \times 10^{10} M_{\odot}$ as observed, and as listed
in Table~\ref{tab:MWLMCparams}.

\cite{Saha10} have detected stars out to 10 disk scale-lengths in the
LMC (at a scale-length of 1.4 kpc, this amounts to stars as far out as
15 kpc). For a bound stellar component to exist at a minimum distance
of 15 kpc from the LMC center-of-mass, the tidal radius must be at
least this large. In a $2\times 10^{12} M_{\odot}$ MW, the LMC's Roche
radius at a MW-LMC separation distance of 49.5 kpc reaches 15.4 kpc if
the LMC total mass is $3 \times 10^{10} M_{\odot}$. This is the reason
for our choice of lower bound in LMC mass. The highest total LMC mass
explored is chosen to match the observed present-day LMC stellar mass
and the relations for the expected infall mass from \cite{Guo10}. For
the evolving MW mass case we only consider LMC mass = $5 \times
10^{10} M_{\odot}$.  For this LMC mass we expect to find solutions
where the LMC has completed an orbit about the MW, and this thus
serves as a fiducial case for testing the plausibility of a non-first
infall scenario.

Whether the SMC is a binary companion to the LMC is primarily
determined by the LMC mass, which provides further motivation for
varying the LMC mass.  If the SMC is a long-term companion of the LMC,
it will have been tidally truncated fairly early on, maintaining a
roughly constant mass within its tidal radius at late
times. Therefore, we do not vary the SMC mass. The SMC is modeled as a
Plummer sphere of total mass $3 \times 10^{9} M_{\odot}$, with a
Plummer softening of 1 kpc so that the mass within 3.5 kpc is $2.4
\times 10^{9} M_{\odot}$ \citep{Stanimirovic04}.  We do not account
for dynamical friction acting on the SMC as it travels through the
dark matter halo of the LMC.  This is not significantly different than
the approach used in B07.

In our study we use the new 3rd epoch values and updated
solar motion for both the LMC and SMC, and their corresponding errors
(see lines 1 and 7 of Table~\ref{tab:vels}).

\subsection{Orbits about the Milky Way}
\label{subsec:MWorbits}

\begin{figure}
\begin{center}
\plotone{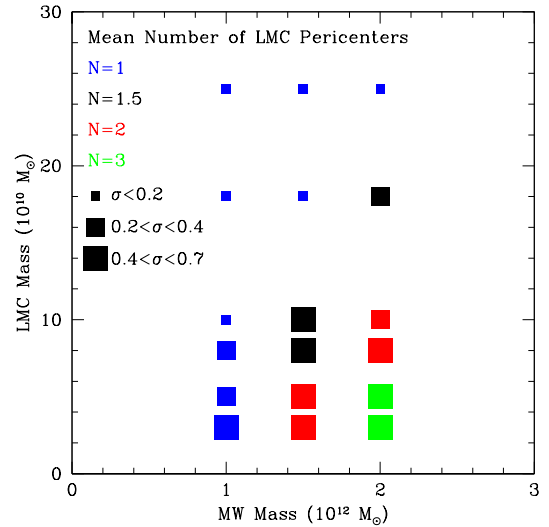}
\caption{The average number of pericentric passages that the LMC has
  made about the MW, as a function of LMC and MW mass, is indicated by
  different colors. Note that in all cases the LMC has at least made
  one pericentric passage since it is just at pericenter now,
  according to our data (N always starts at 1). The N=1.5 case
  represents orbits where there is roughly equal probability that the
  LMC is on first infall (N=1) or has completed one orbit (N=2). The
  point size indicates the RMS errors on the average value, as
  indicated in the legend by black symbols.}
\label{fig:LMCMWorbits}
\end{center}
\end{figure}

As in B07 we follow the orbit of the LMC backward in time, using the
present day velocities and positions and integrating the corresponding
equations of motion\footnote{We implicitly assume here that the LMC
  and SMC are a binary, and that therefore the more massive LMC sets
  the orbital path. We ignore the SMC in this part of the calculation
  (but see $\S$\ref{subsec:LMCSMCorbits}).}. For each combination of
MW and LMC mass, we explore 10,000 Monte-Carlo drawings from the LMC's
velocity error distribution.  We calculate the evolution of the LMC's
Galactocentric radius, and keep track of how many pericentric passages
it makes about the MW over a Hubble time, assuming the MW mass is
static over this time period.  In Figure~\ref{fig:LMCMWorbits} we show
the mean number of pericentric passages that the LMC has made about
the MW, as a function of LMC and MW mass.  Note that in all cases the
LMC has made at least one pericentric passage (N=1): our data indicate
that the LMC is currently just past pericenter, in agreement with
previous works.  This does not, however, imply that the LMC has made a
complete orbit about the MW.  The combination of LMC and MW masses
that yield, on average, solutions where the LMC has completed at least
one orbit are indicated by N$ \ge 2$; first infall scenarios are
indicated by N=1.  N=1.5 represent cases where there is roughly equal
probability that the LMC is on a first infall or that it makes one
orbit about the MW (50\% of the solutions have N=1 and 50\% have N=2).
The size of the square indicates the RMS error on the mean value, as
listed in the legend.

For the majority of the cases we consider here, the LMC makes no
additional pericentric passages about the MW than the one that it is
currently at (it is on a first infall: dark blue squares). As might be
expected, the LMC makes more pericentric passages if the MW mass is
higher (larger binding energy) and the LMC mass is lower (minimal
dynamical friction).  The dispersion in the mean values increases as
the mass of the MW increases; there are more solutions with one or
more complete orbits allowed within the error space within a Hubble
time for higher mass MW models.  The dispersion decreases as the LMC's
mass increases; for high mass LMC models \citep[consistent with $\Lambda$CDM
expectations,][]{Guo10} dynamical friction limits the number of
closed orbit solutions within the error space.  This forces the LMC on
increasingly eccentric orbits with orbital periods longer than a
Hubble time.

Note that we have ignored the mass evolution of the LMC owing to the
MW's tidal field in this analysis.  Because we have constrained all
LMC models to match the total observed mass of the LMC within 9 kpc,
higher mass LMC models will have more material at large radii and
should thus be more affected by the omission of this physical effect
than the lower mass LMC models.  However, Figure~\ref{fig:LMCMWorbits}
illustrates that if the LMC today is well-described by the high mass
models, it could not have made an earlier pericentric approach about
the MW. As such, our argument is self-consistent.

\begin{figure}
\begin{center}
\plotone{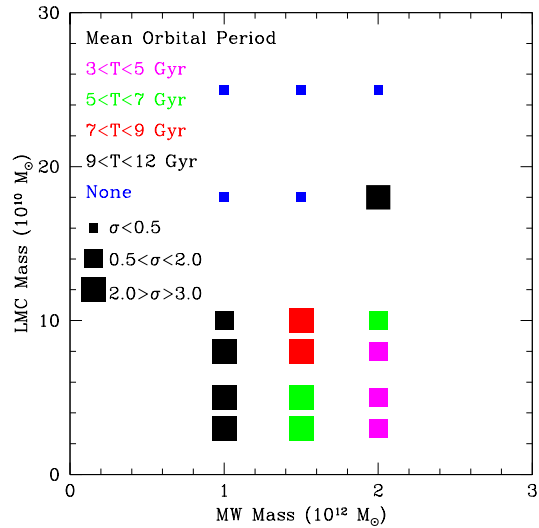}
\caption{The mean period is computed for only those orbits that are
  closed within a Hubble Time, and is indicated using colors that
  correspond to duration. For example, if a particular MW-LMC mass
  combination produces 80\% of LMC orbits with N=1, and 20\% with N=2,
  then we only plot the average of the 20\%. The symbol size indicates
  the RMS in distribution of periods, as indicated in the legend with
  black symbols.}
\label{fig:LMCperiods}
\end{center}
\end{figure}

Figure~\ref{fig:LMCMWorbits} does illustrate that solutions can be
found where the LMC makes more than one complete orbit about the MW
within a Hubble time. For these cases, Figure~\ref{fig:LMCperiods}
illustrates the mean orbital period for the most recent orbit.  Note
that if the LMC has completed more than one orbit, the orbital period
of the previous passage is expected to be larger than the values
quoted for the most recent passage because of dynamical friction.  For
the $1\times 10^{12} M_{\odot}$ MW, the periods are $> 9$ Gyr. For the
$1.5 \times 10^{12} M_{\odot}$ MW, the typical periods are $> 5$ Gyr.
In the most tightly bound scenario (LMC mass $=3 \times 10^{10}
M_\odot$, static MW mass $\sim 2 \times 10^{12} M_\odot$), the LMC
completes on average 2 orbits about the MW where the most recent orbital
period is 3-4 Gyr. This conclusion is no different than in B07 where we showed
that the high MW model recaptures the isothermal sphere orbit from
previous studies (albeit with larger orbital periods).

\begin{figure*}
\centerline{%
\epsfxsize=0.49\hsize
\epsfbox{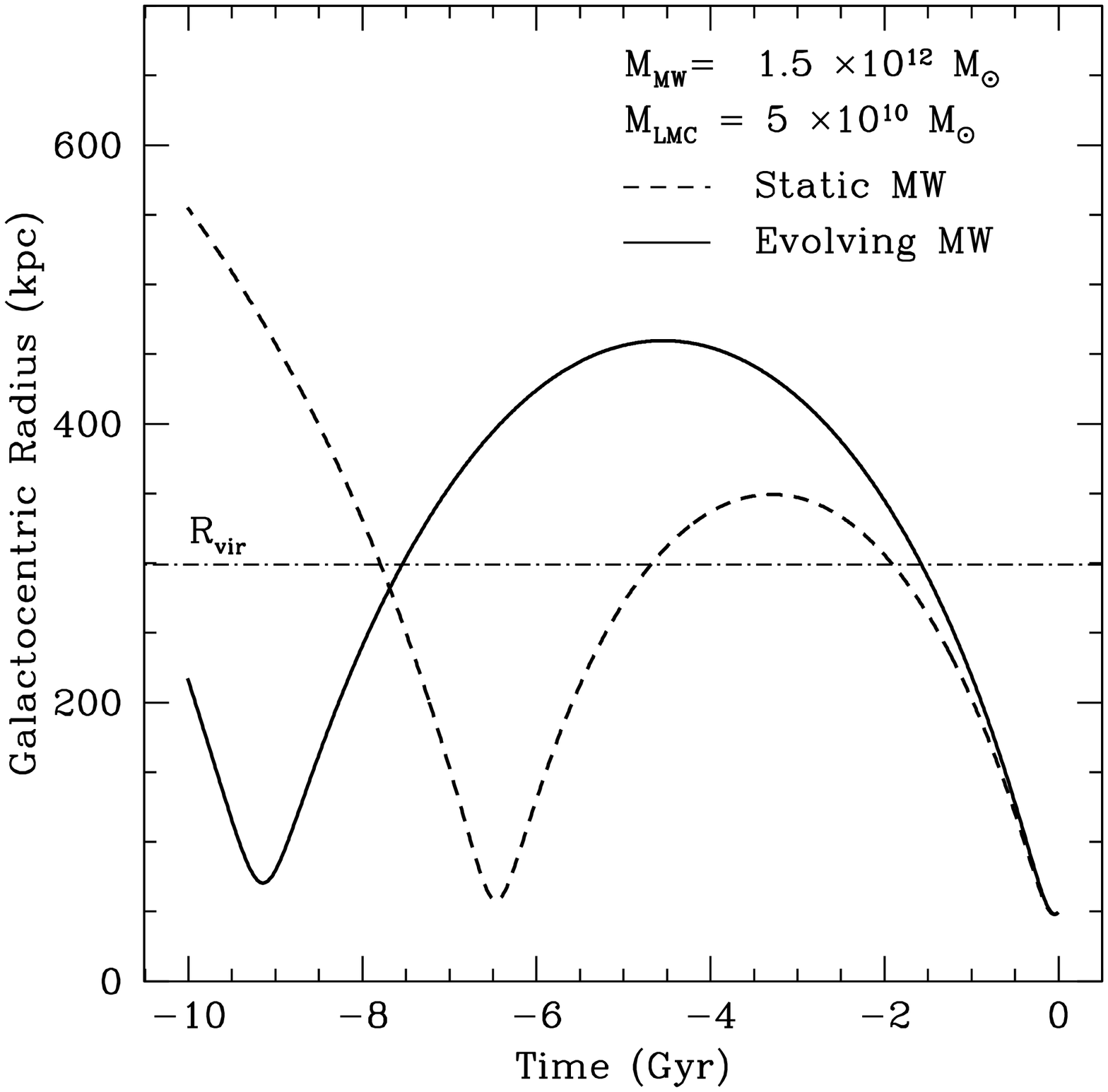}
\epsfxsize=0.49\hsize
\epsfbox{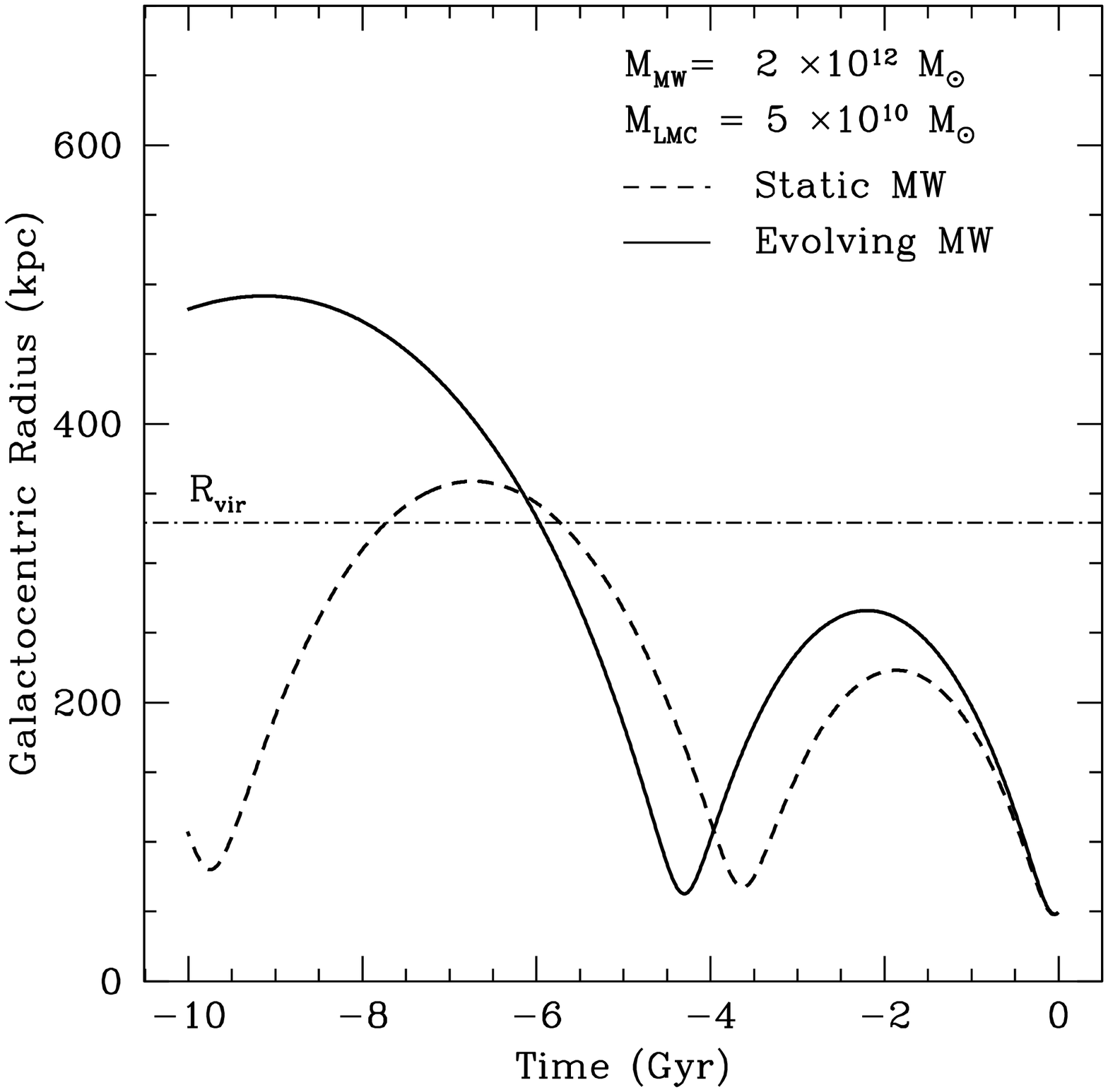}}
\caption{\textit{Left:} the LMC Galactocentric radius as a function of
  time in the past ($t=0$ corresponds to today) for a current day MW
  mass of $1.5 \times 10^{12} M_{\odot}$ and an LMC mass of $5 \times
  10^{10} M_\odot$. The period of the orbit is $\sim 2$ Gyr longer for
  an evolving MW (solid line), and the LMC goes well outside the
  virial radius of the MW, i.e. the orbit is dramatically more
  eccentric than in a static MW (dashed line). \textit{Right:} the
  same for a current day MW mass of $2 \times 10^{12} M_{\odot}$. The
  shortest orbital periods we can obtain are $\sim 4$ Gyr.}
\label{fig:evolvingMW}
\end{figure*}

For the cases where the LMC makes at least one complete orbit about
the MW, i.e. LMC mass $< 5\times10^{10} M_{\odot}$ and MW mass $>1.5
\times 10^{12} M_{\odot}$, we look at the effect of a cosmologically
motivated mass growth history for the MW on the orbital history of the
LMC.  In Figure~\ref{fig:evolvingMW} we show the LMC Galactocentric
radius as a function of time in the past ($t=0$ corresponds to today)
for a current day MW mass of $1.5 \times 10^{12}
M_{\odot}$(\textit{left}) and $2 \times 10^{12}
M_{\odot}$(\textit{right}), and for LMC mass $=5 \times 10^{10}
M_\odot$ in both cases. In the first case, the period of the orbit is
$>8$ Gyr for an evolving MW, which is $\sim 2$ Gyr longer than for a
static MW. The LMC goes well outside the virial radius of the MW,
i.e., the orbit is dramatically more eccentric than in the static MW
case.  In the second case, it is shown that a 3 Gyr orbit would still
on average increase by 1 Gyr if the mass evolution of the MW is taken
into account. The period of the last complete orbit of the LMC about
the MW is critical in figuring out the relative importance of MW tides
in forming the Stream (discussed further in the next section).

\subsection{SMC-LMC Orbits and Implications for the Magellanic Stream}
\label{subsec:LMCSMCorbits}

We are also interested in re-evaluating the orbit of the Clouds about each other. 
A chance three-way encounter between the MW, LMC, and SMC
at $z=0$ has low probability, and therefore we seek to enumerate under
what conditions, given the new velocities, the Clouds constitute a
binary for a significant fraction of a Hubble time. Also, we have 
recently put forth a new Magellanic Stream model that relies on the past
interactions between the Clouds when they are far from the MW virial
radius, rather than on the influence of the MW on the Clouds-system
\citep{Besla10, Diaz11}.  The orbital history of the SMC about the LMC is critical to 
assess the viability of such a model.

We draw 10,000 cases from the LMC and SMC PM error distribibutions in
Monte-Carlo fashion, and orbits are computed for each combination of
LMC and MW masses. The SMC is assumed to be tidally truncated and
hence its mass is kept fixed at $3 \times 10^{9} M_{\odot}$. At each
time-step, we compute the escape velocity of the SMC from the LMC,
$V_{\rm esc}$, and compare this to the relative velocity between the
Clouds, $V_{\rm LS}$.  We do not require them to be bound today, but
rather search for cases at some point in time where $V_{\rm LS} <
V_{\rm esc}$, and keep track of the longest amount of time over which
this condition remains true.

This analysis, therefore, accounts for capture events and the fact
that the MW's tidal field might be disrupting the binary today.  By also requiring
that the escape velocity condition be satisfied, as opposed to
searching for minima in the distance between the Clouds as many previous
authors have done, we are choosing real binary configurations, rather
than chance superpositions caused by cases in which the SMC is on its
own elliptical orbit about the MW, and therefore formally gets closer
and farther away from the LMC even though the LMC may be on a very
different orbit \citep[as illustrated in][]{Ruzicka10}.

In \textit{all} searched cases the SMC has made at least one close
encounter with the LMC in the past.  This is unsurprising, as it is a
direct result of the relative orientation of their three-dimensional
velocity vectors.  The SMC is currently $\sim 20$ kpc from the LMC and
moving away from it.  Integrating the SMC's orbit backwards in time
will thus always yield a close encounter with the LMC within the past
500 Myr.  This conclusion has been drawn by many previous authors
\citep[][GN96]{Ruzicka10} and is the basis of the LMC-SMC collision
theory put forth in \cite{Besla12}.

\begin{figure}
\begin{center}
\plotone{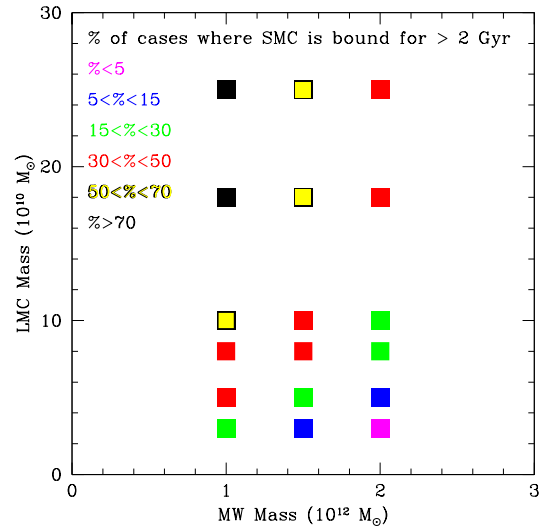}
\caption{The percentage of cases from our 10,000 Monte Carlo drawings
  in which $V_{\rm LS} < V_{\rm esc}$ for the SMC from the LMC for
  more than 2 Gyr in the past, shown as a function of LMC and MW
  mass. The legend shows the color-scheme indicating the range of
  percentages for each combination of LMC and MW mass.}
\label{fig:LMCSMCorbits}
\end{center}
\end{figure}

In Figure~\ref{fig:LMCSMCorbits} we show the outcome of our search for
binary orbits between the Clouds.  The percentage of Monte Carlo
drawings for which $V_{\rm LS} < V_{\rm esc}$ for more than 2 Gyr in
the past is shown as a function of LMC and MW mass. We have chosen the
2 Gyr duration since this is the estimated age of the Magellanic
Stream (as discussed below), and so the LMC and SMC should be a binary for at least this
amount of time in each LMC and MW mass combination. Colored squares
indicate the corresponding percentages, as indicated in the legend.

Unsurprisingly, the percentage of cases for which $V_{\rm LS} < V_{\rm
  esc}$ increases as the mass of the LMC increases (increasing binding
energy). However, we also note a dependence on MW mass. At an LMC mass
of $3\times 10^{10} M_{\odot}$, the percentage of orbits that
satisfies this criterion is 20\% if the MW mass is $1\times 10^{12}
M_{\odot}$. This number drops to 4\% if the MW mass is $2\times
10^{12} M_{\odot}$, implying that bound configurations between the LMC
and SMC are statistically unlikely for low LMC mass and high MW
mass. As the LMC mass increases, these statistics improve, but for a
MW mass of $2\times 10^{12} M_{\odot}$, they only reach 20\% once we
get to a LMC mass of $8\times 10^{10} M_{\odot}$.  The corresponding
percentage for a $1\times 10^{12} M_{\odot}$ MW is 50\%.

The RMS of the distribution of cases matching this escape velocity
criteria increases as the LMC mass increases, representing a larger
spread in the possible durations.  As discussed above, as the MW
mass increases there are much fewer binary LMC-SMC orbits, even for
high LMC mass. The PMs indicate that the current relative velocity
between the Clouds is high. This implies that the SMC must be on an
eccentric orbit about the LMC.  For larger MW mass models, the tidal
field is more effective at disrupting such a wide binary pair.

Therefore, from our statistical analysis we conclude that large LMC
masses ($\gtrsim 1 \times 10^{11} M_{\odot}$) are favored if the
Clouds are to have been in a binary, and that further, a MW mass
$\lesssim 1.5 \times 10^{12} M_{\odot}$ is also required. From
Figure~\ref{fig:LMCMWorbits}, orbits in this mass range almost always
correspond to first infall scenarios. The assumption that the
Magellanic Clouds constitute a long-lived binary pair thus implies
that the Clouds are likely on their first infall about the MW.

Based on our searched parameter space, and the requirement that the
LMC and SMC have been a long-lived binary, we adopt a canonical model
with the highest MW mass and lowest LMC mass that will fulfill these
criteria. This gives (LMC, MW) mass = $1.8 \times 10^{11} M_{\odot},
\ 1.5 \times 10^{12} M_{\odot}$.

\begin{figure*}
\centerline{
\epsfxsize=0.35\hsize
\epsfbox{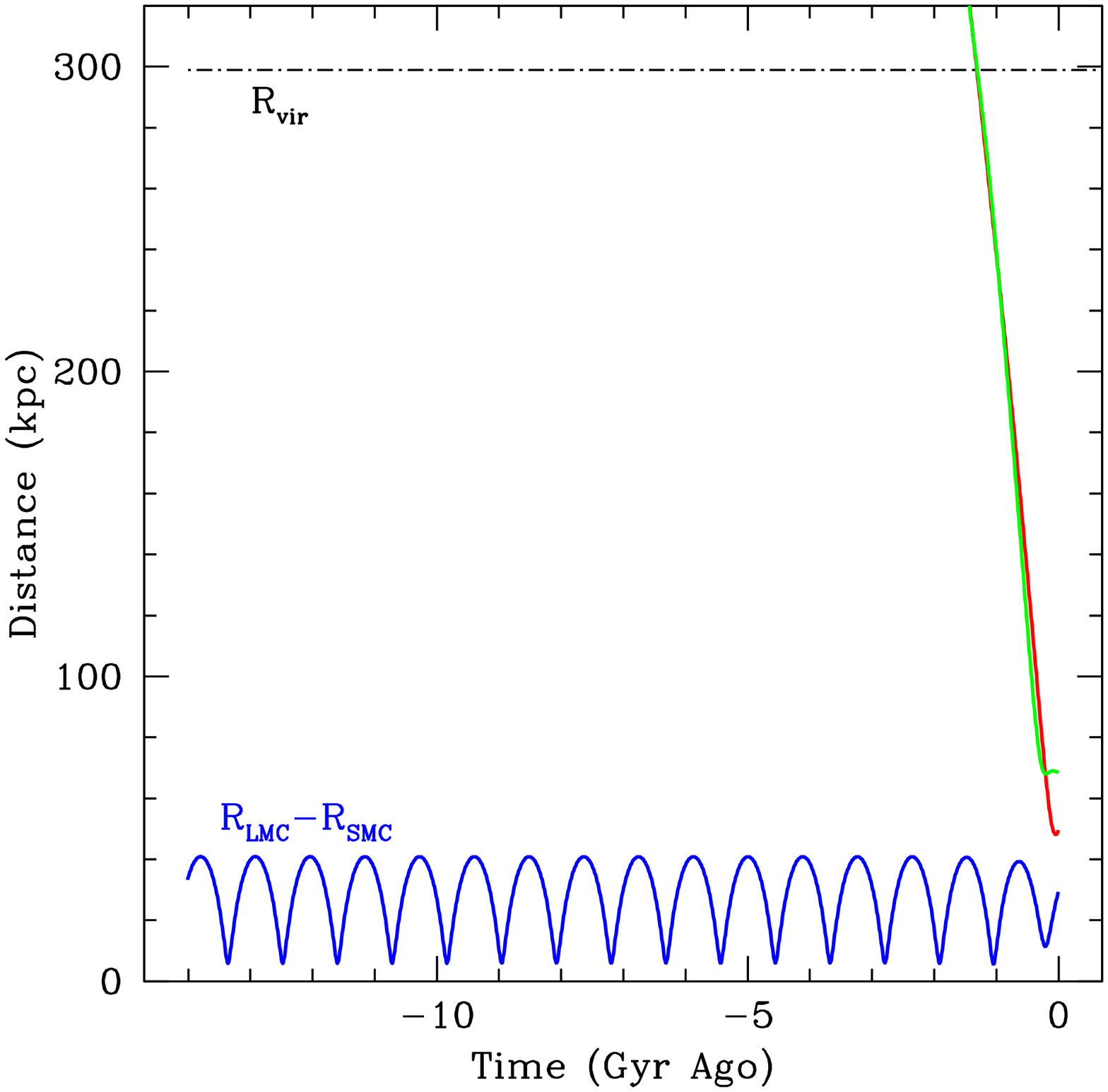}
\epsfxsize=0.35\hsize
\epsfbox{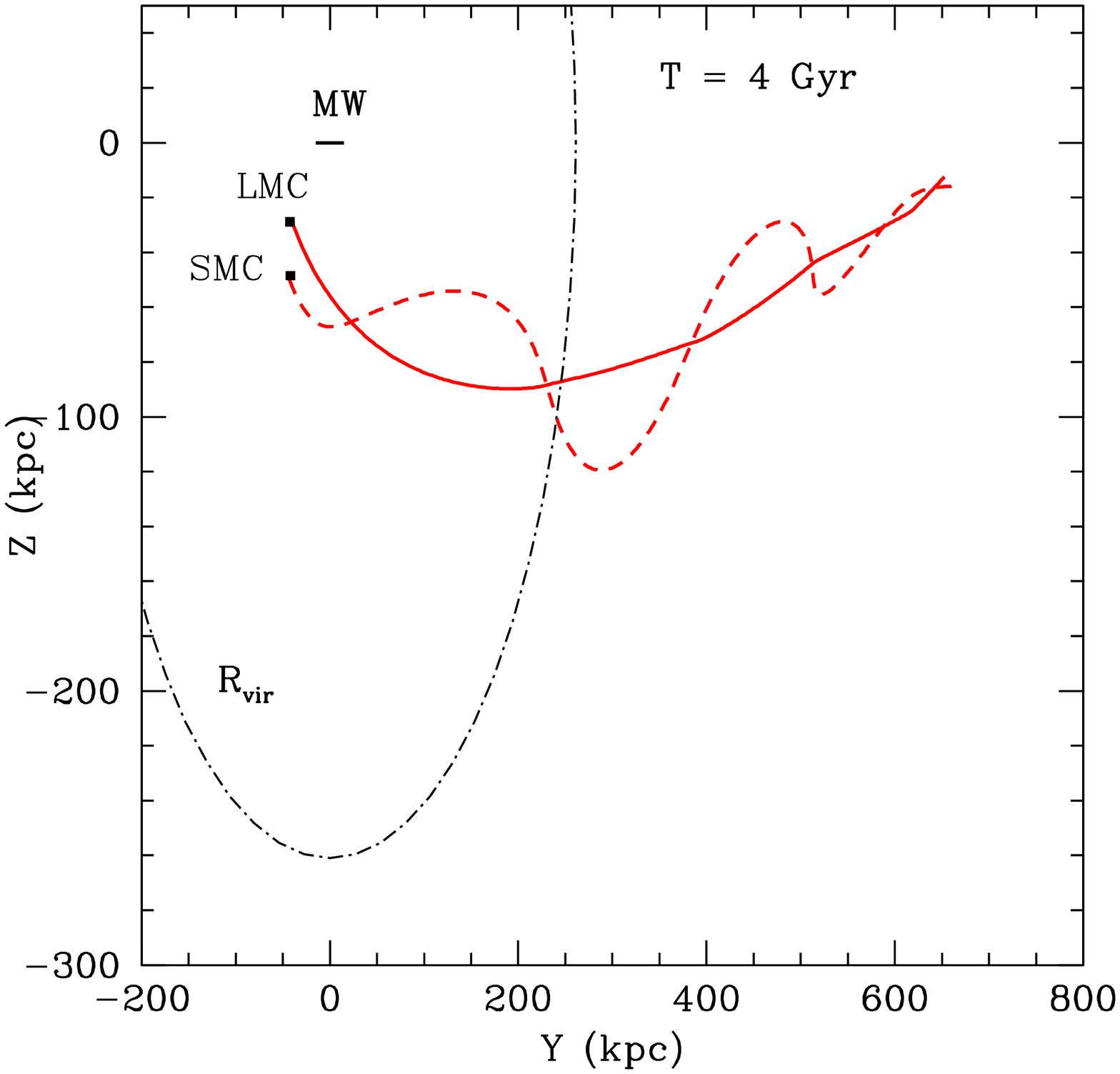}
\epsfxsize=0.35\hsize
\epsfbox{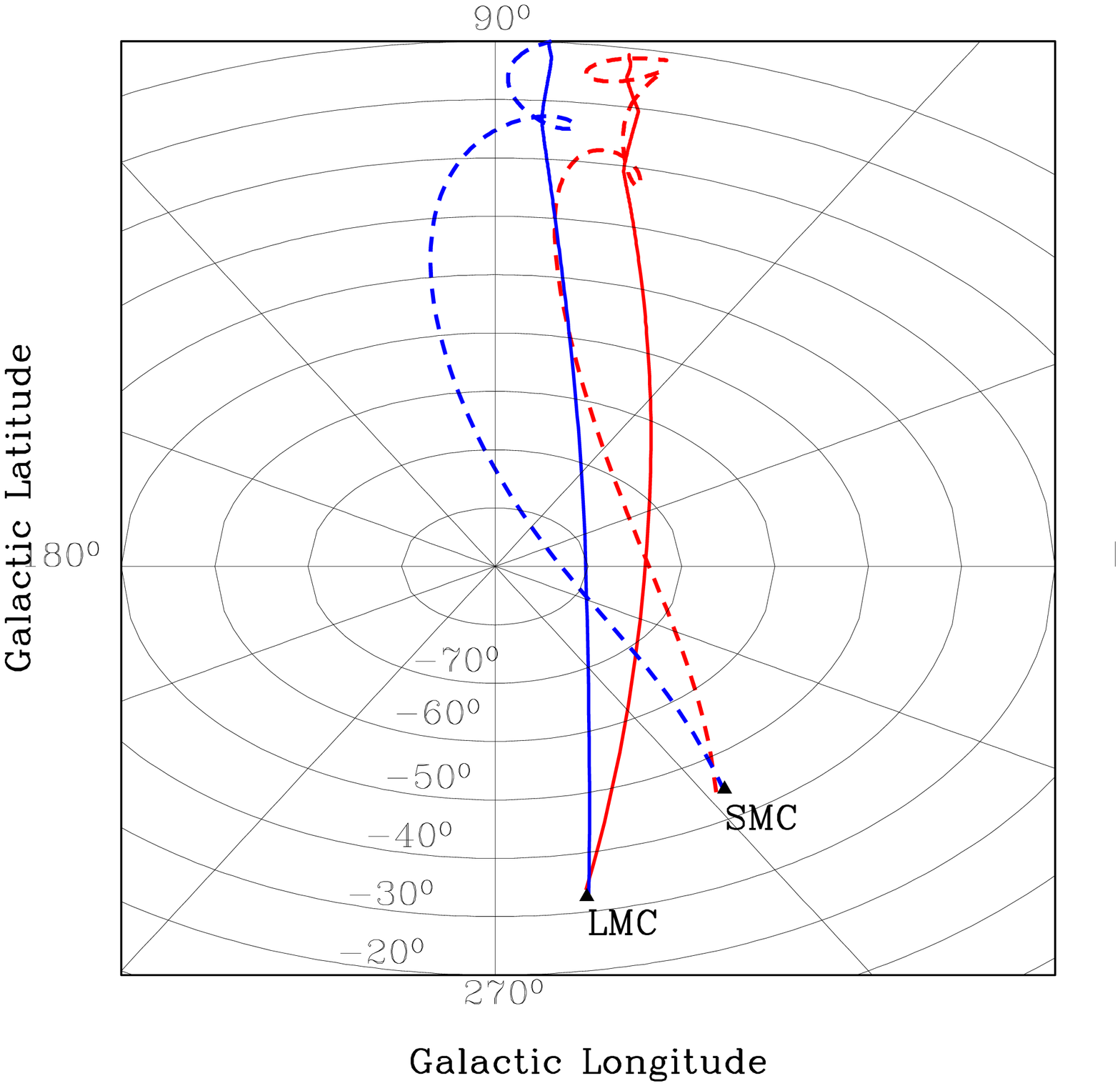}}
\caption{\textit{Left:} Evolution of the LMC (red) and SMC (green)
  Galactocentric radius as a function of time in the past, for our
  canonical orbit, i.e., $M_{\rm vir} = 1.5 \times 10^{12} M_{\odot}$,
  $M_{\rm LMC} = 1.8 \times 10^{11} M_{\odot}$, $M_{\rm SMC} = 3
  \times 10^{9} M_{\odot}$, and the mean third epoch LMC velocity. The
  blue line shows the relative distance between the two Clouds. The
  dot-dash horizontal line shows the virial radius, $R_{\rm vir}$, for
  this MW model. \textit{Middle:} the same shown in the Galactocentric
  $(Y,Z)$-plane. The orbits of the LMC (solid red line) and SMC
  (dashed red line) are followed for the past 4 Gyr, and their present
  locations with respect to the MW are marked with black squares. The
  MW center is marked and its disk orientation is indicated by the
  short dash. $R_{\rm vir}$ is also shown. \textit{Right:} the same
  orbits for the LMC (solid red line) and SMC (dashed red line) are
  shown in the $(l,b)$-plane. The blue lines show the corresponding
  GN96 orbits, which were chosen by those authors to match the
  location of the H{\small I} in the Magellanic Stream (see Figure~9
  of B07), and stands in for the Stream location here.}
\label{fig:bestorbit}
\end{figure*}

In Figure~\ref{fig:bestorbit} we show an example of one of these
long-lived binary states (N=4--6) in the canonical model (note again
that the SMC is assumed to be tidally truncated and therefore kept at
fixed mass in our models). The left-hand panel shows the
Galactocentric radius of the LMC, the Galactocentric radius of the
SMC and the relative distance between the two Clouds as a function of
time. The middle panel shows the past orbits of the Clouds in the
Galactocentric $(Y-Z)$-plane. The right panel shows the past orbit as
a function of $(l, b)$. Interestingly, the new velocities now cause
the past SMC orbit to cross over in the direction of the Stream,
something that we did not achieve with the K2 velocities, but was
advocated for in earlier works like GN96.

We note that with the vdM02 center, the relative LMC-SMC velocity is
$143 \pm 31 \kms$ (Table~\ref{tab:vels}).  With the new center adopted
in this work, it is $128 \pm 32 \kms$. Since the relative velocity is
higher in the former case, the orbit of the SMC would be less bound at
fixed mass, and therefore we expected fewer bound cases.  Suprisingly,
we in fact find \textit{no} binary LMC-SMC orbits when using the
vdM02 velocities. We attribute this not just to the larger absolute
value of relative velocity, but to the fact that the angle between the
velocity vectors is larger. This is an interesting finding that argues
in favor of the new center used here and derived in Paper~II.

While dynamical friction owing to the passage of the
Clouds through the MW's dark matter halo was accounted for, we did not
explicitly account for dynamical friction acting on the SMC owing to
its passage through the LMC's dark matter halo. The inclusion of this
effect would place the SMC on an increasingly eccentric orbit in the
past.  We have recently advocated for a high eccentricity SMC orbit
about the LMC in our simulations of the Magellanic Stream; such
eccentric orbits are needed to explain the large angular extent of the
Stream and to prevent the Clouds from merging \citep{Besla10}.
However, such eccentric orbits are more easily disrupted by the MW's
tidal field and thus work against maintaining long-lived binary
LMC-SMC states.  This further strengthens our argument that the
preferred mass scale for the MW is less than $2\times 10^{12}
M_{\odot}$, since in reality there should be even fewer viable binary
orbits in this model than we currently find.

\begin{figure}
\begin{center}
\plotone{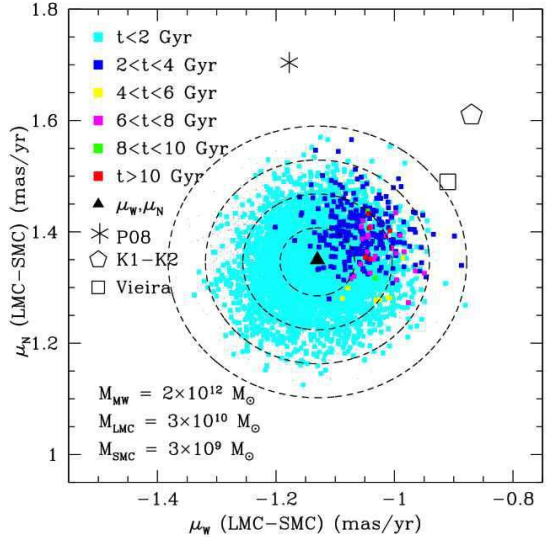}
\caption{A further exploration of LMC-SMC orbits in a low-mass LMC and
  high-mass MW model (masses indicated in the figure). This
  combination of masses gives rise to more traditional LMC orbits
  (albeit with longer orbital periods as can be seen in
  Figure~\ref{fig:evolvingMW}). We did 10,000 Monte Carlo drawings
  from the PM error distributions. We show the relative PM between the
  Clouds with each point color-coded by the amount of time for which
  $V_{\rm LS} < V_{\rm esc}$ in the past (colored squares). The black
  triangle is the mean relative PM value from this study, with
  1$\sigma$, 2$\sigma$, 3$\sigma$, and 4$\sigma$ error ellipses
  indicated as well (dashed). While the vast majority of drawings do
  not give rise to orbits that were bound for more than 2 Gyr (cyan
  squares), there are some orbits that are bound for 2--4 Gyr within
  the $1\sigma$ error ellipse (dark blue squares).}
\label{fig:LMCSMCrelPM}
\end{center}
\end{figure}

Another way to approach these questions is to look in more detail at
the LMC-SMC dynamics in LMC-MW mass combinations that give rise to
more traditional orbital trajectories for the Clouds about the
MW. Specifically, consider the case of a low-mass (tidally truncated)
$3 \times 10^{10} M_{\odot}$ LMC orbiting around a $2.0 \times 10^{12}
M_{\odot}$ MW, with an orbital period of $\sim 4$ Gyr.  We draw 10,000
LMC and SMC PMs in Monte-Carlo fashion from their error distributions,
and plot the relative proper motion between the two Clouds in
Figure~\ref{fig:LMCSMCrelPM}. Each value is color-coded by the length
of time for which the criterion $V_{LS} < V_{\rm esc}$ held true in
the past. While it is clear that the large majority of pairs within
the $1\sigma$ error ellipse do not produce a bound system for $> 2$
Gyr, it is interesting that some orbits with a bound pair in excess of
several Gyr do exist within this error-space.  A larger abundance of
such orbits can be found at relative PMs that are more in the West
direction than our measurements imply. So the combination of galaxy
masses shown in this plot, while unlikely, cannot be ruled out
strictly based on the requirement that the LMC and SMC must have been
bound for at least 2 Gyr. However, it should be noted that this does
not imply that traditional models for the Magellanic Stream (which
have often used masses such as shown in this figure) are tenable,
because such models produce orbital periods that are inconsistent with
the age of the Stream (discussed more below).

In summary, perhaps unsurprisingly, we find that to obtain long-lived
binary LMC-SMC orbits the LMC mass must be relatively high, and the MW
mass must be relatively low. The lower mass of the MW implies that the
Clouds are on a first infall and, as such, the MW's tidal field would
be insufficient to disrupt the binary pair.  We are more readily able
to find such long-lived orbits with the new LMC center derived here
rather than that previously derived by vdM02.  We also find that a
close encounter between the Clouds in the recent past is a generic
result.

Given that the Clouds are a binary pair, the more massive LMC sets the
orbital period for the Clouds about the MW. As shown above, the LMC
mass must be $\gtrsim 1 \times 10^{11} M_{\odot}$ for the Clouds to have
been a long-term binary system. The smallest orbital periods about the
MW that we can obtain for such large LMC masses are in excess of 4
Gyr. This is problematic for models that rely on Milky Way tides to
form the Stream, because age estimates of the Stream imply that the
Stream is a young feature, much younger than 4 Gyr old. In fact, given
the rate at which the Stream is being ablated away, as inferred from
simulations of the observed anomalously high H$\alpha$ emission in the
Stream \citep{Weiner96} by \cite{Bland-Hawthorn07}, and based on
estimates of the survivability of high velocity clouds in MW-type
environments \citep{Heitsch09, Keres09}, it is unlikely that the
Stream could have survived about the MW for more than $\sim$1--2 Gyr.

The past orbit of the LMC and SMC, implied by the new velocities, is
now better aligned with the location of the Stream H{\small I} than it
was with the K1/K2 velocities.  As discussed in B07, this is directly
due to the lower value for $\mu_N$, which in turn is obtained here
directly from the dynamical center of the LMC having changed with
respect to what was used in K1. However, we note that the LMC orbit
still does not trace the Stream closely: there remains a sizeable
offset.  The new SMC PMs, however, do allow the past SMC orbit to cross over the Stream
location, which is now easier to reconcile with models where the Stream is
largely stripped from the SMC \citep[GN96;][]{Ruzicka10, Diaz11,
  Besla12} than it was with the K2 SMC PMs.

\subsection{First Infall?}
\label{subsec:firstpass}

As we learned from B07, the question of first infall is a
model-dependent rather than simply a velocity-dependent one. We
showed in that work that no matter whose velocities were used (K1/GN96/prior
ground-based determinations), the Clouds were always on a first
infall in a $1.2 \times 10^{12} M_{\odot}$ NFW MW halo, and that in
order to recapture something resembling the `traditional' isothermal
sphere orbit, we had to go to a $2\times 10^{12} M_\odot$ NFW halo.

Given that the new Galactocentric velocities derived here are lower
than those in K1, we want to critically re-examine the arguments for
whether the LMC may be on a first passage, which we define as an
orbital solution wherein the LMC has first entered the virial radius
of the MW within the past 1-4 Gyr and has not completed an orbit in
that time.  We have already shown above that with our new velocities,
a low LMC mass can make past pericentric passages about a high mass
MW. Therefore, arguments for a first passage, being still
model-dependent in nature, can be cast as arguments against a low-mass
LMC and a high-mass MW. Our arguments fall into three main catagories:
(1) orbital eccentricity and cosmological expectations, (2) LMC tidal
radius, and (3) LMC-SMC binarity. We also discuss a fourth argument
that attempts to reconcile the high gas content and active star
formation of the Clouds with their current close proximity to the MW.

\smallskip

\noindent {\bf Orbital Eccentricity and Cosmological Expectations:} As
shown in Figure~\ref{fig:LMCperiods} in all cases where the LMC does
complete an orbit about the MW, the corresponding orbital periods are
typically large ($>4$ Gyr).  Such orbits take the LMC to apocentric
distances of order the virial radius of the MW and thus imply large
orbital eccentricities. The computed eccentricities for the $3 \times
10^{10} M_{\odot}$ LMC and the $2 \times 10^{10} M_{\odot}$ evolving
MW model are 0.6--0.7. These eccentricities get higher for larger LMC
masses. Comparing to Figure~5 of \cite{Boylan-Kolchin11}, only 20\% of
LMC analogs accreted at early times ($> 8$ Gyr) have such
orbits. Indeed, in the prevailing $\Lambda$CDM model of hierarchical
structure formation, massive satellites that are accreted at early
times are very rarely found on highly eccentric orbits about MW-type
hosts at z=0 \citep{Boylan-Kolchin11, Wetzel10, Stewart08}.

Generally, subhalos that are accreted on $\sim$radial orbits at early
times are either preferentially destroyed or exist on more
circularized orbits today \citep[e.g.,][]{Benson05}. While 50\% of LMC
analogs accreted more than 4 Gyr ago and 60\% of LMC analogs accreted
more than 8 Gyr ago have eccentricites $< 0.5$ in the Millenium II
sample, only 20\% of LMC analogs accreted within the last 2--4 Gyr
have such modest eccentricites \citep{Boylan-Kolchin11}. A high MW
mass, low LMC mass today isn't ruled out from cosmological
expectations, but the high eccentricities (0.6-0.7) calculated from
our orbits are more typical of LMC analogs accreted within the past 4
Gyr. Therefore, from the eccentricities of our orbits, a first infall
scenario for the Clouds is the favored orbital solution of
$\Lambda$CDM theory.


We can also make a more general timing argument based on the fact that
the Clouds and the MW must have been in close proximity at the time of
the Big Bang (in analogy with similar arguments for e.g., M31 and Leo
I \citep{Li08}).  This implies that the Clouds had a pericentric
approach with the MW at the time of the Big Bang.  So if there has
been more than one complete orbit since, then the period must be $< T_0/2$, where
$T_0=13.73$ Gyr. Hence, orbits with a pericenter $>6.9$ Gyr ago are
not physical, because the Clouds and MW were not together at the Big
Bang.  Of course, this is oversimplified for many reasons, e.g., the
mass of the MW increases with time. But as we show in
Figure~\ref{fig:evolvingMW}, this typically increases the period by
$\sim2$ Gyr. Therefore, finding an orbit that is consistent with the
proper motion data with a pericenter at e.g., 10 Gyr ago, does not
rule out a first infall scenario, since such orbits are not physical
based on such a timing argument.

\smallskip

\noindent{\bf LMC Tidal Radius:} As discussed in
$\S$\ref{subsec:methodology} the LMC total mass must be at least $3
\times 10^{10} M_{\odot}$ from the \cite{Saha10} observations. \cite{Munoz06}
have claimed a detection of LMC stars out to 20 kpc. If stars do indeed
exist out to 20 kpc, the required LMC mass is $\sim 8 \times 10^{10}
M_{\odot}$. It should be pointed out that more traditional Magellanic
Stream models \citep[GN96;][]{Diaz12} assume a total LMC mass of $1
\times 10^{10} M_{\odot}$ which is already ruled out by the
observations. Their choice of LMC mass is important to point out since
this is the reason their orbital periods are so different from ours.

\smallskip

\noindent{\bf LMC-SMC Binarity:} Studies of the star-formation
histories of the Clouds all agree on the fact that the star-formation
rate started to increase $\sim4$--6 Gyr ago, and was quiescent in both
galaxies before this time \citep{Harris09, Cignoni12}. The coincident
increase in star-formation in both galaxies may imply that they were
in a common envelope at the time, and corresponds to the epoch that
the LMC captured the SMC in the \cite{Besla12} model.

If we therefore take the stance that the two Clouds must have fallen
into the MW as a bound system (although they need not necessarily be
bound \textit{at present}) we find that we are able to place limits on
both the MW and LMC galaxy masses. We find that in general an LMC mass
$> 1 \times 10^{11} M_{\odot}$ is needed for the SMC to have been
bound to the LMC for $> 2$ Gyr in the past, and that further a MW mass
$< 1.5 \times 10^{12} M_{\odot}$ is required. Note that LMC orbits at
this particular mass are almost always on first infall regardless of
MW mass.  The LMC mass required to keep the SMC bound to it goes up as
the mass of the MW increases, in order to compensate for the
increasing tidal field of the MW.  But as the LMC mass increases the
orbital eccentricity does as well, making a first infall scenario more
likely based on the plausibility argument above that high
eccentricities and early infall times are relatively rare in
cosmological simulations.

It is harder to maintain a LMC-SMC binary for long periods of time in
high mass MW and low mass LMC models: less than 5\% of the 10,000
searched Monte Carlo cases result in binary configurations, meaning
such models generally require that the LMC capture the SMC while in
orbit about the MW - a statistically improbable event.  This argument
implies that the MW's mass is likely less than $2\times 10^{12}
M_{\odot}$.

\smallskip

\noindent{\bf The Clouds and the Morphology-Distance relation:}
\cite{Vandenbergh06} argued for a recent accretion of the Clouds,
based on a comparison of the morphologies of MW and M31 satellites
that showed that the Clouds are the only gas-rich dIrr galaxies at
small Galactocentric distances, all other dIrrs being at large
distances from their respective hosts. Some recent studies have looked
at this issue from a larger statistical/cosmological 
point-of-view. 

\cite{Tollerud11} study a volume-limited spectroscopic sample of
isolated galaxies in SDSS and find that bright satellite galaxies
around MW-type hosts are significantly redder than typical galaxies in
a similar luminosity range, and argue that this is indicative of
environmental quenching. This is found to be in stark contrast to the
LMC, which is anomalously blue in comparison to other LMC-MW analogs
in SDSS. The authors attribute this to the fact that the LMC may be
undergoing a triggered star formation event upon first infall.

\cite{Geha12} have used the NASA-Sloan Atlas to demonstrate that dwarf
galaxies in the field (with masses in the LMC range) all have active
star-formation ($< 0.06$\% of field-dwarfs in their sample have no
star-formation). By contrast, the majority of quenched galaxies are
all within 2--4 virial radii of a massive host. Therefore, ending
star-formation in such dwarf galaxies appears to require the presence
of a more massive neighbor. Had the Clouds been accreted early in the
Universe then, it is unlikely that these galaxies would currently have
as much gas or star formation as they do.

\cite{Wetzel12} used SDSS galaxy group/cluster catalogs together with
N-body simulations to look at the quenching timescales of satellites
at $z \sim 0$. They find a quenching scenario in which satellite
star-formation histories are unaffected for 2-4 Gyr after infall but
then quench rapidly (with an e-folding time of $< 0.8$
Gyr). Interestingly, because of the time delay before quenching
starts, satellites are found to experience significant stellar mass
growth after infall, which the authors point to as a key reason for
the success of the subhalo abundance matching technique: supporting
the larger LMC masses that we favor in this study.

\smallskip
In conclusion, despite the revised lower estimates for the 3D
velocities of the Magellanic Clouds presented in this work, taken
together, Figures~\ref{fig:LMCMWorbits}, \ref{fig:LMCperiods},
\ref{fig:evolvingMW}, and \ref{fig:LMCSMCorbits} make a strong case
for a first infall scenario. This conclusion draws largely from a
recognition that the fundamental change to our understanding of the
orbital history of the Clouds comes not only from the PMs, but also
from a cosmologically motivated understanding of the mass evolution
and dark matter halo profile of our MW. This picture is thus
consistent with the theory that the origin of the large scale gaseous
structures of the Magellanic System (Stream, Bridge and Leading Arm)
and the internal structure and kinematics of the Clouds are a result
of interactions between the LMC and SMC in a first infall scenario,
rather than interactions with the MW \citep{Besla10, Besla12}.


\section{Conclusions and Discussion}
\label{sec:conc}

We have analyzed a new third epoch of data of 10 QSOs behind the LMC
and 3 behind the SMC with WFC3/UVIS (one LMC field, L22, has only
first epoch ACS and third epoch UVIS data). We combine these data with
previously obtained ACS/HRC data giving a $\sim 7$ yr baseline, in
order to measure PMs for both Clouds, and to refine and validate
measurements that used only two epochs over a 2 yr baseline (K1,
K2, P08). We have also reanalyzed the first two epochs of
data for 21 LMC fields and 5 SMC fields, using a method to account for
the degrading CTE of the HRC as well as an improved approach to linear
transformations. Here we summarize our findings, the implications for
the Clouds' orbits, and provide a look ahead to the future.

\subsection{Conclusions for COM PMs and Galactocentric Velocities}
\label{subsec:conc:COMPMs}
The ACS reanalysis and the third epoch analysis give very consistent
results at the level of the measured per-field PMs. In order to obtain
COM PMs from these field PMs, the geometry and internal motions of the
L/SMC need to be taken into account. The addition of the WFC3 epoch
provides very small per-field PM errors of $\sim 0.03 \masyr$ ($7 \kms$,
or 1.6\% of the total PM). We are therefore able to independently
constrain all parameters of the LMC PM rotation field, including the
inclination, $i$, the position angle of the line-of-nodes, $\theta$,
the dynamical center, rotation velocity amplitude, and even the
distance. This is a big improvement over K1. This procedure and the
results are discussed in detail in Paper~II. The features of import
for the present study are that we obtain a rotation velocity that
agrees very well with that obtained independently by \cite{Olsen11}
using a very large number of LOS velocities of stellar
tracers. Moreover, we find strong evidence that the center of the LMC
(as obtained from the PM fit to the rotation field) is consistent with
the H{\small I} dynamical center, and not the center derived by vdM02,
which agrees with the brightest part of the LMC bar.

For the SMC, the sparse coverage of QSO fields renders it difficult to
gain much insight into its geometry. We therefore fit a relatively
simple model to the SMC PM data, which allows for viewing perspective
and a single overall rotation of the SMC in the plane of the
sky. The old stellar population in the SMC shows little evidence for
rotation \citep{Harris06} so we base our analysis of the SMC COM PM on
the assumption that $V_{\rm rot} = 0 \pm 15 \kms$. Analyses in which
$V_{\rm rot}$ is instead fit to the PM data do not yield a
significantly different answer.

If we analyze the new LMC PM data with the same fixed geometric
parameters as used in vdM02 we obtain a COM PM whose (i) value for
$\mu_W$ is lower, but consistent within $1.6\sigma$ with the K1 result, (ii)
value for $\mu_N$ remains unchanged with respect to K1, (iii) random
errors are smaller by a factor of almost 5 than K1. If we analyze the
PM data with the LMC model derived from the data itself, which is the
preferred approach, then the value of $\mu_W$ and its random error
remain largely unchanged. However, the value for $\mu_N$ changes
significantly (by 4$\sigma$), and its random error increases by a
factor of almost 3. For the SMC, the three-epoch analysis agrees with
K2 for $\mu_N$, while for $\mu_W$ it differs by 2.4$\sigma$. This is
largely due to how fields were combined in K2 and not due to
intra-field differences. Our final SMC COM PM value is in rough
agreement with the results of P08.

We attempt here to give as accurate an estimate as possible of the
observational random errors in the COM PM, by propagating all unknowns
in the geometry as well as the PM determination. This obviously
increases errors with respect to previous studies which only
propagated PM errors.  Remarkably, the PM data are now no longer the
dominant source of uncertainty, but rather uncertainties in the
structure of the Clouds, dominate how well the COM PM can be
established. Our final COM measurement errors, listed in line 1 of
Table~\ref{tab:PMs}, do not represent a huge improvement over what was
listed in previous HST works, despite the significantly improved
accuracy of the present work. However, this is because those previous
works underestimated the true random errors. Regardless, the
uncertainty in the inferred transverse velocities of the Clouds are
now dominated by their distance uncertainties, and not their COM PM
uncertainties.

In order to turn the measured COM PMs into Galactocentric velocities,
we need to know the solar velocity. Our understanding of the solar
velocity has recently been revised upwards, and this works to directly
reduce the Galactocentric velocities of the Clouds by a comparable
amount. Therefore, the Galactocentric $v_{\rm tot} = 321 \pm 24 \kms$ for the LMC
presented here is $57 \kms$ lower than that in K1, owing roughly
equally to the decrease in $\mu_W$ described above and to the increase
in solar velocity. There is a much smaller dependence on the new
dynamical center derived here (see Table~\ref{tab:vels}). The new
center affects mainly the $\mu_N$ value, which determines the location
of the orbit in projection on the sky rather than the tangential
velocity (as described in B07). By contrast, if we use the same
geometric model and solar velocity (IAU value) used in K1, $v_{\rm
  tot}$ decreases by $31 \kms$ compared to the K1 value, which is comparable to the size of the
$1\sigma$ error bar.

The SMC's inferred $v_{\rm tot} = 217 \pm 26 \kms$ is $85 \kms$ lower than in K2. This is due
mainly to the new proper motion derived here (which itself is due to
\textit{how} the fields were combined in K2, rather than differences
in the derived per-field PMs). Roughly $20 \kms$ of the decrease in $v_{\rm tot}$ is
due to the revised solar velocity, and there is also a small
contribution from the different SMC center used here, compared to the K2 analysis.

\subsection{Conclusions for the Clouds' Implied Orbits}
\label{subsec:conc:orbits}

Given the new velocities obtained here, we have re-evaluated the past
orbital histories of the Clouds about the MW. We find that the
dominant unknowns are the MW and LMC masses. However, some reasonable
arguments allow us to narrow down the allowed ranges of MW and LMC
masses. The results continue to make a strong case, as first argued in
B07, that the Clouds are likely on their first infall into the MW.

From the search for bound orbits between the two Clouds (see
Figure~\ref{fig:LMCSMCorbits}), we are able to identify a combination
of masses that satisfy the criterion of long-lived binarity, and also
produces orbits around the MW that are plausible from a cosmological
point of view \citep{Boylan-Kolchin11, Busha11a}. This yields a
preferred MW mass, $\lesssim 1.5 \times 10^{12} M_\odot$. The probability
of a stable LMC-SMC binary configuration decreases as the MW mass
increases. An LMC mass $\gtrsim 1 \times 10^{11} M_\odot$ is needed in order
to keep the SMC bound to the LMC for a reasonable fraction of a Hubble
time. Taken together, this combination of MW and LMC masses imply that
the LMC/the Clouds are on their first infall (see
Figure~\ref{fig:LMCMWorbits}).

From a cosmological point of view, these orbits do not represent a
major alteration to the interpretation given in B07. Here we have
explored a more full set of mass models for the LMC. The inferred LMC
mass agrees with the expected total infall mass of the LMC, as
calculated from the observed present-day stellar mass (vdM02) and the
relations from halo occupation models \citep{Guo10, Sales11}. A
generic outcome of our study is that regardless of what MW mass is
used, large LMC masses are always on a first infall.

We have also implemented a simple model for the expected mass
evolution of the MW over the past $\sim 10$ Gyr. This, as might be
expected, significantly increases the periods of LMC orbits that do
make a previous pericentric passage (higher MW mass models), i.e., the
orbits are highly eccentric. In a recent study, motivated by the
discussion over the higher Solar velocity, \cite{Zhang12} have also
revisited the Clouds' past orbital history, utilizing numerical
simulations to explore the evolution of the MW. Like us here, they
find that it is possible for the LMC to make past pericentric passages
given the observational and theoretical error-space. They do not vary
LMC mass (it is kept fixed at $2\times10^{10} M_{\odot}$), or
investigate whether binarity imposes constraints, so we cannot
directly compare our results with theirs, but were we to use such a
low LMC mass, our studies would likely be consistent.

We put forth four arguments why we think orbits in which the LMC/SMC
make a previous pericentric passage are implausible. One is the
expected eccentricities of orbits from cosmological simulations along
with a simple version of the timing argument (see
$\S$~\ref{subsec:firstpass}), the second is the tidal radius of the
LMC, and the third is the fact that long-lived SMC-LMC binary
configurations have lower probability for such orbits, given the
observed COM PMs of the Clouds. The fourth argument derives from the
work of \cite{Vandenbergh06} who conducted a morphological comparison
of the satellites of the MW and M31, finding that the L/SMC are the
only two gas-rich dIrrs at close Galactocentric distance to a host. He
argues that they may be interlopers from a remote part of the Local
Group rather than true satellites of the MW.

This line of argument is also consistent with the fact that LMC-SMC-MW
analogs are relatively rare today in our local volume \citep{Liu11}
and in cosmological simulations at z=0 \citep{Boylan-Kolchin11}.
MW-type hosts are efficient at tidally disrupting such bound
configurations.  Furthermore, \cite{Tollerud11} find that, compared to
other R-band selected LMC-MW analogs in SDSS, the LMC is unusually
blue in color.  This fact is reconcilable in a first infall scenario
where the LMC is accreted recently, because it has been able to retain
enough gas to maintain a high star formation rate today \citep[see
  also][]{Wetzel12}.  However, \cite{Tollerud11} also find that the
LMC's color is unusual compared to R-band selected LMC analogs in the
{\it field}.  \cite{Besla12} present a theory in which the star
formation rate of both the LMC and SMC has increased recently owing to
interactions between these two galaxies. Given that LMC-SMC
configurations are rare in general, it is thus unsurprising that the
overall star formation rate of the LMC is higher than average isolated
analogs.

Even in the case for the most long-lived LMC-MW configuration
presented here (i.e., low LMC mass and high MW mass) the LMC makes at
best $\sim 2$ past pericentric passages (see
Figure~\ref{fig:LMCMWorbits}).  The long implied orbital periods
exceed the lifetime of the Magellanic Stream (1--2 Gyr)
\citep[e.g.,][]{Heitsch09, Keres09}.  This conclusion is consistent
with our arguments for the need for a new formation mechanism for the
Magellanic Stream that does not rely on a previous pericentric
approach about the MW.  Moreover, such a combination of LMC and MW
masses make it difficult for the LMC and SMC to have been a long-lived
binary system.

The existence of the LMC-SMC system as a binary for more than 2 Gyr in
the past favors a combination of a high mass LMC and an intermediate
or low mass MW. Our work has found that such configurations always
yield first infall orbital configurations.  We show an example of an
orbit that fulfills all the criteria set out above in
Figure~\ref{fig:bestorbit}. This shows that viable orbital solutions
can be found that are consistent with both the $HST$ PMs and
cosmological expectations.  The new SMC PM now allows the past orbit
of the SMC to cross-over the location of the H{\small I} in the
Stream, which was very difficult to obtain with the old K2 PMs and is
consistent with a picture in which the Stream is formed primarily by
the removal of material from the SMC \citep[GN96;][]{ Connors06,
  Besla10, Besla12, Diaz11, Diaz12}.

\subsection{Future Work}
\label{subsec:conc:future}

Recently the sample of QSOs behind the Magellanic Clouds has increased
drastically. There are now well over 200 newly identified QSOs behind
the LMC and 29 behind the SMC.  The new QSOs were selected primarily
based on their mid-IR colors \citep{Kozlowski09}. The majority of them
were spectroscopically confirmed from follow-up data taken with
2DF/AAOMEGA \citep{Kozlowski11, Kozlowski12}, while a smaller subset
(around 40) were confirmed with Magellan/IMACS (by N. Kallivayalil \&
M. Geha).  PM studies of fields around a carefully chosen subset of
these QSOs could vastly improve our knowledge of the structure and
dynamics of the Clouds.

With a larger sample of QSOs homogeneously distributed behind the SMC,
we can investigate its rotation and structure, something we have been
unable to tackle with only 5 QSO fields thus far (and with only 3 QSOs
currently imaged with WFC3/UVIS). A larger number of QSOs will also
improve our knowledge of its COM PM.

For the LMC, a large number of fields with PMs would allow us to
better sample the rotation curve, including its inner slope. The
increased coverage would also allow us to combine our PMs with radial
velocity studies \citep[e.g.,][]{Parisi10,Olsen11}, thereby
constituting a unique 3D data-set. With a large number of QSO fields
we may also be able to distinguish between the kinematics of different
stellar populations within the LMC directly.

Finally, with exposures of sufficient depth, it is possible to measure
absolute PMs with resolved background galaxies \citep{Sohn12}, so we should not be
confined to fields with QSOs. We are in the process of identifying
fields in the Clouds with deep HST/ACS exposures in an earlier epoch,
which should have numerous background galaxies that can be used to
yield an accurate absolute PM of the stars in the field.

\appendix

\section{WFPC2/PC data}

A third epoch of \textit{HST} imaging of our QSO fields was originally due to
execute on ACS/HRC in the period July 2007 to November 2008. However,
due to the failure of ACS at that time, the observations were executed on
the Planetary Camera (PC) of the WFPC2 instrument (PID 11201). All of
the 40 targets (34 LMC \& 6 SMC QSO fields) that had been observed
with ACS/HRC in epoch 1 were observed with WFPC2 in snapshot mode,
providing a typical time baseline of 5 years. The observational
approach was similar to that in epochs 1 \& 2. We used the $V$-band
(F606W) filter and a 5 or 6-point dither pattern.  Total exposure times
ranged from 2.8 to 20 minutes.

\begin{figure*}
\begin{center}
\epsfxsize=0.39\hsize
\epsfbox{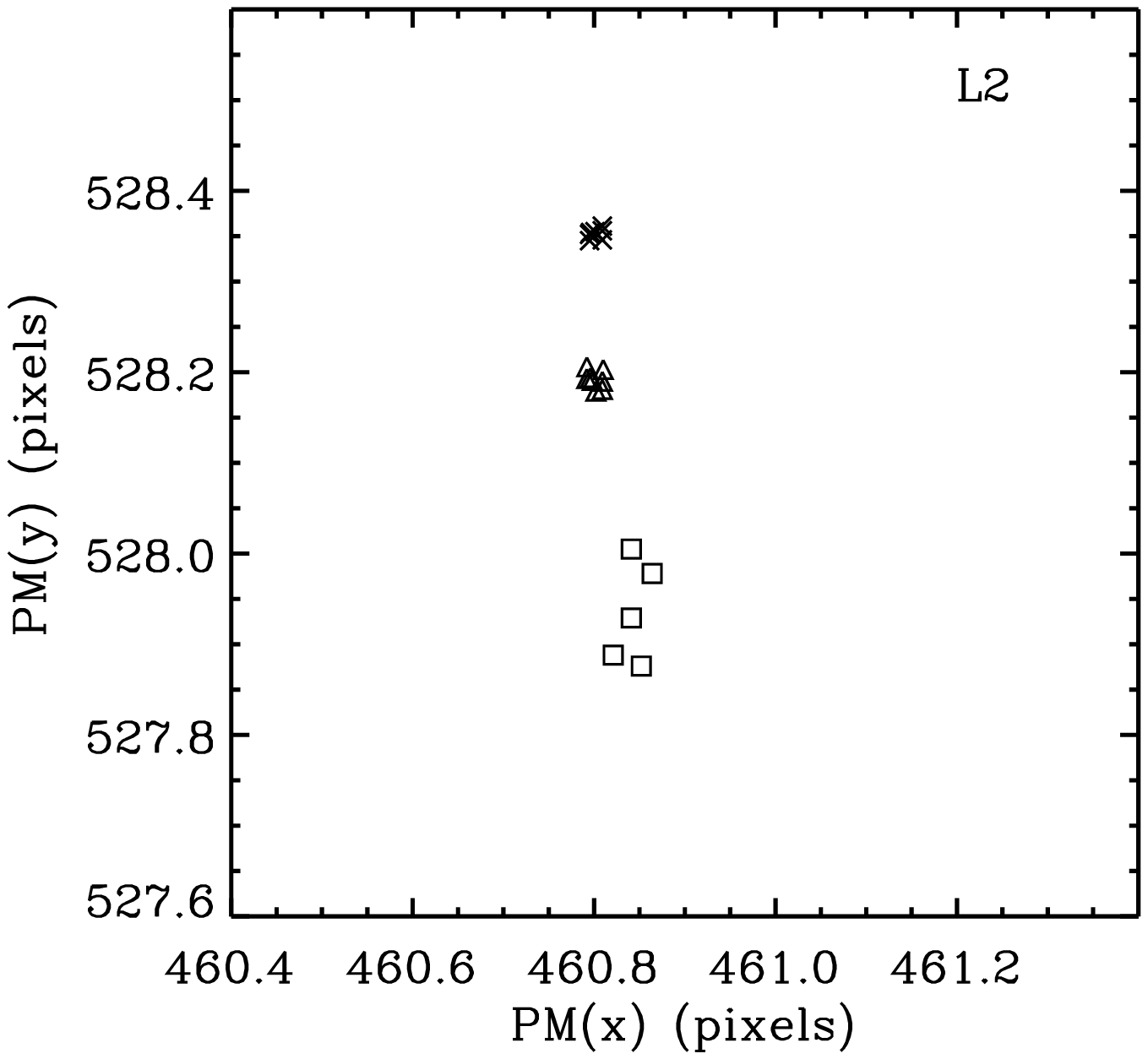}
\epsfxsize=0.39\hsize
\epsfbox{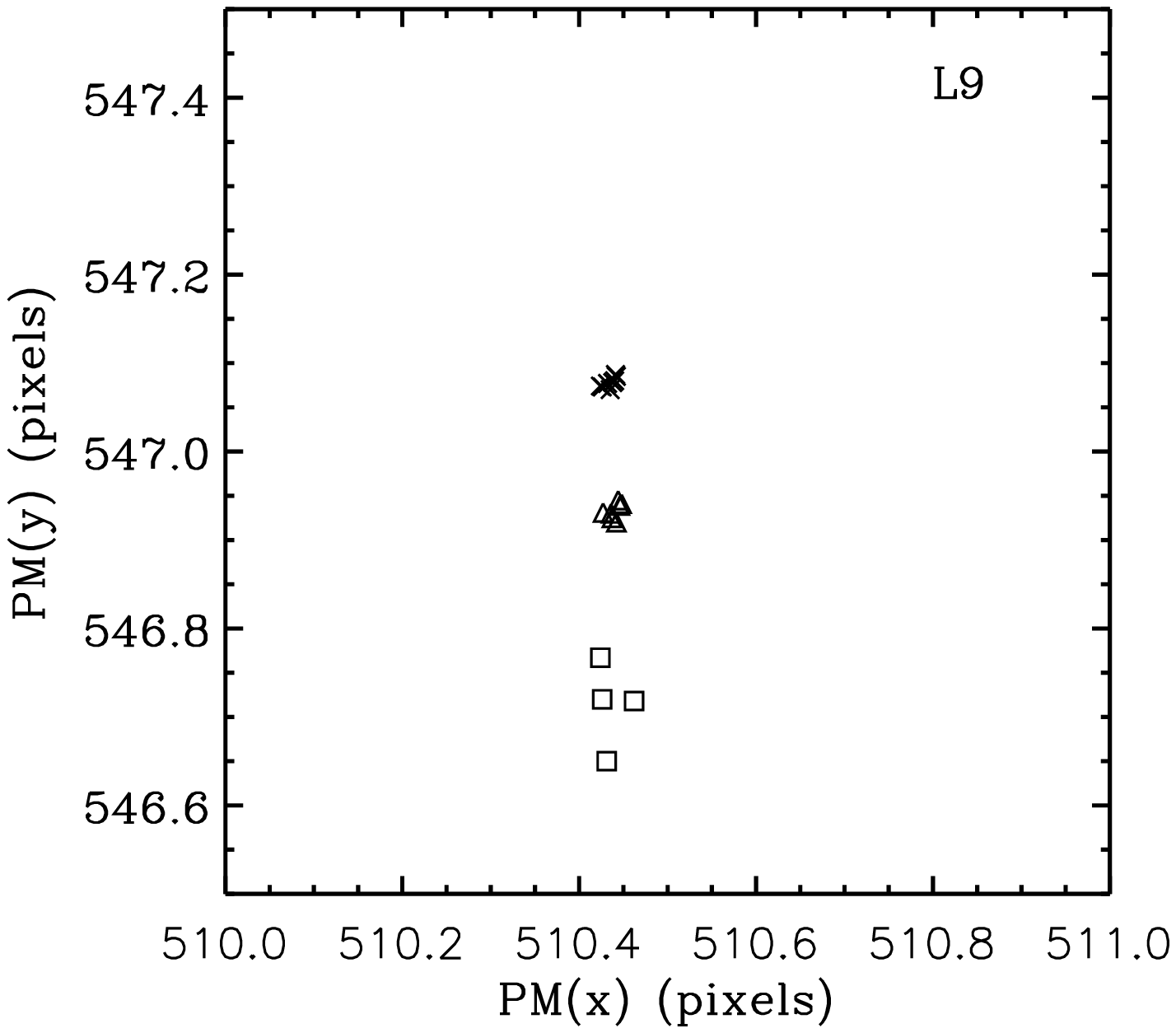}
\end{center}
\vskip -0.7truecm
\begin{center}
\epsfxsize=0.39\hsize
\epsfbox{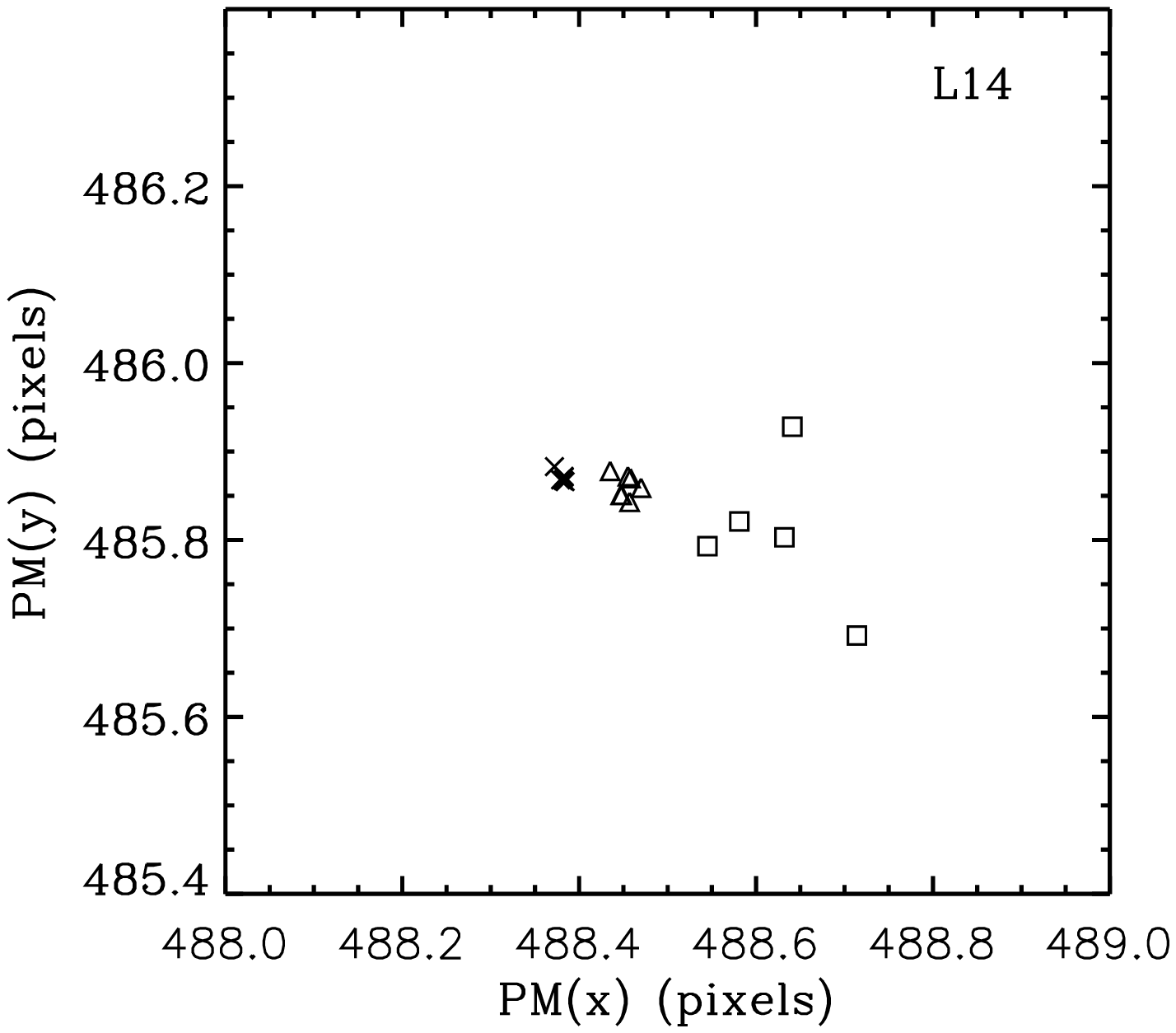}
\epsfxsize=0.39\hsize
\epsfbox{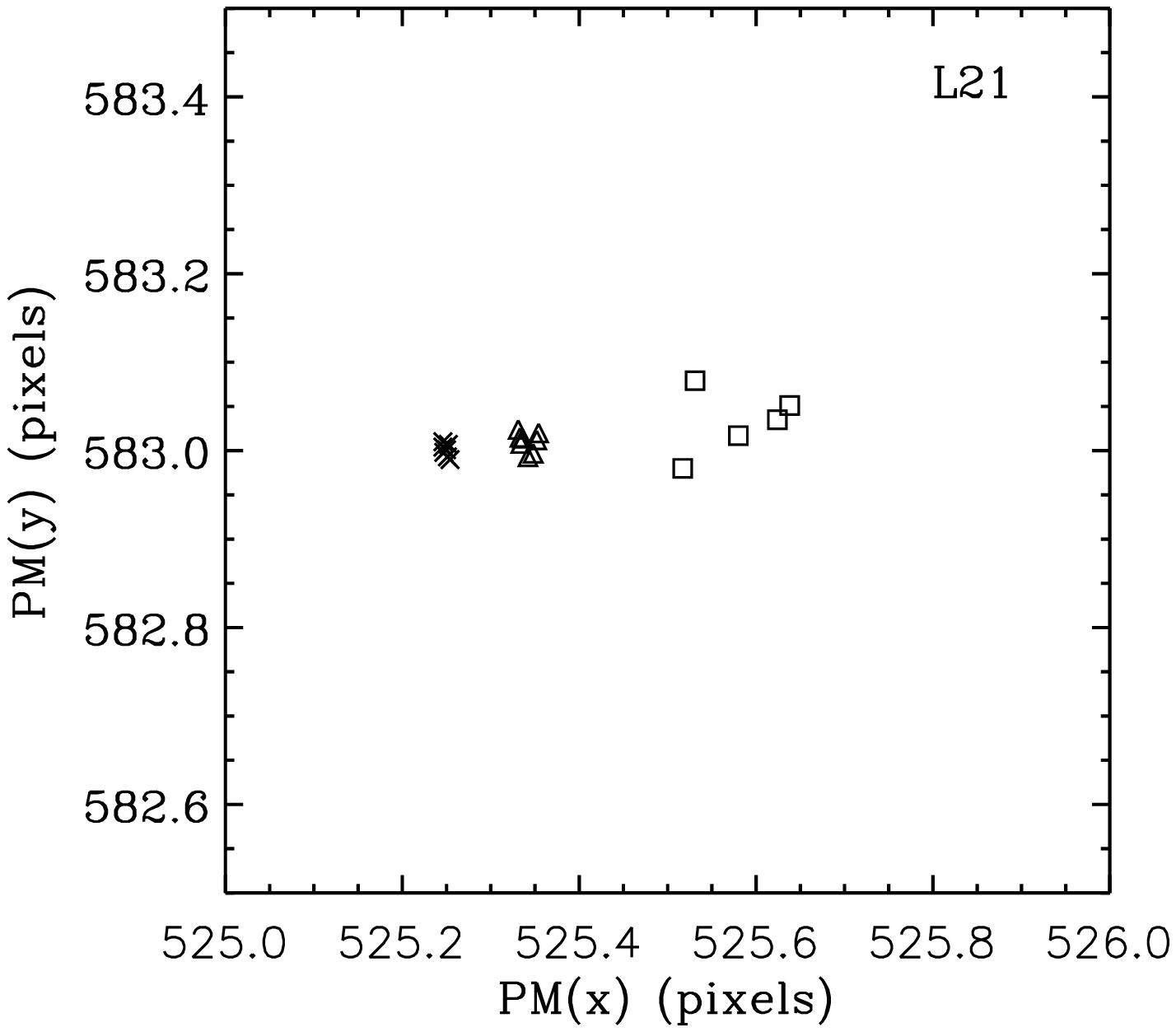}
\end{center}
\caption{Positions, in a reference frame based on the 1st epoch
  images, of the quasar over time for 4 randomly chosen LMC fields
  showing ACS epoch 1 (\textit{crosses}), ACS epoch 2
  (\textit{triangles}) and WFPC2/PC  (\textit{open squares}). The scatter
  per epoch gives the relative size of the RMS errors in the position
  of the quasar relative to the star-field, given that the observation
  has been repeated many times per epoch. Each panel is 
  $1\times1$ pixel (1 pix = 28 mas). North is in a different direction in each panel.}
\label{fig:WFPC2_random_errors}
\end{figure*}

The data were analyzed with similar techniques as used for the first
two ACS epochs. A discussion of the analysis and results for a subset
of the fields was presented in \cite{NK3}. In
Figure~\ref{fig:WFPC2_random_errors} we show the position of the
quasar over time, in a reference frame based on the 1st epoch images,
for 4 randomly chosen LMC fields. The scatter among positions within
the same epoch shows that the RMS error in the position of the quasar
is roughly 3 times as large for WFPC2 as for ACS ($\sim 0.021$ HRC
pixels versus 0.008 HRC pixels, respectively). The lower accuracy for
WFPC2 is due to a combination of several factors. These include the
larger pixel size, lower sensitivity, and less well calibrated and
stable geometric distortions for WFPC2 compared to ACS/HRC.

The 40 fields for which WFPC2 data were obtained fall in three
categories:

\noindent {\bf (1)} 14 Fields for which two epochs of ACS data exist,
but no WFC3 data. For these fields the WFPC2 data provide a useful
consistency check, as evident from
Figure~\ref{fig:WFPC2_random_errors}. The WFPC2 data more than double
the time baseline and fall on the extrapolation of the ACS measured
PMs. The use of a different instrument and the addition of a third
epoch rule out a variety of potential systematic errors. However, as
discussed in \cite{NK3}, the WFPC2 data help very little to reduce the
random PM errors in each field.

\noindent {\bf (2)} 13 Fields for which ACS data exist, as well as one
epoch of WFC3 data. For these fields the WFC3 data provide superior
accuracy and a longer time baseline. Hence, the WFPC2 data do not
provide anything in terms of additional insights or reduced PM errors.

\noindent {\bf (3)} 13 Fields for which only a first epoch of ACS data
exists. For 11 of these fields the data provide the opportunity to
determine a PM where none yet exists (2 fields are unsuitable as
discussed in Appendix B). This might in principle help to obtain
better spatial coverage of measurements within the LMC and
SMC. However, we have found that the PM accuracy that can be achieved
for these fields is relatively low. The random PM errors for 5-year
ACS-WFPC2 measurements are a factor 2--4 higher than for 2-year
ACS-ACS measurements.

These considerations indicate that the WFPC2 data are of insufficient
quality to help better constrain the analysis presented in this paper.
For this reason, we have not attempted to analyze the WFPC2 data to
the same level of detail and completion as for the other instruments,
and we do not report and use the WFPC2 results in this paper.

\section{Observed Fields Unsuitable for PM Determination}

We found that two QSO fields observed with WFC3 proved unsuitable for
PM determination of the Magellanic Clouds. These fields were not part
of the K1 or K2 sample, but only had one epoch of previous ACS data
(which had never been fully analyzed). We briefly discuss here the
reasons for excluding these fields from the sample for the present
paper.

\noindent {\bf RX J0524.0-7011:} This ROSAT X-Ray source at $(\alpha,
\delta) = (05h 24m 02.3s, -70d 11m 09s)$ was optically identified as
an AGN at $z=0.15$ in the LMC background by \cite{Cowley84}.  It was
cross-identified as MACHO ID 006.07059.0207 by \cite{Geha03}.  It has
the lowest redshift in our sample, and is also relatively bright, with
$V=17.7$. In our WFC3 images the AGN is clearly resolved, and a spiral
host galaxy is visible surrounding a point-like nucleus. The
astrometric techniques used in this paper are not designed for
extended sources, and application of these techniques was found not to
produce reliable PM results for this target. We therefore excluded it
from the sample.

\noindent {\bf RX J0111.7-7250:} This ROSAT X-Ray source was included
by \cite{Tinney97} in a study aimed at finding QSOs behind the
SMC. They aligned an optical spectrograph slit along the two bluest
objects within a $\sim 20''$ circle (believed to be a good guess for
the ROSAT positional error) around the source. Neither of the objects
turned out to be a background AGN. However, they detected a
serendipitous background emission-line object along the slit at
redshift $z=0.197$. The object was called QJ0111-7249, and identified
as an optical QSO. This seems tentative at best, given that the
spectrum showed only Balmer emission lines (which leaves open the
possibility for alternative categorizations). The emission-line object
was $23''$ from the ROSAT position at the time, only marginally
consistent within the uncertainties. Subsequently, the X-ray source
was included as \#348 in the ROSAT PSPC catalogue of X-ray sources in
the SMC region compiled by \cite{Haberl00}. They provided an
improved position of $(\alpha, \delta) = (01h 11m 40.9s -72d 50m 28s)$
(J2000.0), with a 90\% confidence random uncertainty of $4.1''$ and an
additional systematic uncertainty of $\sim 7''$. This position is
$41''$ from QJ0111-7249, and the sources therefore are not
associated. Nonetheless, most literature has continued to treat the
sources as one and the same. Our \textit{HST} observations were centered on
QJ0111-7249. Hence, the actual X-ray source did not fall in the
ACS/HRC field of view. Our PM analysis of the field did not reveal any
point-like source with a reflex motion typical of a background
reference source. Therefore, we conclude that QJ0111-7249 is not
suitable for SMC PM determination, and exclude it from the
sample. Interestingly, this same source was used by \cite{Costa11} in
their ground-based study of the SMC PM. They obtained a very discrepant
motion for this SMC field, $(\mu_W, \mu_N) = (-0.06, -0.71) \masyr$,
compared to their average based on the other fields, $(-0.93,
-1.25) \masyr$. This supports our interpretation, although
\cite{Costa11} argue it may be due to streaming motions in the SMC.

Both of the targets discussed in this appendix are atypical in our
sample, given that they were selected as previously known X-ray
sources. The large majority of our sample studied in this paper is
composed of QSOs from \cite{Geha03}, which were identified from their
variability characteristics in the MACHO database, with subsequent
spectroscopic follow-up.

\acknowledgments We would like to thank Lars Hernquist, Mike
Boylan-Kolchin, Marla Geha, Ana Bonaca, Luis Vargas, Nhung Ho, and
Jeremy Bradford for useful discussions about this project as a
whole. William van Altena, Dana Dinescu and Terry Girard provided
valuable insight on per-field errors versus final errors for the
center-of-mass motion. Support for this work was provided by NASA
through grants associated with projects GO-11730 and GO-11201 from the
Space Telescope Science Institute (STScI), which is operated by the
Association of Universities for Research in Astronomy, Inc., under
NASA contract NAS 5-26555. GB acknowledges support from NASA through
Hubble Fellowship grant HST-HF-51284.01-A.


\newpage

\begin{deluxetable}{llllllcccccccccc}
\tabletypesize{\scriptsize}
\tablewidth{0pt}
\tablecolumns{16}
\tablecaption{LMC Observations and PM Results}
\tablehead{\colhead{ID} & \colhead{RA} & \colhead{DEC}  & \colhead{$T$} & 
\colhead{$\Delta time$} & \colhead{$N$} & \multicolumn{8}{c}{PM of FIELD AS OBSERVED} & \multicolumn{2}{c}{LMC PM(CM) estimate}\\
 &  &  &  &  &  & $\mu_W$ & $\mu_N$ & $\delta\mu_W$ & $\delta\mu_N$ &  $\mu_W$ & $\mu_N$ & $\delta\mu_W$ & $\delta\mu_N$ & $\mu_W$ & $\mu_N$\\
 &  &  &  &  &  & \multicolumn{4}{c}{2-epoch PMs} & \multicolumn{4}{c}{3-epoch PMs} & & \\
 & (H,M,S) & (deg, ', '') & (min) & (yr) & & \multicolumn{4}{c}{($\masyr$)} & \multicolumn{4}{c}{($\masyr$)} & \multicolumn{2}{c}{($\masyr$)}}
\startdata
L1  &  5 47 50.2 &  -67 28 1.3  &  2.6  & 7.1 & 8   & -1.491 & 0.689    & 0.076 & 0.069   & -1.604  & 0.714  & 0.029   & 0.032  & -1.820 & 0.209\\
L2  &  4 46 11.1 &  -72 5 9.0   &  6.6  & 7.7 & 13  & -2.163 &  -0.246  & 0.061 & 0.054   & -2.051  & -0.235 & 0.018   & 0.027  & -1.910 & 0.181 \\
L3  &  5 16 28.9 &  -68 37 1.8  &  6.6  & 7.1 & 45  & -1.827 & 0.391    & 0.047 & 0.042   & -1.798  & 0.320  & 0.014   & 0.016  & -1.924 & 0.252\\
L4  &  5 16 36.8 &  -66 34 35.8 &  12.6 & 7.6 & 9   & -1.781 & 0.196    & 0.083 & 0.093   & -1.734  & 0.147  & 0.048   & 0.030  & -1.974 & 0.100\\
L5  &  5 15 36.1 &  -70 54 0.8  &   -   & 1.9 & 41  & -2.008 & 0.112    & 0.079 & 0.079   & -        & -       & -        & -       & -1.767 & 0.143\\
L6  &  6 2 34.3  &  -68 30 41.1 &  13.3 & 7.2 & 8   & -1.750 & 0.897    & 0.158 & 0.177   & -1.664  & 1.058  & 0.045   & 0.055  & -1.873 & 0.392\\
L7  &  5 16 26.3 &  -69 48 19.0 &  -    & 1.8 & 127 & -1.942 &  0.226   & 0.053 & 0.053   & -        & -       & -        & -       & -1.824 & 0.207\\
L8  &  5 0 54.0  &  -66 43 59.8 &  -    & 2.0 & 39  & -1.883 & -0.0232  & 0.062 & 0.062   & -        & -       & -        & -       & -2.068  & 0.247\\
L9  &  4 53 56.5 &  -69 40 35.4 &  -    & 2.0 & 73  & -1.985 &  -0.058  & 0.079 & 0.079  & -        & -       & -        & -       & -1.892 & 0.394\\
L10 &  5 49 41.6 &  -69 44 15.1 &  -    & 1.9 & 65  & -1.704 &  0.648   & 0.087 & 0.087   & -        & -       & -        & -       & -1.802 & 0.062\\
L11 &  5 57 22.4 &  -67 13 21.5 &  2.9  & 6.6 & 8   & -1.591 &  0.895   & 0.174 & 0.078   & -1.664  & 0.814 & 0.041   & 0.031 & -1.899 & 0.233\\
L12 &  5 1 46.7  &  -67 32 39.8 &  13.3 & 6.4 & 16  & -1.903 & -0.201   & 0.083 & 0.121   & -1.817  & -0.065 & 0.033   & 0.020  & -1.980 & 0.231\\
L13 &  5 20 57.0 &  -70 24 52.6 &  -    & 1.5 & 85  & -2.199 & 0.543    & 0.093 & 0.093   & -        & -       & -        & -       & -1.980 & 0.409\\ 
L14 &  5 2 53.7  &  -67 25 45.0 &  13.3 & 7.1 & 18  & -1.664 & -0.091   & 0.108 & 0.140   & -1.707  & -0.054 & 0.034   & 0.031  & -1.885 & 0.212\\
L15 &  5 14 12.1 &  -70 20 25.8 &  -    & 1.1 & 75  & -2.512 & 0.553    & 0.127 & 0.127   & -        & -       & -        & -       & -2.273 & 0.628\\
L16 &  5 4 36.0  &  -66 24 15.7 &  10.7 & 7.1 & 13  & -1.561 &  0.166   & 0.063 & 0.077   & -1.684  & 0.079  & 0.031   & 0.035  & -1.890 & 0.270\\
L17 &  5 25 14.4 &  -65 54 45.7 &  -    & 1.2 & 17  & -1.773 &  1.247   & 0.290 & 0.290   & -        & -       & -        & -       & -2.009 & 1.081\\
L18 &  5 30 26.8 &  -66 48 52.9 &  -    & 1.6 & 22  & -1.767 &  0.716   & 0.150 & 0.150   & -        & -       & -        & -       & -2.001 & 0.434\\
L19 &  5 31 59.7 &  -69 19 51.5 &  -    & 1.1 & 114 & -1.636 &  0.780   & 0.150 & 0.150   & -        & -       & -        & -       & -1.706 & 0.340\\
L20 &  4 56 14.3 &  -67 39 9.0  &  -    & 2.5 & 48  & -1.826 &  -0.139  & 0.057 & 0.057   & -        & -       & -        & -       & -1.939 & 0.260\\
L21 &  5 10 32.5 &  -69 27 15.5 &  -    & 1.1 & 129 & -2.042 &  -0.113  & 0.118 & 0.118   & -        & -       & -        & -       & -1.964 & 0.043\\
L22 &  5 20 56.5 &  -65 39  4.8 &  17.7 & 6.9 & 23  & -       &  -        & -      & -        & -1.681  & 0.298  & 0.020   & 0.036  & -1.914 & 0.206\\  
\enddata
\tablecomments{The LMC field number, using the same notation as in
   K1, and RA/DEC for the QSOs (columns 1, 2 \& 3). Column 4
  lists the total exposure times in minutes for e3. Column 5 lists the time
  baseline, in years, between e1 and e3. Column 6 lists the number of
  stars used in the final transformations after all cuts and
  iterations have been applied. Columns 7--10 list the observed PMs
  and errors for the ACS reanalysis (see $\S$\ref{sec:ACS}).  Columns
  11--14 list the corresponding observed PMs and errors obtained from
  the 3-epoch analysis (see $\S$\ref{sec:WFC3-ACS}). Columns 15 \& 16
  list the final COM PM estimates from each QSO field obtained from
  the procedure described in $\S$\ref{subsec:LMCPM}. Note that L22 is
  a new field, with only e1 and e3 data.}
\label{tab:LMCobs}
\end{deluxetable}


\begin{deluxetable}{llllllcccccccccc}
\tabletypesize{\scriptsize}
\tablewidth{0pt}
\tablecolumns{16}
\tablecaption{SMC Observations and PM Results}
\tablehead{\colhead{ID} & \colhead{RA} & \colhead{DEC}  & \colhead{$T$} & 
\colhead{$\Delta time$} & \colhead{$N$} & \multicolumn{8}{c}{PM of FIELD AS OBSERVED} & \multicolumn{2}{c}{SMC PM(CM) estimate}\\
 &  &  &  &  &  & $\mu_W$ & $\mu_N$ & $\delta\mu_W$ & $\delta\mu_N$ &  $\mu_W$ & $\mu_N$ & $\delta\mu_W$ & $\delta\mu_N$ & $\mu_W$ & $\mu_N$\\
 &  &  &  &  &  & \multicolumn{4}{c}{2-epoch PMs} & \multicolumn{4}{c}{3-epoch PMs} & & \\
 & (H,M,S) & (deg, ', '') & (min) & (yr) & & \multicolumn{4}{c}{($\masyr$)} & \multicolumn{4}{c}{($\masyr$)} & \multicolumn{2}{c}{($\masyr$)}}
\startdata
S1 & 00 51 17.0 &  -72 16 51.3  & -    & 1.9 & 42  & -0.682 & -1.288  & 0.100  &  0.100  & -         &  -        & -      &  -      & -0.738  & -1.244\\
S2 & 00 55 34.7 &  -72 28 33.9  & 17.7 & 7.6 & 25  & -0.748 & -1.246  & 0.070  &  0.068  & -0.722   &  -1.214  & 0.032 &  0.024 & -0.760  & -1.185\\
S3 & 01 02 14.5 &  -73 16 26.6  & 17.7 & 7.7 & 36  & -0.893 & -1.397  & 0.097  &  0.101  & -0.679   &  -0.974  & 0.026 &  0.028 & -0.691  & -0.973\\
S4 & 00 36 39.7 &  -72 27 42.0  & -    & 2.8 & 10  & -0.460 & -1.114  & 0.109  &  0.109  & -         &  -        & -      &  -      & -0.579  & -1.031\\
S5 & 01 02 34.7 &  -72 54 23.8  & 13.3 & 6.8 & 30  & -1.046 & -1.072  & 0.084  &  0.083  & -0.806   &  -1.199  & 0.017 &  0.038 & -0.815   & -1.195\\
\enddata
\tablecomments{Same columns as in Table~\ref{tab:LMCobs} but for the SMC.}
\label{tab:SMCobs}
\end{deluxetable}

\begin{deluxetable}{lcccc}
\tabletypesize{\small}
\tablewidth{0pt}
\tablecolumns{5}
\tablecaption{Kinematical determinations of LMC geometry and rotation parameters}
\tablehead{
\colhead{} & \colhead{} & \colhead{Paper II} & \colhead{vdM02} & \colhead{Olsen11}\\
\colhead{(1)} & \colhead{(2)} & \colhead{(3)} & \colhead{(4)} & \colhead{(5)}}
\startdata
inc                 &  deg    &  $ 39.6 \pm  4.5$                    & $34.7 \pm 6.2$\tablenotemark{[1]}    &   $34.7 \pm 6.2$\tablenotemark{[1]}\\
theta               &  deg    &  $147.4 \pm  10.0$                   & $129.9 \pm 6.0$                     &   $142  \pm 5$\\
RA                  &  deg    &  $78.76 \pm  0.52$                   & $81.91 \pm 0.98$                    &   $81.91 \pm 0.98$\tablenotemark{[2]}\\
DEC                 &  deg    &  $-69.19 \pm  0.25$                  & $-69.87 \pm 0.41$                   &   $-69.87 \pm 0.41$\tablenotemark{[2]}\\
$R_0$\tablenotemark{[3]} &  kpc    &  $1.18 \pm  0.48$                    &   $4.0 \pm 0.3$\tablenotemark{[5]}             &   $2.4 \pm 0.1$\\
$V_0$\tablenotemark{[3]} &  $\kms$ &  $76.1 \pm  7.6$\tablenotemark{[4]}                     &   $49 \pm 2$\tablenotemark{[5]}             &   $87 \pm 5$\tablenotemark{[6]}\\ 
$v_{sys}$                &  $\kms$ &  $262.2 \pm  3.4$\tablenotemark{[7]}  & $262.2 \pm 3.4$                     &   $263 \pm 2$\\
$m-M$               &  mag    &  $18.50 \pm  0.10$\tablenotemark{[8]} & $18.50 \pm  0.10$\tablenotemark{[8]} &   $18.50 \pm 0.10$\tablenotemark{[8]}\\
\enddata
\tablecomments{Column~(1) lists the following quantities: inclination of the
LMC disk plane; position angle of the line of nodes; (RA,DEC) of the
rotation center; turnover radius $R_0$ and asymptotic velocity amplitude
$V_0$ of the rotation curve; systemic LOS velocity, $v_{sys}$; and distance
modulus, $m-M$. Column~(2) lists the corresponding units. Column~(3) lists
the values inferred from the new PM data as described in
Paper~II. Columns~(4) and (5) list the values inferred from the LOS
velocity studies of vdM02 and \cite{Olsen11}.}
\tablenotetext{[1]}{Value from \cite{vdM01}, used but not
independently determined by vdM02 or \cite{Olsen11}.}
\tablenotetext{[2]}{Value from vdM02, used but not independently determined by \cite{Olsen11}.}
\tablenotetext{[3]}{The rotation curve is parameterized so that it rises linearly to
velocity $V_0$ at radius $R_0$, and then stays flat at larger radii.}
\tablenotetext{[4]}{Applies to a mix of stellar populations (each with different
asymmetric drift) as shown in Figure~6 of K1.}
\tablenotetext{[5]}{Applies to (old) carbon stars
(with large asymmetric drift). Determined from Table~2 of vdM02. Used
an LMC COM PM value that pre-dates the now current HST values.}
\tablenotetext{[6]}{Applies to (young) red
supergiants (with little asymmetric drift). Used the LMC COM PM from
P08.}
\tablenotetext{[7]}{Value from vdM02, used but not independently determined in Paper~II.}
\tablenotetext{[8]}{Value from \cite{Freedman01}, used but not independently
determined from kinematics.}
\label{tab:fit}
\end{deluxetable}

\begin{deluxetable}{lccccl}
\tabletypesize{\scriptsize}
\tablewidth{0pt}
\tablecolumns{6}
\tablecaption{Summary of Recent LMC and SMC PM Measurements}
\tablehead{
\colhead{} & \multicolumn{2}{c}{LMC} & \multicolumn{2}{c}{SMC} & \colhead{}\\
\colhead{} & \colhead{$\mu_W$} & \colhead{$\mu_N$} & \colhead{$\mu_W$} & \colhead{$\mu_N$} & 
\colhead{}\\
\colhead{Work} & \colhead{$(\masyr)$} & \colhead{$(\masyr)$} & \colhead{$(\masyr)$} & 
\colhead{$(\masyr)$} & \colhead{Data}}
\startdata
{\bf This Paper} & {\bf -1.910} $\pm$ {\bf 0.020} & {\bf 0.229} $\pm$ {\bf 0.047} & {\bf -0.772} $\pm$ {\bf 0.063} & {\bf -1.117} $\pm$ {\bf 0.061} & {\bf HST 3-epoch}\\
{\it \qquad (vdM02 model)} & $-1.899 \pm 0.017$ & $0.416 \pm 0.017$ & $\dots$ &$\ldots$ & HST 3-epoch\\
K1 \& K2 & $-2.03 \pm 0.08$ & $0.44 \pm 0.05$ & $-1.16 \pm 0.18$ & $-1.17 \pm 0.18$ & HST 2-epoch\\
P08 & $-1.956 \pm 0.036$ & $0.435 \pm 0.036$ & $-0.754 \pm 0.061$ & $-1.252 \pm 0.058$ & HST 2-epoch\\
Costa et al. & $-1.72 \pm 0.13$ & $0.50 \pm 0.15$ & $-0.93 \pm 0.14$ & $-1.25 \pm 0.11$ & 2.5m du Pont\\
Vieira et al. (2010) &  $-1.89 \pm 0.27$ & $0.39 \pm 0.27$ & $-0.98 \pm 0.30 $ & $-1.10 \pm 0.29$ & SPM\\
\enddata

\tablecomments{PMs of the LMC (columns 3 and 4) and SMC (columns 5 and
  6) COM. Column~(1) indicates the source of the result, and
  column~(6) the type of data that was used. The first line (labeled
  ``This Paper'') is the final result from the present paper, which
  uses the fit to the PM rotation field from Paper II. Ours is the
  first study that propagates the uncertainties in the geometry,
  center and rotation of each Cloud into the COM PM estimates. For
  this reason, other studies have generally underestimated their error
  bars. The second line shows the LMC COM PM estimate that is obtained
  when the LMC orientation and center are are kept fixed to the same
  vdM02 values that were used by most other authors (see
  Section~\ref{subsec:compareHST}). For Costa et al. we list for the
  LMC the average of their 2009 results for $V_{\rm rot} = 50$ km/s
  and 120 km/s, respectively; for the SMC we list their 2011 result.}
\label{tab:PMs}
\end{deluxetable}

\begin{deluxetable}{lcclcccccc}
\tabletypesize{\scriptsize}
\tablewidth{0pt}
\tablecolumns{10}
\tablecaption{Galactocentric Velocities from \textit{HST} measurements}
\tablehead{\colhead{Line} & \colhead{Galaxy} & \colhead{solar} & \colhead{PM} &
\colhead{$v_X$} & \colhead{$v_Y$} & \colhead{$v_Z$} & \colhead{$v_{\rm tot}$} &
\colhead{$v_{\rm rad}$} & \colhead{$v_{\rm tan}$}\\
 & & & & \colhead{km/s} & \colhead{km/s} & \colhead{km/s} & \colhead{km/s} & \colhead{km/s} & \colhead{km/s}}
\startdata
{\bf (1)}  & {\bf LMC} & {\bf new} & {\bf 3-epoch} & {\bf -57} $\pm$ {\bf 13} & {\bf -226} $\pm$ {\bf 15} & {\bf 221} $\pm$ {\bf 19} & {\bf 321} $\pm$ {\bf 24} & {\bf 64} $\pm$ {\bf 7} & {\bf 314} $\pm$ {\bf 24}\\
(2)  &  LMC    &  IAU  & 3-epoch       &  $-59 \pm 12$ &  $-252 \pm 15$ &  $221  \pm 19$ &  $340 \pm 23$  & $86  \pm 5$  &  $329 \pm 24$\\
(3)  &  LMC    &  new  & 3-epoch vdM02 &  $-77 \pm 8$  &  $-224 \pm 14$ &  $227  \pm 18$ &  $328 \pm 23$  & $65  \pm 5$  &  $322 \pm 24$\\
(4)  &  LMC    &  IAU  & 3-epoch vdM02 &  $-78 \pm 8$  &  $-250 \pm 14$ &  $227  \pm 18$ &  $347 \pm 23$  & $87  \pm 5$  &  $336 \pm 24$\\
(5)  &  LMC    &  IAU  & P08           &  $-83 \pm 11$ &  $-258 \pm 15$ &  $238  \pm 20$ &  $361 \pm 25$  & $88  \pm 5$  &  $350 \pm 26$\\
(6)  &  LMC    &  IAU  & K1            &  $-86 \pm 14$ &  $-268 \pm 18$ &  $252  \pm 25$ &  $378 \pm 31$  & $89  \pm 5$  &  $367 \pm 31$\\
 & \\
{\bf (7)} & {\bf SMC} & {\bf new} & {\bf 3-epoch} &  {\bf 19} $\pm$ {\bf 18} & {\bf -153} $\pm$ {\bf 21} & {\bf 153} $\pm$ {\bf 17} & {\bf 217} $\pm$ {\bf 26} & {\bf -11} $\pm$ {\bf 5} & {\bf 217} $\pm$ {\bf 26}\\
(8)  &  SMC    &  IAU  & 3-epoch       &  $18   \pm 17$ &  $-179 \pm 21$ &  $153  \pm 17$ &  $236 \pm 26$  & $6  \pm 4$  &  $236 \pm 26$\\
(9)  &  SMC    &  IAU  & P08           &  $23   \pm 16$ &  $-197 \pm 22$ &  $166  \pm 17$ &  $259 \pm 26$  & $7  \pm 4$  &  $259 \pm 26$\\
(10) &  SMC    &  IAU  & K2            &  $-86  \pm 49$ &  $-248 \pm 46$ &  $150  \pm 39$ &  $302 \pm 57$  & $23 \pm 7$  &  $301 \pm 57$\\
 & \\
{\bf (11)} & {\bf SMC-LMC} & {\bf ...} & {\bf 3-epoch} & {\bf 76} $\pm$ {\bf 22} & {\bf 73} $\pm$ {\bf 26} & {\bf -68} $\pm$ {\bf 25} & {\bf 128} $\pm$ {\bf 32} & {\bf 112} $\pm$ {\bf 32} & {\bf 61} $\pm$ {\bf 16}\\
(12) & SMC-LMC & ...   & 3-epoch vdM02 &  $81  \pm 19$ &  $73 \pm 26$   &  $-89 \pm 25$  &  $143 \pm 31$  & $134 \pm 32$ &  $50 \pm 15$ \\
(13) & SMC-LMC & ...   & P08           &  $106 \pm 20$ &  $61   \pm 26$ &  $-72  \pm 26$ &  $145 \pm 30$  & $136 \pm 33$ &  $46  \pm 18$\\
(14) & SMC-LMC & ...   & K2            &  $0   \pm 51$ &  $20   \pm 49$ &  $-103 \pm 46$ &  $127 \pm 46$  & $77  \pm 50$ &  $88  \pm 45$\\
\enddata
\tablecomments{The lines (1)--(6) list the LMC velocity, lines (7)--(10) the SMC
velocity, and lines (11)--(14) the relative velocity of the SMC with
respect to the LMC. Column~(1) lists a line identifier. Column~(2)
lists the galaxy name. Column~(3) lists the source of the velocities
used to correct for solar reflex motion: ``IAU'' uses the IAU value of
$V_0 = 220 \kms$ and the \cite{Dehnen98} solar peculiar
velocity; ``new'' uses the improved \cite{McMillan11} value of $V_0 =
239 \pm 5\kms$ and the improved \cite{Schonrich10} 
solar peculiar velocity. Column~(4) lists the assumed PM value, taken
from the list of observations in Table~\ref{tab:PMs}: ``3-epoch''
corresponds to the line labeled ``This Paper'' (which uses the new
data with the LMC geometry fit from Paper~II, with uncertainties
included); ``3-epoch vdM02'' corresponds to the line labeled ``(vdM02
model)'' (which uses the new data with the fixed geometry parameters
from vdM02). Columns~(5)--(7) list the Galactocentric velocity
coordinates $(v_X, v_Y, v_Z)$. Columns~(8)--(10) lists the total
length of the velocity vector, the radial component, and the
transverse component, respectively. Uncertainties were calculated
using a Monte-Carlo scheme that propagates all relevant uncertainties
in the position and velocity of both the Clouds and the Sun. Distance
uncertainties are based on $\Delta m-M = 0.1$. Velocity uncertainties
in the Galactocentric frame are highly correlated, because
uncertainties in the LOS direction are much smaller than in the
transverse direction.}
\label{tab:vels}
\end{deluxetable}

\begin{deluxetable}{cccccc}
\tablewidth{0pt}
\tablecolumns{6}
\tablecaption{MW and LMC Properties}
\tablehead{\colhead{Galaxy} & \colhead{Total Mass} & \colhead{$c_{\rm vir}$} & \colhead{$R_{\rm vir}$}
& \colhead{$M_{\rm disk}$} & {Plummer Softening}\\
\colhead{} & \colhead{[$M_{\odot}$]} & \colhead{} & \colhead{[kpc]} & \colhead{[$M_{\odot}$]} & \colhead{[kpc]}}
 \startdata
MW  & $1\times10^{12}$   & 9.86 & 261 & $6.5\times10^{10}$ & -\\
MW  & $1.5\times10^{12}$ & 9.56 & 299 & $5.5\times10^{10}$ & -\\
MW  & $2\times10^{12}$   & 9.36 & 329 & $5.0\times10^{10}$ & -\\
LMC & $3\times10^{10}$   & -    & -   & -      & 8\\
LMC & $5\times10^{10}$   & -    & -   & -      & 11\\
LMC & $8\times10^{10}$   & -    & -   & -      & 14\\
LMC & $1\times10^{11}$   & -    & -   & -      & 15\\
LMC & $1.8\times10^{11}$ & -    & -   & -      & 20\\
LMC & $2.5\times10^{11}$ & -    & -   & -      & 22.5\\
\enddata
\tablecomments{Different MW and LMC mass models used in the orbital
  calculations. $c_{\rm vir}$ is the halo concentration, and $R_{\rm
    vir}$ is the virial radius of the MW in each case. The
  mass of the MW bulge is kept fixed at $1\times10^{10} M_{\odot}$ and the
  Hernquist scale radius for the MW bulge is kept fixed at 0.7 kpc.  The
  MW's exponential disk scale radius is kept fixed at 3.5 kpc. The mass of
  the MW disk is varied to get the observed circular velocity at the
  solar circle as discussed in the text.}
\label{tab:MWLMCparams}
\end{deluxetable}


\end{document}